\documentclass[12pt]{article}
\usepackage[T1]{fontenc}
\usepackage[utf8]{inputenc}
\usepackage[english]{babel}
\usepackage{geometry, booktabs}
\usepackage{setspace}
\usepackage{amsmath,amssymb,mathtools}
\usepackage{bm}
\usepackage{graphicx}
\usepackage{microtype}
\usepackage{enumitem}
\usepackage{xcolor}
\usepackage{hyperref}
\hypersetup{colorlinks=true,linkcolor=blue,citecolor=blue,urlcolor=blue}
\usepackage{times}
\usepackage{natbib}

\usepackage{amssymb}

\usepackage{bbm}

\usepackage{amsmath}
\usepackage{amsmath}
\usepackage{graphicx}
\usepackage{subcaption}
\usepackage{tabularx}
\usepackage{booktabs}
\usepackage{placeins}

\newtheorem{remark}{Remark}
\newtheorem{corollary}{Corollary}
\newtheorem{proposition}{Proposition}
\newtheorem{proof}{Proof}

\newtheorem{theorem}{Theorem}[section]
\newtheorem{lemma}[theorem]{Lemma}

\newcommand{\indicator}{\mathbf{1}}
\newcommand{\indic}{\indicator}

\newcommand{\prob}{\mathbb P}
\renewcommand{\P}{\prob}
\newcommand{\Mmax}{M_{\max}}

\begin{document}

\title{Confidence intervals for maximum unseen probabilities, with application to sequential sampling design}

\author{
\Large{Alessandro Colombi} \\
Bocconi Institute for Data Science and Analytics, \\
Bocconi University \\
\Large{Mario Beraha} \\
Department of Economics, Management and Statistics, \\
University of Milano-Bicocca\\
\Large{Amichai Painsky} \\
School of Industrial Engineering and 
Depertment of Statistics and Operations Research, \\
Tel Aviv University\\
\Large{Stefano Favaro}\\
Department of Economics and Statistics, University of Torino and\\ Collegio Carlo Alberto,
}

\date{}



\maketitle

\begin{abstract}
Discovery problems often require deciding whether additional sampling is needed to detect all categories whose prevalence exceeds a prespecified threshold. We study this question under a Bernoulli product (incidence) model, where categories are observed only through presence--absence across sampling units. Our inferential target is the \emph{maximum unseen probability}, the largest prevalence among categories not yet observed. We develop nonasymptotic, distribution-free upper confidence bounds for this quantity in two regimes: bounded alphabets (finite and known number of categories) and unbounded alphabets (countably infinite under a mild summability condition). We characterise the limits of data-independent worst-case bounds, showing that in the unbounded regime no nontrivial data-independent procedure can be uniformly valid. We then propose data-dependent bounds in both regimes and establish matching lower bounds demonstrating their near-optimality. We compare empirically the resulting procedures in both simulated and real datasets. Finally, we use these bounds to construct sequential stopping rules with finite-sample guarantees, and demonstrate robustness to contamination that introduces spurious low-prevalence categories.
\end{abstract}


\section{Introduction}

\subsection{Background and motivation}

Missing probabilities, the probability mass assigned to as-yet-unobserved outcomes, are central inferential objects in many domains, including ecology \citep{Coverage2014}, optimization \citep{dick2014how}, genomics \citep{ionita2009estimating, masoero2022more}, security analysis \citep{bubeck2013optimal}, debugging \citep{ChaoMaYang1993}, and language modeling \citep{noorani2025conformal}. A statistical question in these settings is \emph{when to stop sampling}: when does additional effort cease to be justified by the information gained on rare or unobserved symbols? Stopping decisions are often guided by coverage-based diagnostics, species-accumulation curves, or rules of thumb. For instance, in ecology it is common to rely on coverage-based rarefaction and extrapolation to standardise sampling effort, and to deem a sample ``complete enough'' once estimated coverage exceeds a fixed threshold or the species-accumulation curve has flattened \citep{ChaoJost2012,Beguinot2015,Montes2021}. In clinical safety and reliability, zero-event rules argue that an unobserved adverse event must be sufficiently rare \citep{HanleyLippmanHand1983}. Similarly, in recapture debugging of computer code, stopping rules balance the expected number of remaining bugs against testing cost \citep{ChaoMaYang1993}. In uncertainty quantification for language models, \cite{noorani2025conformal} asked how many times a query should be repeated to observe all answers the model can generate, casting it as an optimal stopping rule driven by missing probabilities.

Such stopping rules provide useful guidance in practice, but their justification is typically asymptotic or oracle-based, and their performance can be fragile in the presence of contamination mechanisms that generate many artificial low-prevalence symbols. For example, in high-throughput sequencing, PCR artefacts can produce large numbers of rare OTUs (operational taxonomic units), thereby inflating the apparent tail of the distribution \citep{Schloss2011Artifacts}. Similarly, in metabarcoding studies, barcode/tag misidentification and index swapping can misassign reads across samples, yielding many artefactual taxa that appear only a handful of times \citep{Eisenhofer2019Contamination}.

In this paper we focus on settings where the basic observation for each category is an \emph{incidence} indicator across sampling units (patients, sites, or samples), that is, for each symbol we only record presence or absence within each unit. This setting arises naturally in several applications: in high-throughput genomics and somatic mutation studies, data record whether a given mutation is present in each patient \citep{ionita2009estimating}, with each patient presenting multiple mutations; in community ecology and biodiversity monitoring, presence--absence matrices are a standard format for site-by-species data \citep{Magurran2004, ghilotti2025bayesian, stolf2025infinite}; and in user-activity modelling for online A/B testing, one records whether each user performs certain actions \citep{beraha2025online}. A convenient formalisation is provided by a \emph{Bernoulli product model}: for each symbol $j \in \{1,\ldots,M\}$ and each sampling unit $i \in \{1,\ldots,n\}$, we observe independent Bernoulli random variables $X_{ij}$ indicating presence or absence of symbol $j$ in unit $i$. The model is completed by assuming that each symbol has an incidence probability $p_j$, constant across sampling units. We allow $M \leq \infty$, where the extreme case $M=\infty$ represents the possibility of countably many symbols; in the aforementioned contexts, the underlying alphabet is large or effectively unbounded, and contamination can generate many low-prevalence artefactual symbols.

\subsection{Preview of our contributions}

A natural design principle for stopping rules under incidence sampling is to fix a prevalence threshold $\varepsilon>0$, representing the minimal effect size or frequency of interest, and to stop once all symbols with prevalence at least $\varepsilon$ have been observed at least once. Letting $N_j=\sum_{i=1}^n X_{ij}$ be the count of symbol~$j$, we seek an upper bound $U_n=U_n(\alpha)$ such that, with probability at least $1-\alpha$,
\begin{equation}\label{eq:joint_interval}
    p_j \le U_n(\alpha) \quad \text{for every symbol } j \text{ with } N_j = 0,
\end{equation}
and such that $U_n(\alpha)\le \varepsilon$. Equivalently, we seek a simultaneous post-selection upper confidence bound for all unobserved probabilities. To formalise this goal, define the \emph{maximum unseen probability}
\begin{equation}\label{eq:mmax_def}
    M_{\max} = M_{\max}(N) := \max_{j : N_j = 0} p_j,
\end{equation}
namely the largest probability among symbols not observed in a sample of size $n$. Then $\Mmax \le U_n(\alpha)$ is equivalent to \eqref{eq:joint_interval}.

The main contribution of this paper is a non-asymptotic, distribution-free analysis of confidence intervals for the maximum unseen probability $M_{\max}$ in the Bernoulli product model. We develop a unified theory across two complementary regimes: a bounded-alphabet regime, where the total number of categories $M$ is finite and known, and an unbounded-alphabet regime, where the alphabet is countable and potentially infinite, under a mild summability condition on the prevalences.

In the bounded-alphabet regime, where $M$ is known, a multiplicity-adjusted \emph{rule of three} suggests that $\log(M/\alpha)/n$ is a valid upper confidence bound for $M_{\max}$ at level $1-\alpha$. While appealingly simple, this construction is conservative and does not exploit the realised pattern of observed and unobserved categories. We first show how to improve upon this benchmark in a worst-case, data-independent manner. We then introduce a data-dependent upper confidence bound that adapts to the empirical incidence counts and yields a further tightening. Finally, we establish a matching lower bound, showing that the resulting rate is minimax-optimal over bounded alphabets.

We then turn to unbounded alphabets under a mild summability assumption on the prevalences. In this regime we first establish an impossibility result: no non-trivial worst-case (data-independent) upper bound for $M_{\max}$ can attain valid coverage uniformly over the model class. This motivates the development of data-dependent procedures. We propose such an upper confidence bound, based on a suitable smoothing of the maximum functional, and show that it is essentially tight by proving a matching lower bound. A simulation study confirms the theoretical findings and suggests that the unbounded-alphabet procedure can outperform the bounded-alphabet construction even when the true alphabet is finite but contains many low-incidence symbols. Building on these experiments, we propose a simple rule of thumb for choosing between the two intervals in practice.

Finally, we use our confidence intervals as building blocks for principled stopping rules in sequential sampling. Given a prevalence threshold $\varepsilon$ and confidence level $1-\alpha$, we consider the procedure that stops at the first sample size $n$ such that the upper endpoint $U_n(\alpha)$ of one of our intervals satisfies $U_n(\alpha)\le \varepsilon$. We evaluate the resulting stopping rules against coverage- and extrapolation-based criteria from the ecological literature, under both clean and contaminated sampling scenarios in which spurious low-prevalence categories are introduced to mimic sequencing and amplification errors. The results indicate that our $\Mmax$-based procedures achieve the desired risk control and remain stable and robust under contamination, whereas coverage-based criteria may stop too early or require substantially more sampling effort in the presence of many artefactual rare categories.

\subsection{Related work}

The closest work to ours is \citet{Painsky2025}, who studies confidence intervals for parameters of unobserved events under multinomial sampling. He derives worst-case, dimension-free confidence bounds for unseen-category probabilities, which can be used to control extremal quantities such as $M_{\max}$, with applications to large-alphabet inference and $\ell_\infty$ distribution estimation \citep{KontorovichPainsky2025}. Our focus is instead on the Bernoulli product (incidence) model, and on deriving tighter, data-dependent bounds for $M_{\max}$ in both bounded- and unbounded-alphabet regimes; we expect that some of our techniques could also sharpen the worst-case multinomial bounds of \citet{Painsky2025}.

Stopping criteria based on coverage, richness estimators, and species-accumulation curves are widely used in ecology. In particular, coverage-based rarefaction and extrapolation \citep{ChaoJost2012} are commonly employed to standardise effort and to assess whether sampling has become sufficiently exhaustive; see also \citet{Beguinot2015,Montes2021} and references therein. Related approaches include \citet{chao2009}, who propose a method to estimate the additional sampling effort required to achieve a fixed target fraction of the asymptotic richness.

Our work is also connected to the extensive literature on the \emph{missing mass} \citep{Good1953}, and to classical tools such as the Good--Turing estimator. Missing-mass ideas play a prominent role in sequential discovery and exploration problems, including modern applications to large language models \citep{noorani2025conformal}, global optimization \citep{dick2014how}, and adaptive decision-making \citep{bubeck2013optimal}. These contributions provide global diagnostics for unseen outcomes, but do not distinguish between many ultra-rare unobserved categories and a single moderately common unseen category. Our analysis of $M_{\max}$ targets precisely this extremal aspect and yields finite-sample, distribution-free guarantees that translate directly into stopping rules.

\subsection{Other applications}

Although sequential stopping is a primary motivation, the confidence intervals for $M_{\max}$ apply whenever one needs finite-sample control on the prevalence of unseen categories. A seminal discussion is given by \citet{HanleyLippmanHand1983}, who stress that observing zero events does not justify concluding that the underlying risk is zero. This point is made concrete in the medical literature by \citet{eypasch1995probability}, who advocate reporting upper confidence bounds for complications that have not yet occurred. Such situations are common in clinical trials, diagnostic testing, and post-marketing surveillance, where safety claims are often based on the absence of observed adverse events.

These concerns have motivated a substantial literature on confidence intervals and risk bounds under zero or near-zero counts. In emergency medicine, \citet{jovanovic1997safety} emphasize that standard normal approximations can fail when events are rare or unobserved, and advocate exact binomial confidence intervals and simple rules of thumb such as the \emph{rule of three}. In comparative settings, \citet{moller2021estimating} study inference for relative risks when one study arm exhibits zero events, highlighting that valid inference is still possible but becomes delicate in small samples.

Closely related problems arise in reliability engineering, environmental risk assessment, and safety certification, where large collections of components or failure modes are monitored and zero failures are common. \citet{quigley2011estimating} emphasize that maximum likelihood assigns probability zero to unobserved failures, which is inadequate for risk-averse decision-making, and develop principled alternatives. Across these domains, the common statistical structure is inference on many rare-event probabilities for outcomes that do not appear in the sample, motivating finite-sample procedures that account for multiplicity and provide meaningful guarantees for unobserved events.

\section{Bounded alphabet sizes}
\label{sec:finite_M}
We begin with the setting where the alphabet size is known and finite. That is,
we observe independent counts $N=(N_1, \ldots, N_M)$
\[
N_j \sim \mathrm{Bin}(n,p_j), \quad j=1,\ldots,M\,,
\]
with \(M<\infty\) and unknown success probabilities \( (p_1,\ldots,p_M) \in [0,1]^M\).
For short, we call $N$ a sample of size $n$ from the Bernoulli product model with alphabet size $M$.
Recall that $M_{\max}$ defined in \eqref{eq:mmax_def}
is the maximum probability among categories that did not appear in the sample.
Our aim is to construct non-asymptotic upper confidence bounds for \(M_{\max}\),
uniformly over \((p_1,\ldots,p_M)\).

A natural benchmark is given by the multiplicity-corrected \textit{rule of three}:
for a single binomial variable with \(N \sim \mathrm{Bin}(n,p)\) and \(N=0\), the
classical upper \((1-\alpha)\)-level confidence bound (approximaetly) satisfies
\(p \leq -\log(\alpha)/n\). Applying a Bonferroni correction over
\(M\) independent coordinates yields the worst-case, data-independent bound
\begin{equation}
    \label{eqn:ROT}
    T_{\mathrm{Bonf}}(N)=\frac{\log(M/\alpha)}{n},
\end{equation}
which guarantees \(\mathbb{P}_p\left(M_{\max}(N)\le T_{\mathrm{Bonf}}(N)\right)\ge 1-\alpha\)
for all \((p_1,\ldots,p_M)\). This construction is simple and robust, but also
conservative, as it implicitly assumes that all \(M\) categories are unseen.
It might be then tempting to simply correct for multiplicity using the actually unseen categories in the sample. Unfortunately, this leads to a loss of coverage due to selective inference.
In this section we show how to improve on this benchmark.

\subsection{A worst-case, data independent, bound}\label{sec:worst_case}

Our first result refines the Bonferroni benchmark while remaining completely
data-independent. The construction follows \cite{Painsky2025} and proceeds in two steps. First, we bound from above the maximum functional \(M_{\max}(N)\) by an
\(\ell_r\)-norm and apply Markov's inequality to the corresponding smoothed
quantity. Then, we take the supremum of the resulting bound over all
underlying prevalence vectors \(p=(p_1,\ldots,p_M)\), which yields a worst-case
bound that depends only on \((n,M,\alpha)\). The technical details are described
in the Appendix.

\begin{theorem}\label{thm:worstcase}
Let $N$ be a sample of size $n$ from the Bernoulli product model with alphabet size $M$. Let $M_{\max}(N)$ be as in \eqref{eq:mmax_def}, fix \(\alpha \in (0,1)\) and assume that
\(\log M=o(n)\). Define
\[
r^\star=\log(M/\alpha),
\qquad
T_{\mathrm{wc}}(n,M,\alpha)
=\left(\frac{M}{\alpha}\right)^{1/r^\star}
\frac{r^\star}{r^\star+n}
\exp\left(-\frac{n}{n+r^\star}\right).
\]
Then, for every \(p=(p_1,\ldots,p_M)\),
\[
\mathbb{P}_p\bigl(M_{\max}(N)\le T_{\mathrm{wc}}(n,M,\alpha)\bigr)\ge 1-\alpha.
\]
\end{theorem}

The construction in Theorem~\ref{thm:worstcase} can be given a more concrete
interpretation as follows.
Fix \(r\ge 1\) and suppose that all prevalences coincide at $p_j = p^\star= \tfrac{r}{n+r} $.
Then, for this choice of \(p\), with probability at least \(1-\alpha\) (see the proof of Theorem \ref{thm:worstcase}),
\[
    M_{\max}(N)\le T_{\mathrm{hom}}(r) \coloneqq \left(\frac{M}{\alpha}\right)^{1/r} \frac{r}{n+r}\left(\frac{n}{n+r}\right)^{n/r}  
\]
Observing
\[
\left(\frac{n}{n+r}\right)^{n/r} =\exp\left(-\frac{n}{n+r}\right)\bigl(1+o(1)\bigr) \quad \text{as } n\to\infty,
\]
and setting $r^\star = \log(M/\alpha)$ yields precisely $T_{\mathrm{wc}}$.
That is, the worst-case upper bound in Theorem \ref{thm:worstcase} corresponds to constant $p_i$.

Moreover we observe that
\[
T_{\mathrm{wc}}(n,M,\alpha)
=\frac{\log(M/\alpha)}{n}
+O\left(\frac{1}{n^2}\right)
\quad \text{as } n\to\infty,\ \log M=o(n),
\]
so that the leading term matches the classical Bonferroni bound
\(\log(M/\alpha)/n\) while improving its non-asymptotic constant.

\subsection{Data-dependent bounds}\label{subsec:data_dep}

We now refine the worst-case analysis by introducing a data-dependent bound
that adapts to the empirical configuration of the counts. 
Technically, this is achieved by replacing the crude worst-case bound
\[
    \sum_{j=1}^M p_j^r(1-p_j)^n \le M \max_{p \in [0, 1]} p_j^r(1-p_j)^n = M \left(\frac{r}{r+n}\right)^r\left(\frac{n}{r+n}\right)^n,
\]
underpinning the proof of Theorem \ref{thm:worstcase}, which depends only on \(M\) and yields the choice of $p^\star$, with a finer upper
bound that depends on the underlying prevalences through
\[
m_b(p)\coloneqq \sum_{j=1}^M (1-p_j)^b,
\]
for a tuning parameter \(b\ge 1\). We then estimate \(m_b(p)\) from the data via
\[
m_b(\hat{p})\coloneqq \sum_{j=1}^M (1-\hat{p}_j)^b,
\qquad
\hat{p}_j=\frac{N_j}{n},
\]
and control the deviation \(m_b(\hat{p})-m_b(p)\) using McDiarmid's inequality.
The details are given in the Appendix; here we state the resulting bounds and
briefly discuss their interpretation.

\begin{theorem}\label{T2}
Let $N$ be a sample of size $n$ from the Bernoulli product model with alphabet size $M$, with \(\log M=o(n)\). Let
\(b_n\) be a sequence such that \(b_n=o(n)\), and fix
\(\alpha,\delta\in(0,1)\). Define
\[
m_{b_n}(\hat{p})=\sum_{j=1}^M (1-\hat{p}_j)^{b_n},
\qquad
\epsilon_{M,n,b_n,\delta}
=b_n\sqrt{\frac{M}{n}\log\frac{1}{\delta}}.
\]
Then, for all \(p=(p_1,\ldots,p_M)\),
\[
M_{\max}(N)\le \frac{\log\bigl(m_{b_n}(\hat{p})/\alpha+\epsilon_{M,n,b_n,\delta}/\alpha\bigr)}{n-b_n}
+O\left(\frac{1}{n^2}\right)
\]
with probability at least \(1-\alpha-\delta\). In particular, taking
\(b_n=\log n\) yields
\[
M_{\max}(N)\le \frac{\log\bigl(m_{b_n}(\hat{p})/\alpha+\epsilon_{M,n,\delta}/\alpha\bigr)}{n-\log n}
+O\left(\frac{1}{n^2}\right)
\]
with probability at least \(1-\alpha-\delta\), where
\(\epsilon_{M,n,\delta}=\sqrt{\frac{M\log^2 n}{n}\log\frac{2}{\delta}}\).
\end{theorem}

Observe that, as \(b_n\) increases, the quantity \(m_{b_n}(\hat{p})=\sum_j (1-\hat{p}_j)^{b_n}\)
concentrates on categories with \(\hat{p}_j\) close to zero and, in the limit,
approaches the number of unseen categories
\[
K=\sum_{j=1}^M \mathbbm{1}\{N_j=0\}.
\]
Theorem~\ref{T2} can therefore be interpreted as an automatic refinement of
Theorem~\ref{thm:worstcase}, in which the worst-case factor \(M\) is replaced
by a data-dependent effective alphabet size \(m_{b_n}(\hat{p})\), which is
typically much smaller when many categories have already been observed.

Under a mild condition on the underlying prevalences, the dependence on the
error term \(\epsilon_{M,n,b_n,\delta}\) becomes negligible and the leading
constant is determined entirely by \(m_{b_n}(\hat{p})\).

\begin{corollary}\label{C1}
Let $N$ be a sample of size $n$ from the Bernoulli product model with alphabet size $M$. Suppose there exists \(c>0\) and a
sequence \(b_n=o(\sqrt{n})\) such that
\[
m_{b_n}(p)=\sum_{j=1}^M (1-p_j)^{b_n}\ge c
\]
for all sufficiently large \(n\). Then, for every \(\alpha\in(0,1)\) and
sufficiently large \(n\),
\[
M_{\max}(N)\le \frac{\log\bigl(m_{b_n}(\hat{p})/\alpha\bigr)}{n-b_n}
+o\left(\frac{1}{n}\right)
\]
with probability at least \(1-\alpha\).
\end{corollary}

In Section~\ref{sec:lower_bound_finite} below we show that, in the bounded
alphabet setting, no uniformly valid upper confidence bound can improve on the
order \(\log(K/\alpha)/n\), where \(K\) is the number of unseen categories.
Combined with Corollary~\ref{C1}, this implies that the data-dependent bounds
above are essentially rate-optimal.

\subsection{A matching lower bound}\label{sec:lower_bound_finite}

We now show that, in the bounded-alphabet setting, the order \(\log(K/\alpha)/n\)
cannot be improved in general, where $K$
denotes the number of unseen categories. In particular, any uniformly valid
\((1-\alpha)\)-level upper confidence bound \(T(N)\) for \(M_{\max}(N)\) must,
for some choice of parameters \(p\), be at least of order \(\log(K/\alpha)/n\).
Combined with Corollary~\ref{C1}, this implies that the data-dependent bounds
from Section~\ref{subsec:data_dep} are essentially rate-optimal.

\begin{theorem}\label{thm:lower_bound_finite}
Let $N$ be a sample of size $n$ from the Bernoulli product model with alphabet size $M$, and let \(T(N)\) be any upper confidence bound for \(M_{\max}(N)\) at level
\(1-\alpha\):
\begin{equation}\label{eq:CI-general}
\mathbb{P}_p\bigl(M_{\max}(N)\le T(N)\bigr)\ge 1-\alpha
\quad \text{for all } p.
\end{equation}
Then there exists a choice of parameters \(p\) such that, for sufficiently large
\(n\) and small \(\alpha\),
\[
T(N)\gtrsim \frac{\log(K/\alpha)}{n}
\]
with positive probability. In particular, no uniformly valid upper confidence
bound can have order smaller than \(\log(K/\alpha)/n\).
\end{theorem}

The proof of Theorem~\ref{thm:lower_bound_finite} is constructive and rests on
an explicit family of least favourable parameters for which the distribution of
\(M_{\max}(N)\) can be computed exactly. The idea is to identify a regime where
\(M_{\max}(N)\) is essentially a Bernoulli random variable that takes values
\(0\) and some \(q\in(0,1)\), and then characterise the shortest possible upper confidence
bound at level \(1-\alpha\). The following proposition formalises this
construction.

\begin{proposition}\label{prop1}
Let \(N_j\sim\mathrm{Bin}(n,p_j)\) be independent and fix a subset
\(\mathcal{K}\subseteq\{1,\ldots,M\}\) of size \(k_0=|\mathcal{K}|\).
Assume that \(p_j=q\in(0,1)\) for all \(j\in\mathcal{K}\) and \(p_j=1\) for
all \(j\notin\mathcal{K}\). Let \(M_{\max}(N)=\max_{j:N_j=0}p_j\) and define
\[
t\coloneqq 1-\left(1-\alpha^{1/k_0}\right)^{1/n}.
\]
Consider the rule
\[
T_{\mathcal{K}}(N)=
\begin{cases}
0, & \text{if } K=0 \text{ (no zeros among } \mathcal{K}),\\
t, & \text{otherwise}.
\end{cases}
\]
Then \(T_{\mathcal{K}}(N)\) is an exact upper confidence bound for
\(M_{\max}(N)\) with confidence level \(1-\alpha\), and the bound is tight in
the sense that
\[
\inf_{q\in(0,1)} \mathbb{P}_p\bigl(M_{\max}(N)\le T_{\mathcal{K}}(N)\bigr)
=1-\alpha.
\]
Moreover, for fixed \(k_0\) and small \(\alpha\),
\[
t\sim \frac{\log(k_0/\alpha)}{n}
\quad \text{as } n\to\infty.
\]
\end{proposition}

Under the parameter configuration of Proposition~\ref{prop1}, the random
quantity \(M_{\max}(N)\) can only take the values \(0\) and \(q\): if every
category in \(\mathcal{K}\) appears at least once, then \(M_{\max}(N)=0\),
whereas the appearance of a single zero within \(\mathcal{K}\) forces
\(M_{\max}(N)=q\). The threshold \(t\) in \(T_{\mathcal{K}}(N)\) is precisely
the smallest value for which the coverage constraint \(1-\alpha\) can be met in
this model, and its asymptotic behaviour \(t\sim \log(k_0/\alpha)/n\) shows
that the \(\log(k_0/\alpha)/n\) scale is unavoidable.


\section{Unbounded alphabet sizes}\label{sec:unbounded}
\label{sec:infinite_M}
The lower bound in Theorem~\ref{thm:lower_bound_finite} shows that, in the bounded-alphabet case, the shortest possible upper confidence bound for \(M_{\max}(N)\) scales as \(\log(K/\alpha)/n\), where \(K\) is the number of unseen categories. If the alphabet size is unbounded and no further structural assumptions are imposed on the underlying prevalences, no nontrivial bound is possible: in the worst case \(M_{\max}(N)\) may be arbitrarily close to \(1\).

To obtain meaningful confidence bounds in the unbounded case, we therefore restrict attention to the class of summable prevalence sequences
\begin{equation}\label{eqn:fea_Fcaltilde_def}
\mathcal F = \Big\{ p = (p_j)_{j\ge 1} \colon p_j\in[0,1],\ \sum_{j\ge 1} p_j < \infty \Big\}.
\end{equation}
Clearly, every finite alphabet model belongs to \(\mathcal F\). Moreover, if \(p\notin\mathcal F\) then \(\sum_j p_j = \infty\), and by the Borel–Cantelli lemma this implies \(\mathbb{P}_p(N_j>0\ \text{infinitely often})=1\); in such regimes new categories continue to appear forever and the notion of exhausting the set of “relevant” categories becomes ill posed.
Throughout this section we assume \(p\in\mathcal F\), and we write
\[
S \coloneqq \sum_{j\ge 1} p_j < \infty,
\qquad
\hat S \coloneqq \frac{1}{n} \sum_{j\ge 1} N_j
\]
for the total mass and its natural unbiased estimator.

\subsection{Impossibility of worst-case, data-independent bounds}

The worst-case construction in Section~\ref{sec:worst_case} relied on controlling the expectation of the functionals
\[
    M_r(N) \coloneqq \sum_{j\ge 1} p_j^r \mathbbm{1}\{N_j = 0\}, \quad r\ge 1,
\]
uniformly over the parameter space. In the bounded-alphabet case, the supremum \(\sup_p \mathbb{E}_p[M_r(N)]\) is finite and leads to an explicit data-independent bound. The next result shows that this is no longer true when the alphabet is unbounded, even if we restrict to \(\mathcal F\).

\begin{proposition}\label{prop:fea_worstcase_fails}
Let \(N\) be a sample of size \(n\) from the Bernoulli product model with parameter \(p\in\mathcal F\) as in Equation \eqref{eqn:fea_Fcaltilde_def}. 
Then \(\sup_{p\in\mathcal F} \mathbb{E}_p[M_r(N)] = \infty\) for every fixed \(r\ge 1\).
\end{proposition}

Thus, the Markov-type step that underpins Theorem~\ref{thm:worstcase} cannot be recycled in the unbounded alphabet case.
The next theorem strengthens this conclusion by showing that any nontrivial data-independent bound fails in this setting.

\begin{theorem}\label{thm:fea_no_U_n}
Let \(n\ge 1\) and \(\alpha\in(0,1)\) be fixed, and consider any function \(U_n(\alpha\in(0,1)\) which depends only on $n$ and $\alpha$.
For each \(p\in\mathcal F\), let $M_{\max}(N, p)$ be as in \eqref{eq:mmax_def}, where we made explicit the dependence on $p \in \mathcal F$.
Then, for every such candidate upper bound \(U_n(\alpha)\), there exists some \(p^\star\in\mathcal F\) such that
\[
\mathbb{P}_{p^\star}\bigl(M_{\max}(N,p^\star) \ge U_n(\alpha)\bigr) > \alpha.
\]
In particular, no deterministic bound of the form \(T(N)\equiv U_n(\alpha)\) can define a valid \((1-\alpha)\)-level upper confidence interval for \(M_{\max}(N,p)\) simultaneously over all \(p\in\mathcal F\).
\end{theorem}

Propostion~\ref{prop:fea_worstcase_fails} and Theorem~\ref{thm:fea_no_U_n} together motivate a shift from worst-case, data-independent approach to genuinely data-dependent procedures. In the remainder of this section, we develop an \(r\)-norm-based bound that depends on the total mass \(S\) and on its estimate \(\hat S\), and we show that it is rate-optimal.

\subsection{A tight, data-dependent, \(r\)-norm bound}

We start from an analogue of the \(r\)-norm construction used in the bounded alphabet case, but now exploit the finiteness of the total mass \(S\) rather than a bound on the alphabet size.

\begin{lemma}\label{lem:rnorm-infinite}
Let \(N\) be a sample of size \(n\) from the Bernoulli product model with parameter \(p = (p_j)_{j\ge 1}\in\mathcal F\), and let \(S = \sum_{j\ge 1} p_j < \infty\). For any fixed \(r\ge 1\) and \(\alpha\in(0,1)\) define
\[
U_n(r,S;\alpha) \coloneqq \left(\frac{S}{\alpha}\right)^{1/r} \left(\frac{r-1}{n+r-1}\right)^{(r-1)/r} \left(\frac{n}{n+r-1}\right)^{n/r}.
\]
Then \(\mathbb{P}_p\bigl(M_{\max}(N) \le U_n(r,S;\alpha)\bigr) \ge 1-\alpha\).
\end{lemma}

We remark two things. First, the bound in Lemma \ref{lem:rnorm-infinite} depends on $S$, which is typically unknown, so that we will need to upper-bound $U_{n}(r, S; \alpha)$ with a data-dependent bound on $S$.
Second, for a fixed $r$, the rate of convergence to zero of $U_n(r, S; \alpha)$ is $O(n^{-1+1/r})$ which is slower than $\log(n)/n$ for any fixed $r$.
However, a careful choice of $r = r(S, n, \alpha)$ might improve the rate as shown in the following Lemma.
\begin{lemma}\label{lem:Ur-asymptotic}
Let \(r = r_n\) be a sequence such that \(r_n \to \infty\) and \(r_n = o(n)\) as \(n\to\infty\). Then, for fixed \(S>0\) and \(\alpha\in(0,1)\),
\begin{equation}\label{eq:asymp_U}
U_{n}(r_n,S;\alpha) = \frac{r_n}{en} \left(\frac{S}{\alpha}\right)^{1/r_n} \bigl(1 + o(1)\bigr) \quad \text{as } n\to\infty.
\end{equation}
\end{lemma}
Lemma~\ref{lem:Ur-asymptotic} shows that, once the factor \((S/\alpha)^{1/r}\) has been controlled, the leading term is \(r/n\). In particular, choosing
\begin{equation}\label{eq:r_star}
r^\star = \log\left(\frac{S}{\alpha}\right) + \log n - \log\log n
\end{equation}
yields
\[
U_{n}(r^\star,S;\alpha) = \frac{\log n - \log\log n}{n} \bigl(1 + o(1)\bigr),
\]
which matches the optimal rate obtained in Theorem \ref{thm:lower_bound_infinite} up to lower-order terms.

We now turn Lemma~\ref{lem:rnorm-infinite} into a fully data-dependent procedure by replacing \(S\) with an upper confidence bound \(S^\star\) and choosing a suitable data-dependent \(R^\star\) in place of \(r^\star\).

\begin{theorem}\label{thm:rnorm_unbounded}
Let \(N\) be a sample of size \(n\) from the Bernoulli product model with parameter \(p\in\mathcal F\), and define \(\hat S = n^{-1} \sum_{j\ge 1} N_j\). Fix \(\alpha\in(0,1)\) and \(0 < \beta < \alpha\), and set
\[
S^\star = \left( \sqrt{\frac{\log(1/\beta)}{2n}} + \sqrt{\frac{\log(1/\beta)}{2n} + \hat S} \right)^2, \quad R^\star = \log\left(\frac{S^\star}{\alpha - \beta}\right) + \log n - \log\log n,
\]
and let \(U_n(r,S;\cdot)\) be as in Lemma~\ref{lem:rnorm-infinite}. Suppose that for the corresponding oracle value \(r^\star\) in Equation \eqref{eq:r_star} we have
\begin{equation}\label{eq:cond_rstar}
(r^\star - 1) + \log(r^\star - 1) \ge 1 + \log\log n.
\end{equation}
Then
\[
\mathbb{P}_p\bigl(M_{\max}(N) \le U_{n}(R^\star,S^\star;\alpha - \beta)\bigr) \ge 1-\alpha.
\]
\end{theorem}

\begin{remark}
Condition \eqref{eq:cond_rstar} is very mild, since \(r^\star \asymp \log n - \log\log n\). It is satisfied for all sufficiently large \(n\) and for practically relevant values of \(S\).
\end{remark}

We conclude by establishing a matching lower bound on the attainable length of any uniformly valid confidence interval in the unbounded alphabet regime. The result shows that the \(n^{-1}(\log n - \log\log n)\) rate achieved by the \(r\)-norm-based bound in Theorem~\ref{thm:rnorm_unbounded} is essentially optimal.

\begin{theorem}\label{thm:lower_bound_infinite}
Let \(N\) be a sample of size \(n\) from the Bernoulli product model with parameter \(p\in\mathcal F\), and let \(S = \sum_{j\ge 1} p_j\). Suppose that \(T(N)\) is an upper confidence bound for \(M_{\max}(N)\) at level \(1-\alpha\), in the sense that
\[
\mathbb{P}_p\bigl(M_{\max}(N) \le T(N)\bigr) \ge 1-\alpha \quad \text{for all } p\in\mathcal F.
\]
Then, for sufficiently large \(n\), there exists \(p^\star\in\mathcal F\) with total mass \(S\) such that, with respect to \(\mathbb{P}_{p^\star}\),
\begin{equation}\label{eq:lower_unbounded_rate}
T(N) \ge \frac{1}{n} \left( \log\frac{S}{-\log(1-\alpha)} + \log n - \log\log n \right) + o\left(\frac{1}{n}\right)
\end{equation}
with probability at least \(1-\alpha\). In particular, no uniformly valid upper confidence bound can have order smaller than \(n^{-1}(\log n - \log\log n)\).
\end{theorem}

\section{Numerical illustrations}

\subsection{Simulated data with bounded and unbounded alphabets}
\label{section:sim_study_1}
We assess the performance of the proposed methodology through a simulation study with two complementary designs, in which either the sample size $n$ or the alphabet size $M$ is held fixed while the other varies over a grid. In the first design we fix $n=2000$ and let $M$ range from $100$ to $10{,}000$, whereas in the second we fix $M=5000$ and consider sample sizes $n \in [5000,10000]$. For each pair $(n,M)$, we generate benchmark collections of binomial parameters $(p_1,\ldots,p_M)$ spanning a broad range of distributional regimes, from light-tailed to heavy-tailed. We consider these scenarios:
\begin{itemize}
    \item[(a)] a Zipf-like distribution, defined by $p_j = (j+1)^{-\gamma}$ with $\gamma\in\{0.25,0.5,1.02\}$;
    \item[(b)] a geometric-like distribution, defined by $p_j = a^{j}$ with $a\in\{0.005,0.1,0.25\}$;
    \item[(c)] a homogeneous community, with constant probabilities $p_j \equiv 1/c$, where $c\in\{2,20,1000\}$.
\end{itemize}

\begin{figure}
    \centering
    \includegraphics[width=0.325\linewidth]{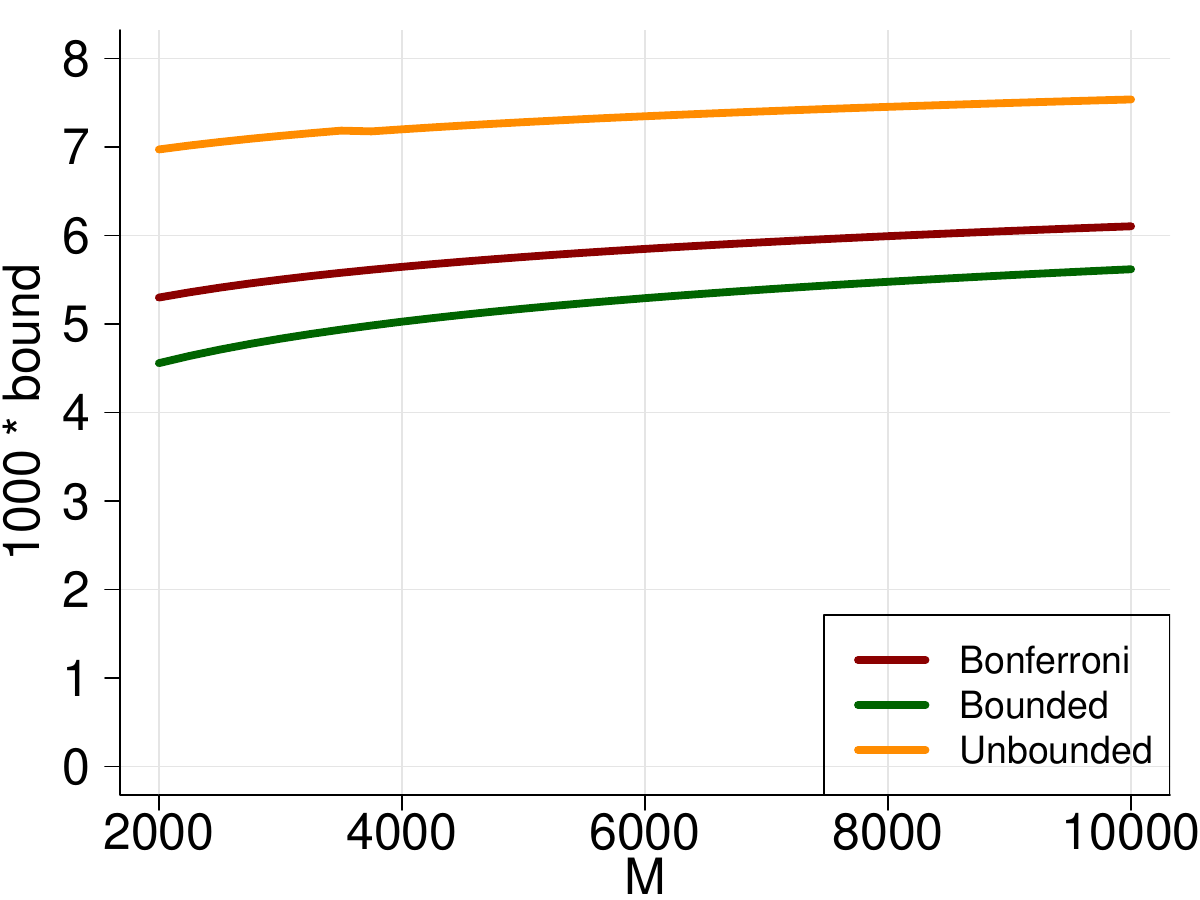}
    \hfill
    \includegraphics[width=0.325\linewidth]{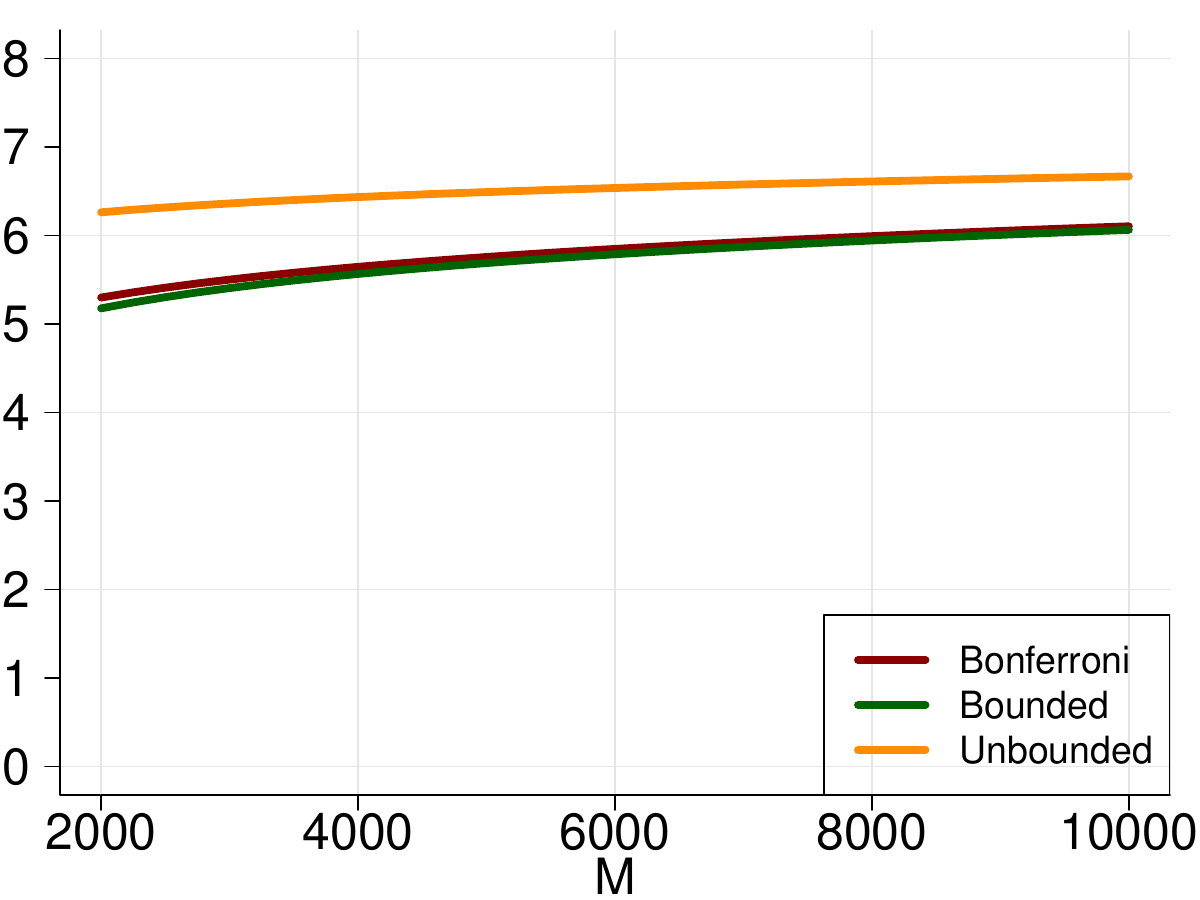}
    \hfill
    \includegraphics[width=0.325\linewidth]{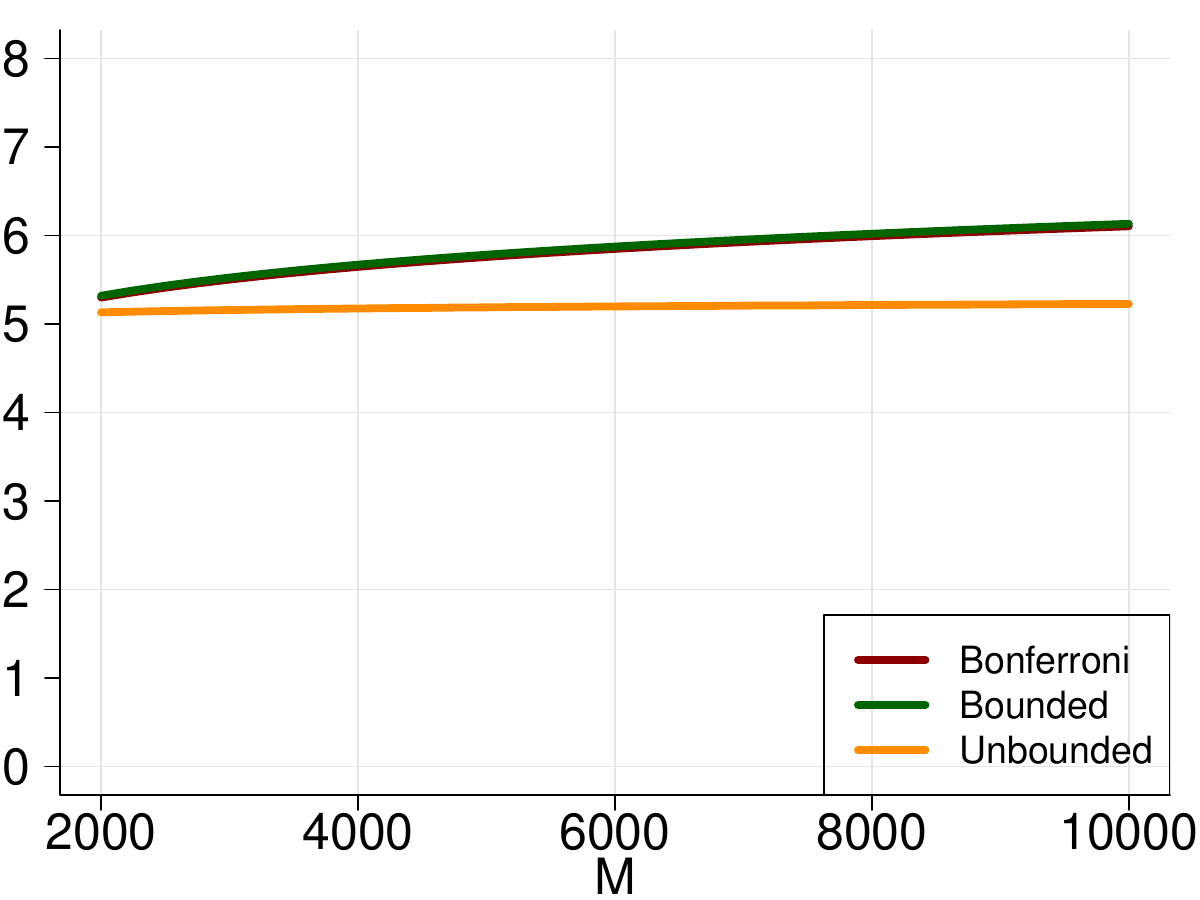}\\
    \includegraphics[width=0.325\linewidth]{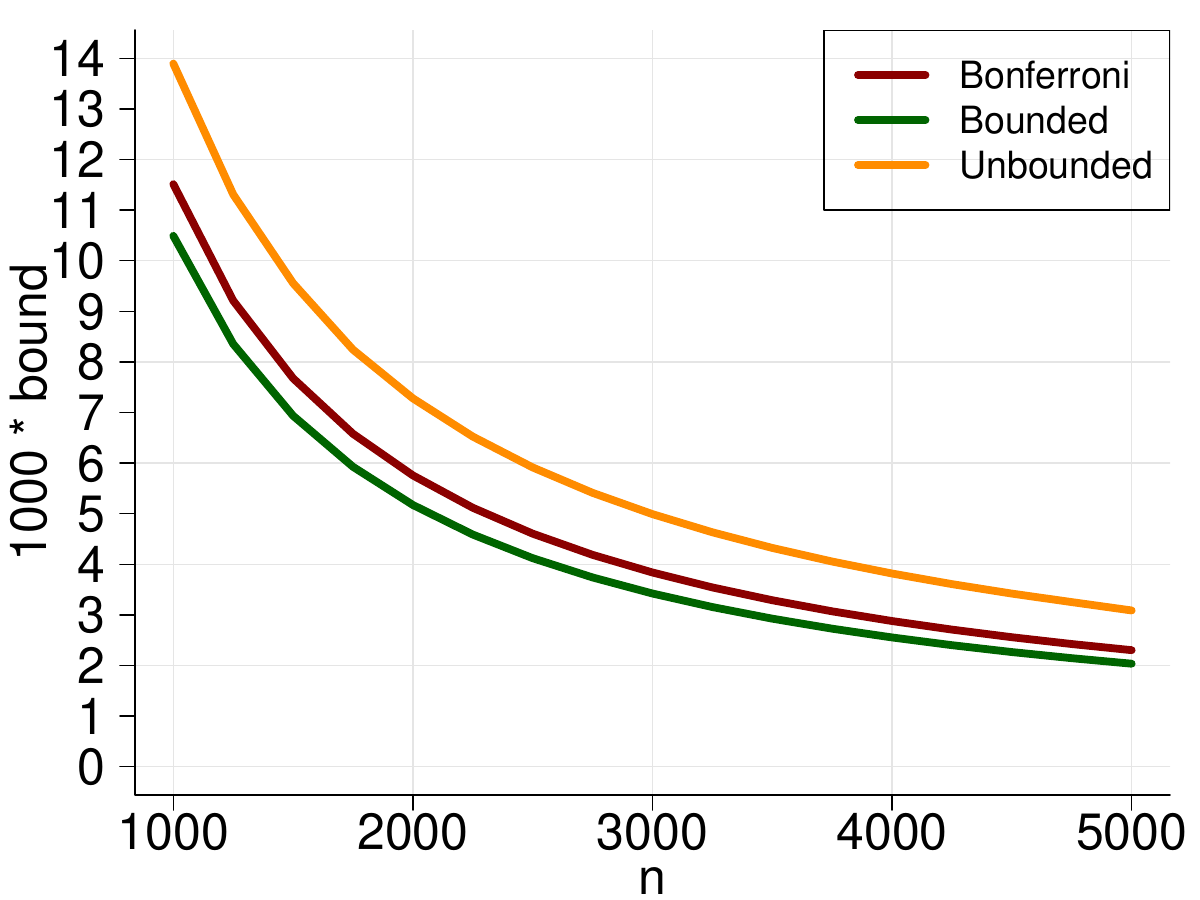}
    \hfill
    \includegraphics[width=0.325\linewidth]{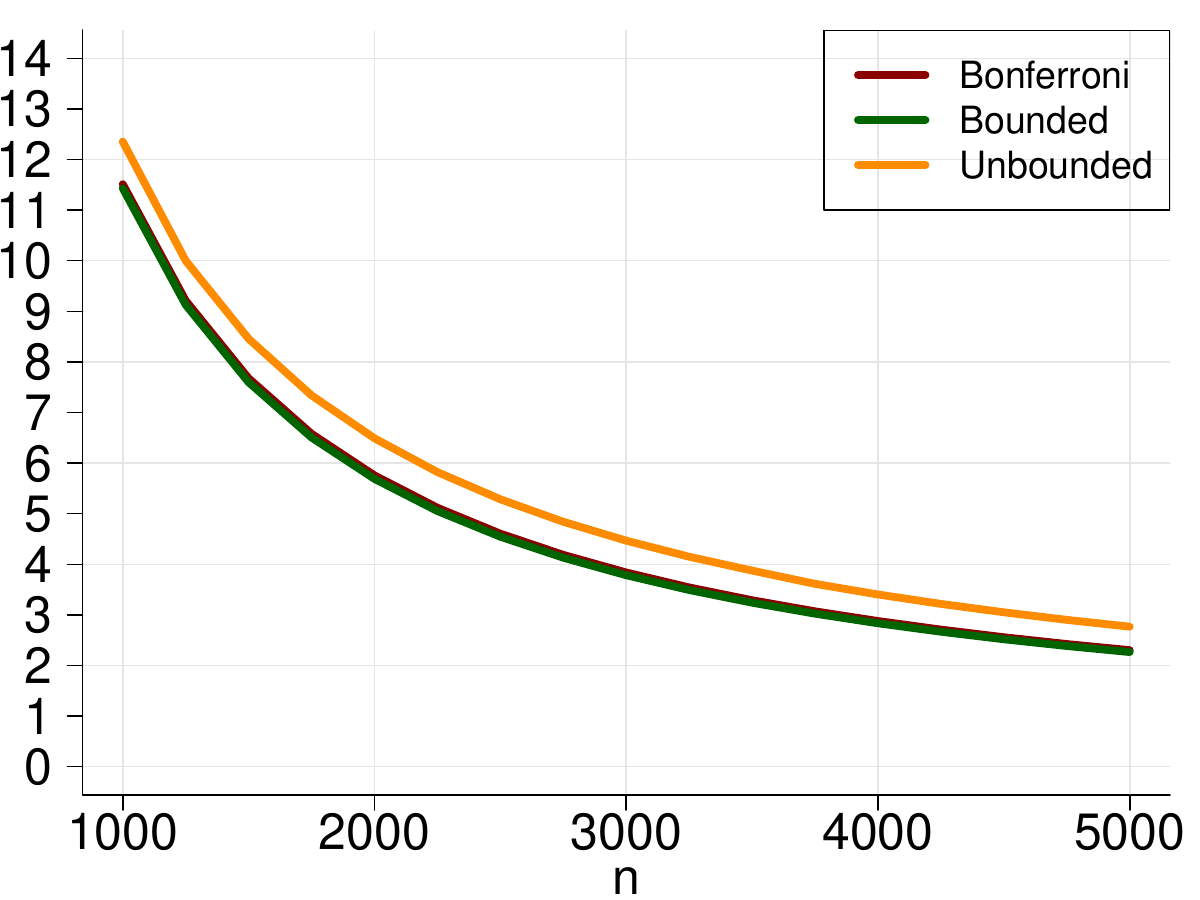}
    \hfill
    \includegraphics[width=0.325\linewidth]{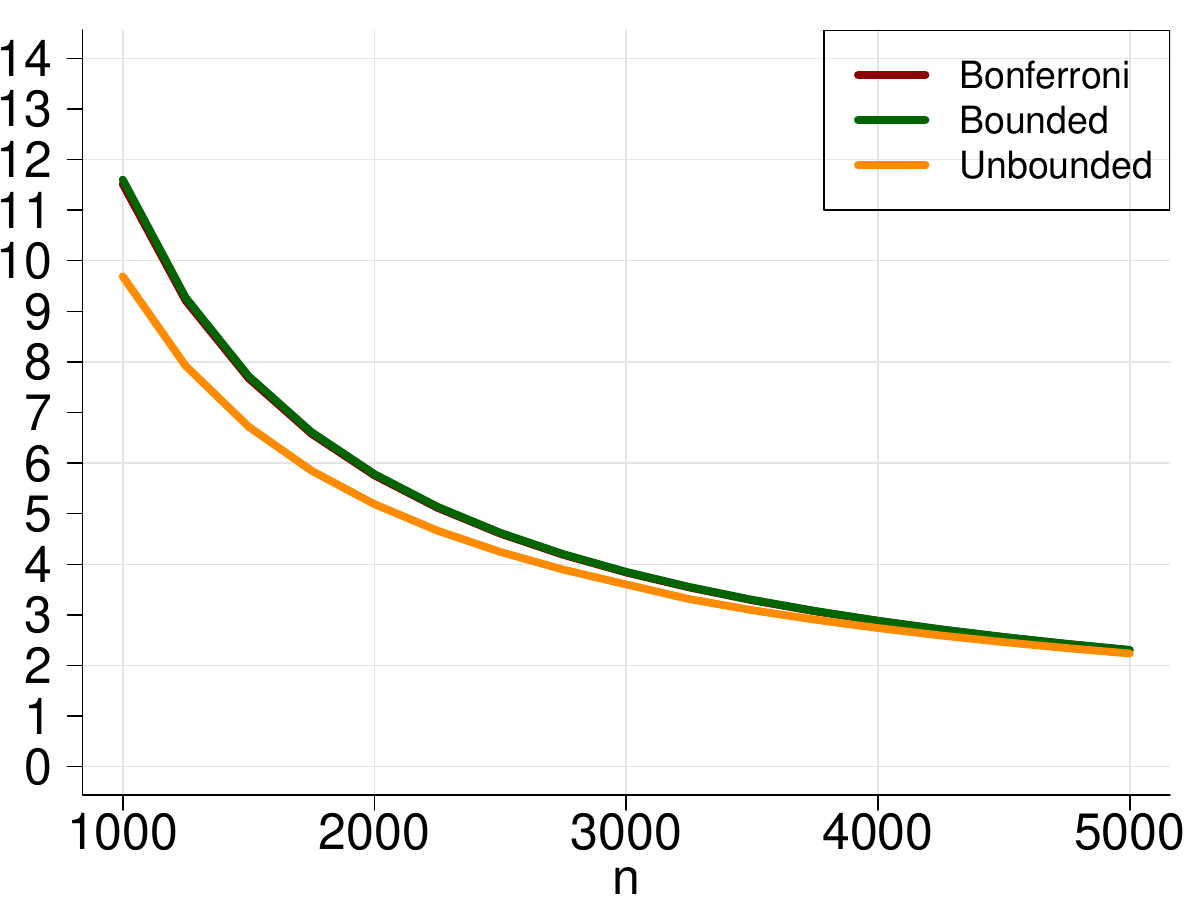}\\
    \caption{Confidence interval length under Zipf-like Bernoulli probabilities.
    First row: the sample size is fixed while the alphabet size $M$ varies.
    Second row: the alphabet size is fixed while the sample size $n$ increases.
    Each column corresponds to a different value of $\gamma$: $\gamma = 0.25$ (left), $\gamma = 0.5$ (middle), and $\gamma = 1.02$ (right). 
    The y-axis is scaled by a factor of $10^3$ for readability.
    }
    \label{fig:Bdd_Zipfs}
\end{figure}

For each configuration, we repeat the sampling procedure $n_{\text{reps}}=100$ times to estimate average performance metrics and empirical coverage. We compare three competing upper bounds: the bounds of Theorem~\ref{T2} and Theorem~\ref{thm:rnorm_unbounded}, corresponding to the bounded- and unbounded-alphabet regimes and denoted \textit{Bounded} and \textit{Unbounded}, together with the benchmark bound in \eqref{eqn:ROT}, denoted \textit{Bonferroni}. We compute the \textit{Bounded} bound with $b_n=\log(n)$ and $\delta=0.01\alpha$, and the \textit{Unbounded} bound with $\beta=10^{-5}$.

Figure~\ref{fig:Bdd_Zipfs} reports the resulting confidence interval lengths in the Zipf-like setting, as $n$ and $M$ vary, while Figures~\ref{fig:Bdd_Geom} and~\ref{fig:Bdd_Const} in the Appendix show the results for the geometric-like and homogeneous settings, respectively.
Focusing on the Zipf-like settings, Figure \ref{fig:Bdd_Zipfs} shows that with light-tailed distributions i.e., smaller values of $\gamma$,  the \textit{Bounded} upper bound consistently outperforms the \textit{Bonferroni} benchmark, while the \textit{Unbounded} bound is not competitive and can even yield wider intervals than the benchmark. This behavior is expected, as in such settings most symbols are observed and the unbounded-alphabet assumption introduces unnecessary conservatism. In contrast, as the tail becomes heavier -- corresponding to scenarios in which a few symbols are observed repeatedly while many others remain unseen or extremely rare -- the \textit{Bounded} and \textit{Bonferroni} bounds become nearly indistinguishable, whereas the \textit{Unbounded} bound produces substantially shorter intervals.
Similar conclusions can be drawn by looking at the geometric-like and homogeneous settings in the Appendix. Namely, when there are many infrequent symbols in the alphabet, the \textit{Unbounded} interval is the shortest.
Although not reported here for brevity, the nominal $95\%$ coverage level is attained by all methods, with empirical coverage typically close to $100\%$ across all settings. 

In Section \ref{app:m_upperbound} of the Appendix we also explore the effect of $M$ on the \textit{Bounded} interval. Specifically, what happens if we don't know $M$ but only a reasonable upper bound? Letting $M = M_{\operatorname{true}} + M_{\operatorname{add}}$, where $M_{\operatorname{true}}$ is the actual size of the alphabet, we find that the interval is very robust to $M_{\operatorname{add}}$ and starts to deteriorate noticeably only when $M_{\operatorname{add}} > 2 M_{\operatorname{true}}$, i.e., when the upper bound is extremely loose.


\vspace{-0.1cm}

\subsection{Bounded vs unbounded bound: when to choose one over the other?}\label{section:bdd_vs_ubd}
The experiments in Section~\ref{section:sim_study_1} show that the \textit{Unbounded} upper bound can yield shorter confidence intervals than its \textit{Bounded} counterpart, even when the alphabet size is known and finite. However, this behavior is not universal: in several regimes the \textit{Unbounded} bound performs poorly relative to \textit{Bounded}. Intuitively, the unbounded-alphabet assumption should be advantageous in heavy-tailed settings, where many symbols remain undetected. In this section we propose empirical and heuristic tools to guide the choice between the bounded and unbounded regimes. These tools are intentionally simple and practical, providing fast guidance without requiring detailed tail modelling or power-law analysis.

The first tool we propose is a simple visual inspection of the accumulation curve. Accumulation curves, widely used in ecology \citep{colwell2004}, plot the number of distinct species observed as a function of sample size. Since the data are typically recorded without preserving the order of appearance, the curve is usually obtained by averaging over multiple random permutations of the sample. Such curves often exhibit rapid initial growth for small sample sizes, followed either by stabilization around an apparent asymptote or by continued growth at a slower rate. Persistent growth suggests that new symbols are still being discovered, consistent with many low-prevalence categories, whereas a flattened curve indicates that most symbols have already been observed.

\begin{figure}[t]
    \centering
    \includegraphics[width=0.325\linewidth]{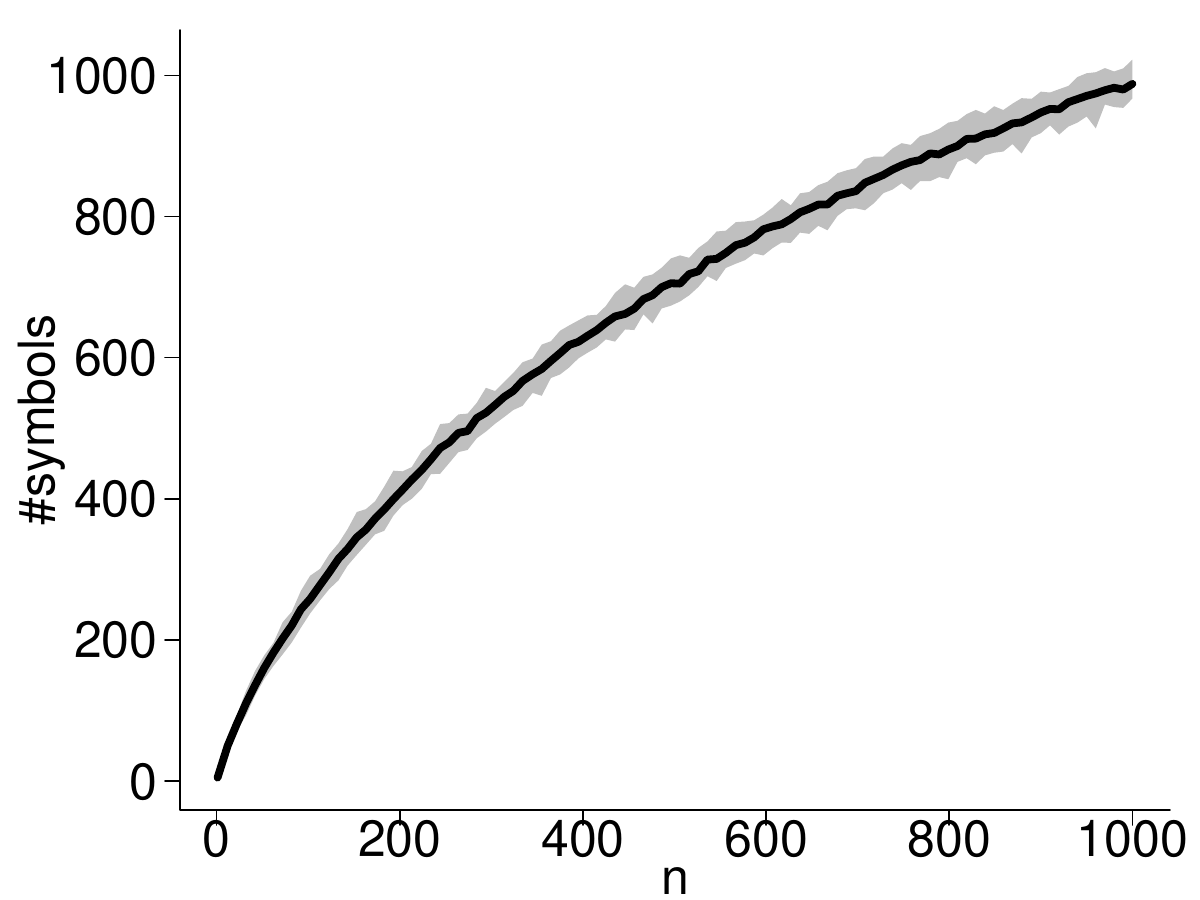}
    \hfill
    \includegraphics[width=0.325\linewidth]{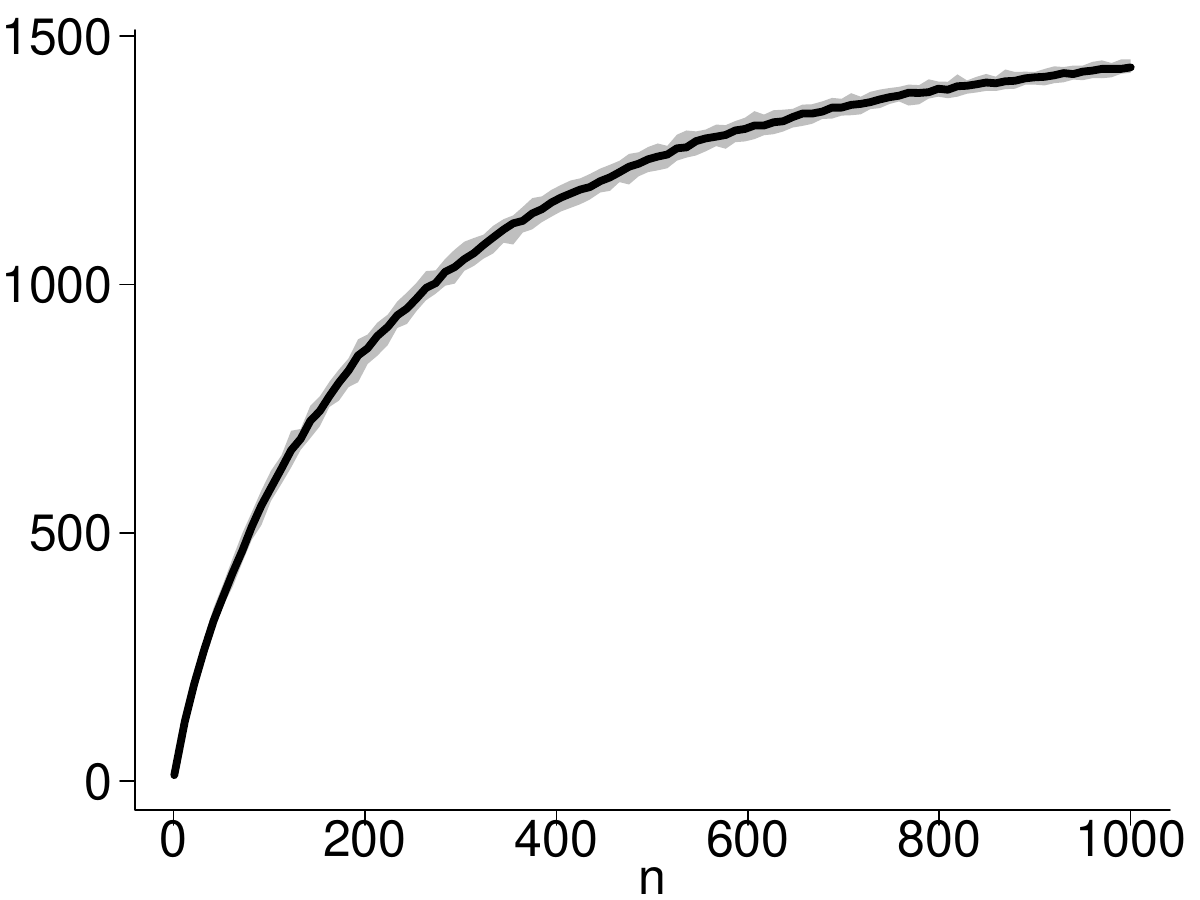}
    \hfill
    \includegraphics[width=0.325\linewidth]{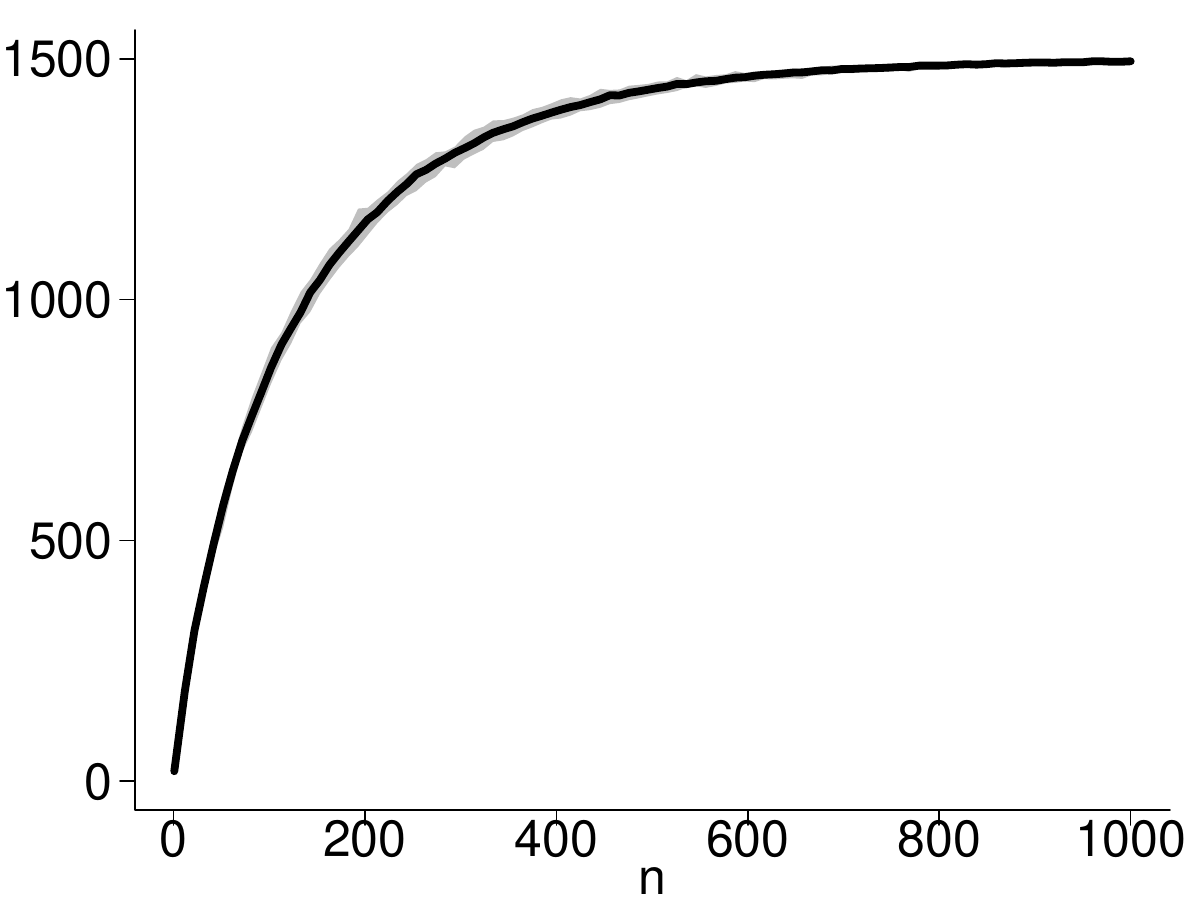}
    \caption{Accumulation curves under Zipf-like Bernoulli probabilities. Panels correspond to $\gamma = 1.05$ (left), $\gamma = 0.85$ (center), and $\gamma = 0.75$ (right). Each curve shows the number of distinct symbols observed as a function of the sample size $n$.}
    \label{fig:ExtCrv_Zipfs}
\end{figure}

As a practical guideline, we recommend inspecting the accumulation curve of the data. If a clear horizontal asymptote emerges, the \textit{Bounded} upper bound should be preferred; otherwise, the \textit{Unbounded} version is likely more appropriate. An illustrative example is provided in Figure~\ref{fig:ExtCrv_Zipfs}.

We now introduce a quantitative heuristic that complements the visual inspection. Specifically, we propose to favor the \textit{Unbounded} bound over the \textit{Bounded} one whenever
\begin{equation}
    \label{eqn:threshold_S}
    S < \frac{-\log(1-\alpha)}{\alpha}\,\frac{M}{n}\,\log\left(\frac{M}{\alpha}\right).
\end{equation}
For $\alpha=0.05$, the prefactor is close to $1$, which leads to the simplified heuristic rule
$$
    S < \frac{M}{n}\log(20M)\,.
$$

The derivation of the threshold in Equation~\eqref{eqn:threshold_S} is deferred to the Appendix. The threshold is admittedly heuristic: it is obtained by comparing an oracle version of the \textit{Unbounded} upper bound with the \textit{Bonferroni} bound, and therefore ignores both the data-dependent nature of $\hat S$ and the analytical differences between \textit{Bonferroni} and the \textit{Bounded} bound. Nevertheless, the threshold is informative in practice. Empirically, whenever \textit{Unbounded} outperforms \textit{Bonferroni}, it also outperforms \textit{Bounded}, up to negligible differences. Moreover, although the threshold depends on the unknown quantity $S$, this is typically well approximated by its empirical estimate $\hat S$. Finally, the threshold admits a simple interpretation as a diagnostic of whether the total probability mass is small relative to the effective alphabet size, formalizing the intuition that the \textit{Unbounded} approach is preferable in regimes with many very small probabilities.

\begin{figure}[t]
    \centering
    \includegraphics[width=0.325\linewidth]{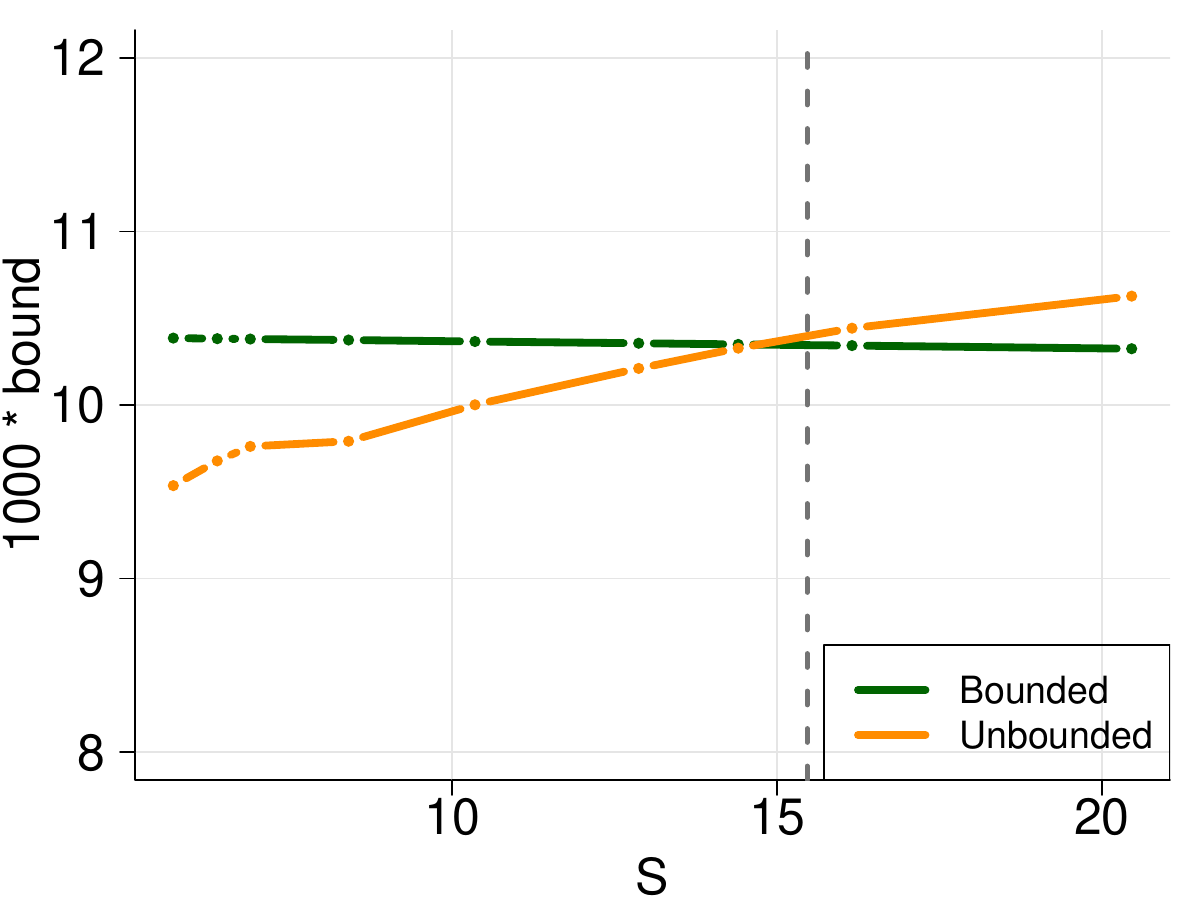}
    \hfill
    \includegraphics[width=0.325\linewidth]{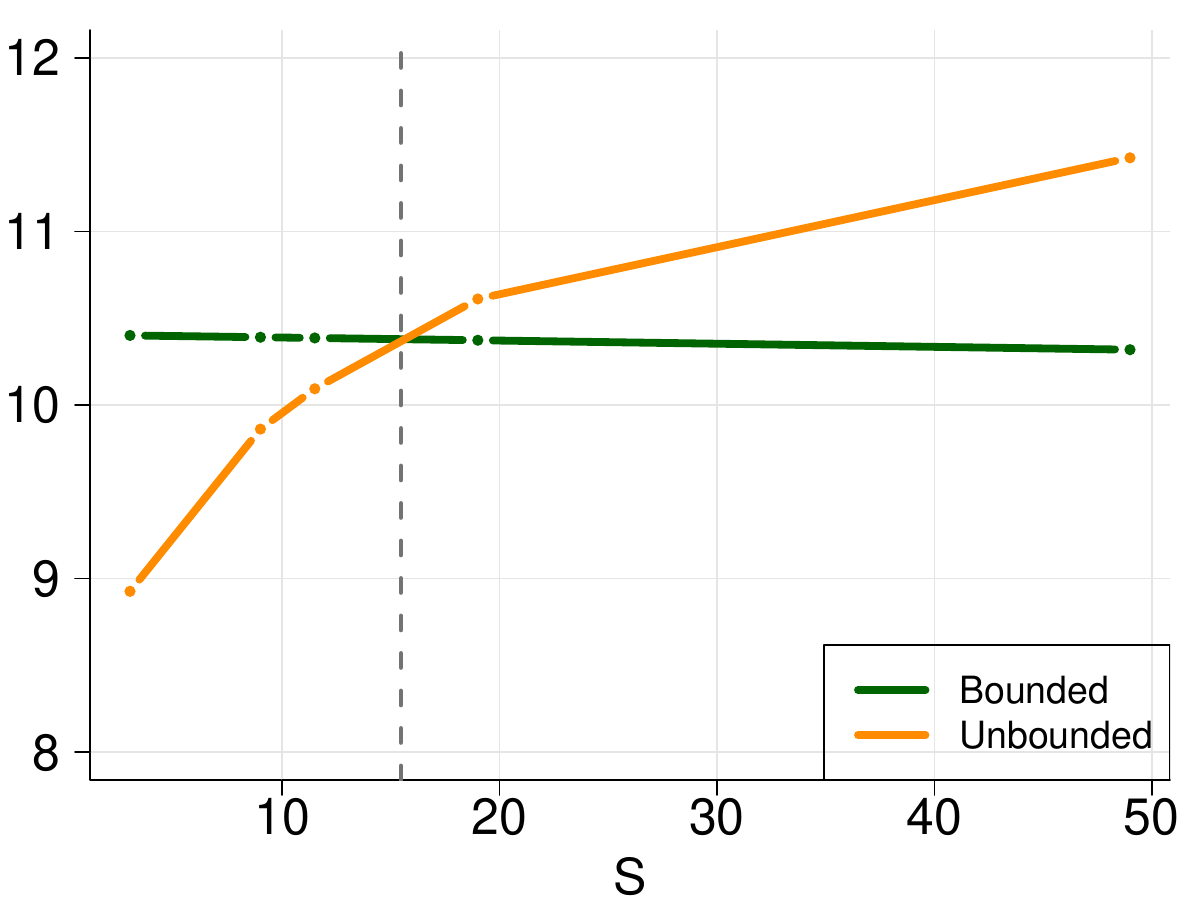}
    \hfill
    \includegraphics[width=0.325\linewidth]{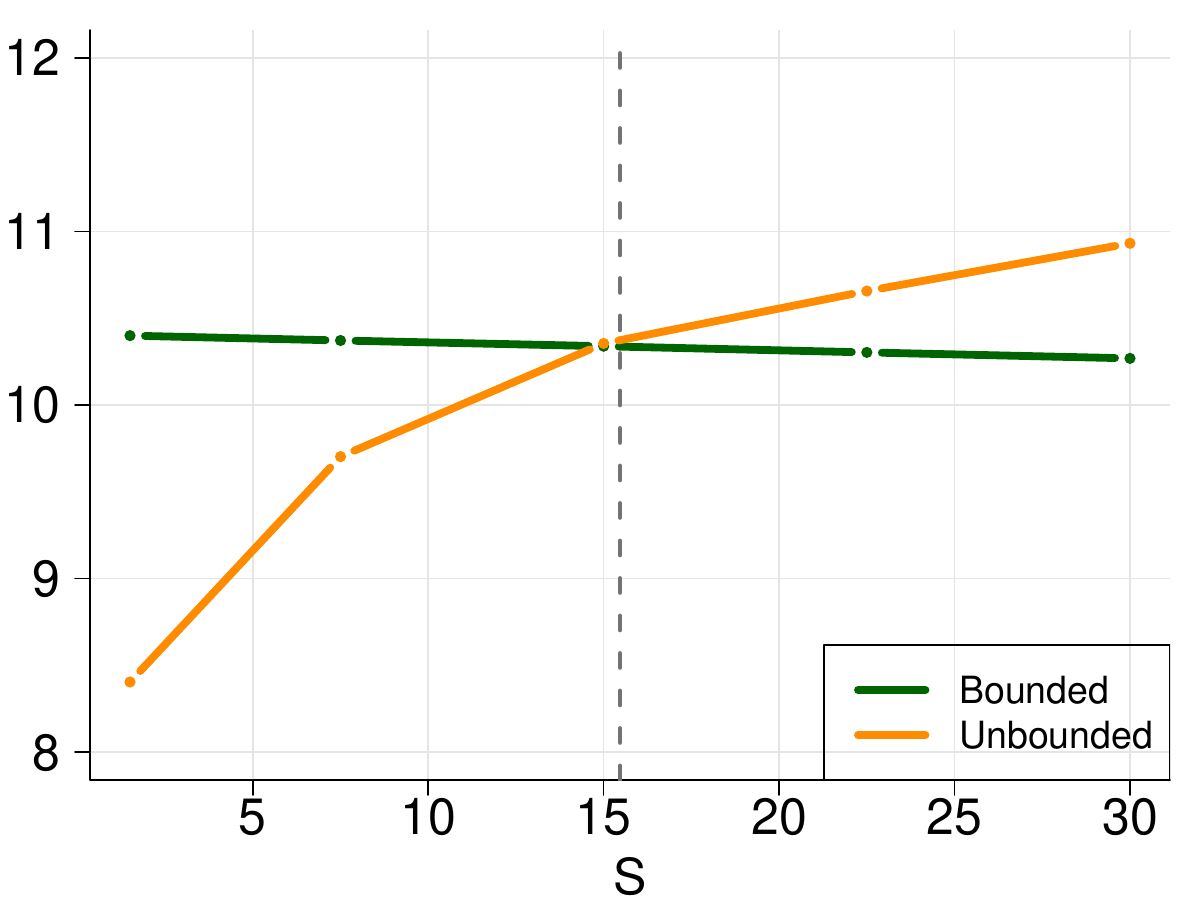}
    \caption{Confidence interval length for Zipf-like (left), geometric-like (center), and homogeneous (right) Bernoulli probabilities.
    The vertical dotted line indicates the heuristic threshold in Equation \eqref{eqn:threshold_S}.
    The y-axis is rescaled by a factor of $10^3$.
    }
    \label{fig:Bdd_vs_Unb}
\end{figure}

We empirically assess the validity of the heuristic threshold through the following experiment. We fix $n=1000$ and $M=1500$, yielding a threshold value of $15.46$. The Bernoulli probabilities are generated as in Section~\ref{section:sim_study_1}. For the Zipf-like distribution we consider $\gamma\in\{1.05,1.02,1,0.95,0.9,0.85,0.825,0.8,0.75\}$; for the geometric-like case we take $a\in\{0.25,0.1,0.08,0.05,0.02\}$; and for the homogeneous case we set $c\in\{1000,200,100,200/3,50\}$. In each configuration, we compute the total mass $S=\sum_{j=1}^M p_j$ and evaluate the corresponding upper bounds. The results are summarised in Figure~\ref{fig:Bdd_vs_Unb}, where the vertical dotted line marks the heuristic threshold.

Across all three settings, the vertical dotted line provides a clear separation between the regimes in which the \textit{Bounded} and \textit{Unbounded} bounds are preferable, confirming the usefulness of the proposed threshold. The Zipf-like case (left panel of Figure~\ref{fig:Bdd_vs_Unb}) highlights most clearly the heuristic nature of the criterion: the two bounds already coincide for $\gamma=0.825$, corresponding to $S$ slightly below $15$, even though the threshold exceeds this value. However, in this range the resulting confidence intervals are nearly indistinguishable, making either choice practically equivalent.

Turning back to the visual diagnostic, Figure~\ref{fig:ExtCrv_Zipfs} displays the average accumulation curves in the Zipf-like setting. For clarity, we report only $\gamma\in\{1.05,0.85,0.75\}$. In the left panel, the number of discovered symbols increases slowly with $n$, indicating very small probabilities and rare discoveries; indeed, for $n=1000$, nearly $500$ symbols remain unseen. In this regime, $S$ is small relative to $M$, and accordingly the \textit{Unbounded} estimator outperforms \textit{Bounded}, consistent with Figure~\ref{fig:Bdd_vs_Unb}. The right panel shows the opposite case: probabilities are larger, the accumulation curve quickly plateaus, and the entire alphabet is observed after only a few hundred samples. Here, $S$ exceeds the threshold in \eqref{eqn:threshold_S}, making \textit{Bounded} the preferable choice. Finally, the middle panel represents an intermediate regime in which all symbols are eventually detected but the curve grows more gradually, reflecting smaller probabilities. In this case, \textit{Bounded} and \textit{Unbounded} have comparable performance. Similar conclusions hold in the other distributional settings.

\subsection{Application to the TCGA data}
\label{sec:TCGA_analysis}
The Cancer Genome Atlas (TCGA) is a large, publicly available cancer genomics resource that catalogs somatic mutations from whole-exome sequencing data in tumor samples across $32$ cancer types \citep{Ellrott2018}. These data are of major scientific importance, as identifying somatic mutations plays a central role in precision oncology. However, while traditional analyses focus on well-known cancer genes with relatively frequent mutations, large-scale sequencing studies have shown that somatic variants exhibit a highly heterogeneous, long-tailed distribution: a small subset is repeatedly observed across tumors, whereas most variants are extremely rare, often appearing only once or not at all. 

Previous statistical analyses on the TCGA data \citep{Chakraborty2019,masoero2022more,MasoeroCamScaled} have addressed the problem of estimating the number of new variants expected in future, unobserved tumor samples, based on information from existing patient cohorts.
In particular, \citet{masoero2022more,MasoeroCamScaled} analyze each cancer type separated, and model data for each cancer via a Bernoulli product model, where the different symbols in the alphabet corresponds to the genomic variants. 
Even if such a model assumes independence across variants,  which is known to be violated in practice due to local correlations along the genome \citep{Leiserson2015}, the predictions obtained are very accurate, see also \citet{Chakraborty2019,masoero2022more} for discussion. Figure~\ref{fig:TCGA_sizes} in the appendix reports the sample size for each cancer type and the corresponding number of discovered somatic variants .

While accurate predictions of this quantity can be highly valuable, for example, in inferring the primary site of tumors of unknown origin or in interpreting evidence of a tumor detected in circulating tumor DNA, here we take a complementary approach by addressing a related question: how likely is it to encounter previously unseen variants? 
Our main contribution is to provide finite-sample, distribution-free guarantees by constructing confidence intervals that simultaneously cover the prevalences of all unseen variants at a prescribed confidence level. This goes beyond estimating expected discovery counts, and yields rigorous uncertainty quantification for rare and unobserved variants in cancer genomic datasets.

To answer this question, we first determine whether the \textit{Bounded} or the \textit{Unbounded} upper bound is more appropriate for the TCGA application. Both diagnostic tools in Section~\ref{section:bdd_vs_ubd} provide consistent evidence in favor of \textit{Unbounded}. 
In particular, from \eqref{eqn:threshold_S}, the \textit{Unbounded} interval should be preferred whenever $M > W(-\log(1-\alpha)S n/\alpha)$, where $W(\cdot)$ denotes the Lambert $W$ function \citep[Section 4.13]{Olver2010}. Setting $\alpha=0.05$ and plugging in $\hat S$ in place of $S$,
we have that $M$ should be smaller than $5000$ in order to prefer the \textit{Unbounded} interval.
This is orders of magnitude smaller than any realistic estimate of the number of possible somatic variants in the human genome, and existing databases already catalog billions of variants of potential interest \citep{Bryan2025}. This comparison supports the use of \textit{Unbounded}.

This conclusion is reinforced by a qualitative inspection of accumulation curves. Figure~\ref{fig:TCGA_ExtCurves} in the Appendix reports accumulation curves for four representative cancer types. In all cases, the curves grow approximately linearly and are far from saturation, indicating that new variants continue to be discovered at an almost constant rate as sample size increases. As discussed in Section~\ref{section:bdd_vs_ubd}, this behavior is characteristic of regimes with many rare and unobserved symbols, and provides further evidence against a bounded-alphabet model.

\begin{figure}[t]
    \centering
    \includegraphics[width=0.24\linewidth]{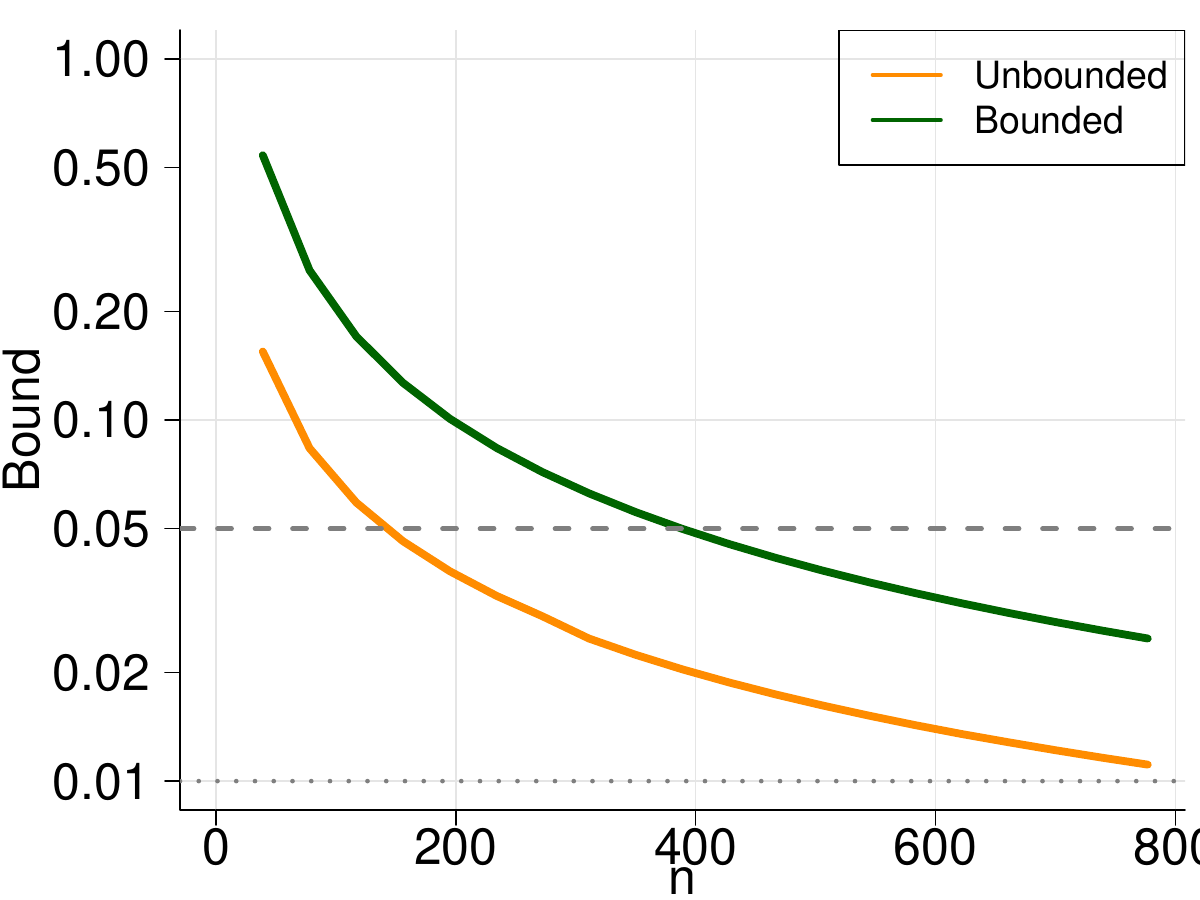}
    \hfill
    \includegraphics[width=0.24\linewidth]{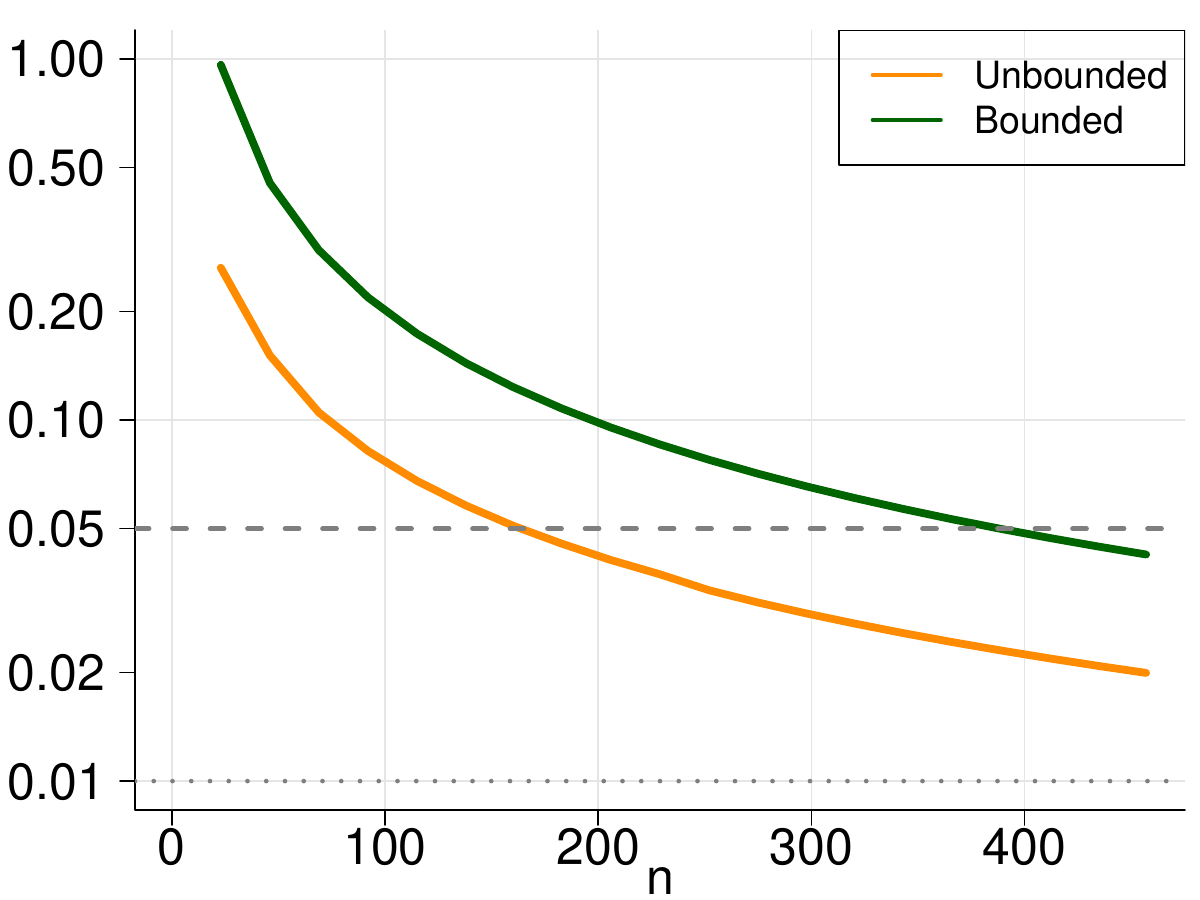}
    \hfill
    \includegraphics[width=0.24\linewidth]{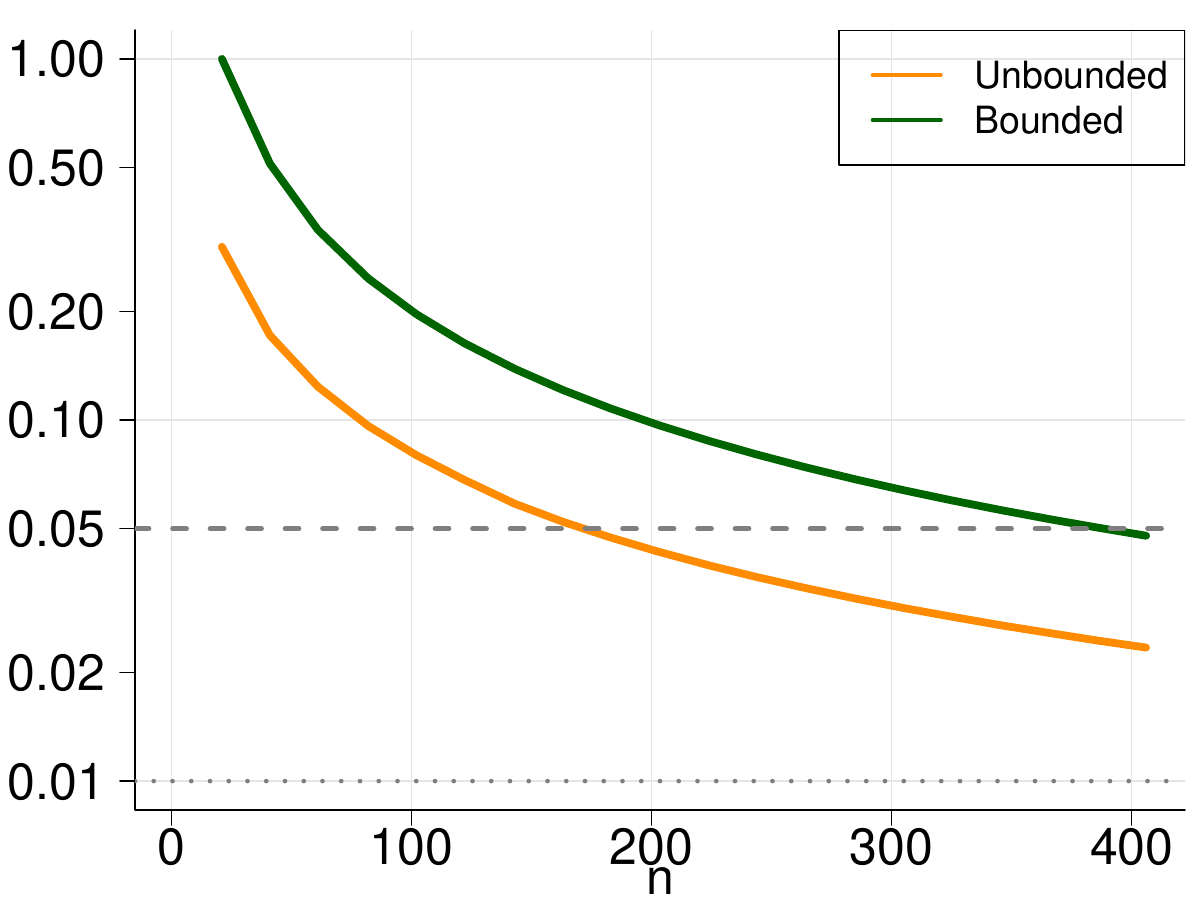}
    \hfill
    \includegraphics[width=0.24\linewidth]{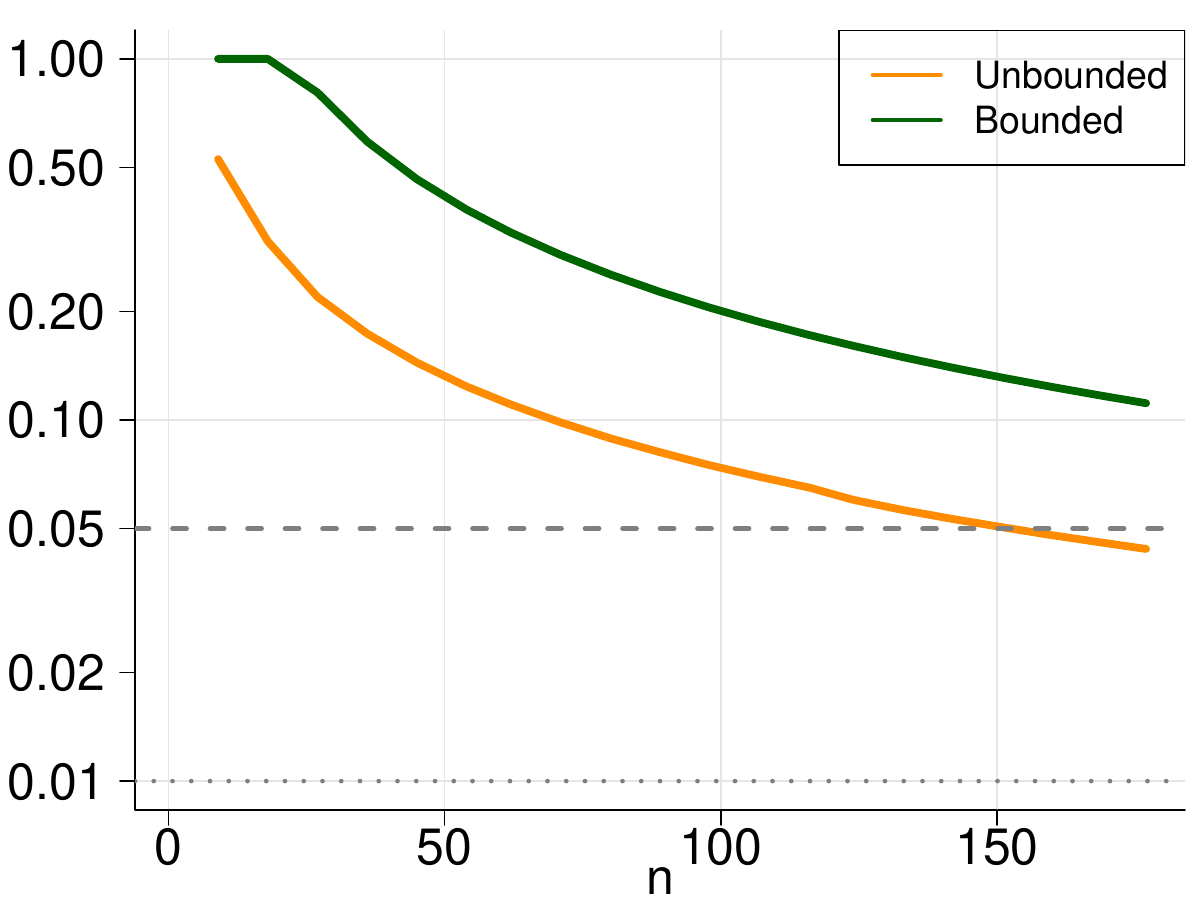}
    \caption{Length of confidence intervals for four selected  TCGA cancer types as a function of the sample size $n$.
    Panels correspond to BRCA, LUSC, SKCM, and ESCA, from left to right. The dashed and dotted grey horizontal lines refer to probability levels $0.05$ and $0.01$, respectively.}
    \label{fig:TCGA_fit}
\end{figure}

Finally, Figure \ref{fig:TCGA_fit} shows the interval length for four selected datasets, that are Breast Invasive Carcinoma (BRCA), Lung Squamous Cell Carcinoma (LUSC), Skin Cutaneous Melanoma (SKCM) and Esophageal Carcinoma (ESCA), as a function of the sample size $n$. It is clear that the \textit{Unbounded} interval is always substantially smaller that the \textit{Bounded} one. 
While looking at the accumulation curves in Figure \ref{fig:TCGA_ExtCurves} one expects that new variants will continue to appear at a constant rate as the sampling effort increases, our bounds in Figure \ref{fig:TCGA_fit} show that such variants will have rather low, but not neglibile, prevalences across the population, always below $5\%$, and close to $1\%$ in the case of BRCA, probably warranting further sequencing.

\vspace{-0.5cm}
\section{Stopping rules based on \texorpdfstring{$M_{\max}$}{Mmax}}\label{sec:stopping}

\subsection{Design principle and interpretation}

We view the Bernoulli product model as a sequential sampling framework. At each step we collect one more sampling unit (patient, site, sample), update the incidence matrix, and face a sample size determination problem: is the current sample size $n$ large enough that all species with prevalence at least some target threshold $\varepsilon>0$ have been seen at least once, or should we continue sampling?

Formally, fix $\varepsilon>0$ representing the minimal prevalence of scientific or practical interest and a confidence level $1-\alpha$. The event
\[
A_n(\varepsilon) \coloneqq \{M_{\max}\le \varepsilon\}
= \bigl\{p_j\le\varepsilon\ \text{for all}\ i\ \text{with}\ N_j=0\bigr\}
\]
is exactly the event that all species with $p_j>\varepsilon$ have appeared at least once in the sample. Thus the sample size determination problem can be phrased as choosing a stopping time $N_{\mathrm{stop}}$ such that, with high probability, $A_{N_{\mathrm{stop}}}(\varepsilon)$ holds.

Hence, any $(1-\alpha)$-level upper confidence bound $U_n(\alpha)$ for $M_{\max}$ provides a direct way to construct such a stopping time. By definition,
\[
\mathbb{P}\bigl(M_{\max}\le U_n(\alpha)\bigr)\ge 1-\alpha,
\]
and the inequality can be equivalently rewritten in simultaneous post-selection form as
\[
\mathbb{P}\bigl(p_j\le U_n(\alpha)\ \text{for all}\ i\ \text{with}\ N_j=0\bigr)\ge 1-\alpha.
\]
This suggests the following rule: given $(\varepsilon,\alpha)$, continue sampling until the first $n$ such that $U_n(\alpha)\le\varepsilon$, and then stop. At that random sample size $N_{\mathrm{stop}}$ we have, by construction,
\[
\mathbb{P}\bigl(A_{N_{\mathrm{stop}}}(\varepsilon)\bigr)
= \mathbb{P}\bigl(M_{\max}\le\varepsilon\bigr)
\ge 1-\alpha,
\]
so that, with probability at least $1-\alpha$, every species that remains unseen has prevalence at most $\varepsilon$, or equivalently every species with $p_j>\varepsilon$ has been detected at least once.

Sections~\ref{sec:finite_M} and \ref{sec:infinite_M} give two concrete choices of $U_n(\alpha)$: In the next subsection we evaluate the induced stopping times $N_{\mathrm{stop}}$ in a simulation study designed to mimic metabarcoding-like data with and without contamination.

\subsection{Simulations for the stopping rule}
\label{sec:SS4stopping}

We consider a Bernoulli product model for metabarcoding-like data, where observations are arranged in a site--by--species incidence matrix, a standard format in community ecology and biodiversity assessment \citep[e.g.][]{Magurran2004,ChaoJost2012}. Within each sample $i$, the entry $X_{ij}$ indicates whether species $j$ is observed. For a fixed community, we specify prevalences $(p_j)_{j=1}^M$ and generate samples independently from the Bernoulli product model. We fix a target threshold $\varepsilon=0.005$ and treat species with $p_j\ge\varepsilon$ as biologically relevant targets for discovery. To span light- and heavy-tailed regimes, we consider four choices of $(p_j)$ with $M=1500$, corresponding to the Zipf-like distribution (with $\gamma=1.05$), the homogeneous community with $p_j \equiv c$ and $c=0.006$ (prevalences close to $\varepsilon$) or $c=0.05$ (where prevalences are well above $\varepsilon$), and a truncated geometric decay with $p_j=(1-a)^{i-1}$ for $a=0.05$, yielding an intermediate tail between the uniform and Zipf cases.

To mimic sequencing errors in metabarcoding pipelines, we corrupt the data at the level of individual entries. Independently for each realised $1$ in the matrix, we declare it to be an error with probability $q\in\{0,10^{-4},5\times10^{-4},10^{-3},2.5\times10^{-3},5\times10^{-3}\}$. If $X_{ij}=1$ is flagged as an error, we set $X_{ij}=0$ and create a new ``error species'' with a unique label that occurs exactly once in the entire dataset (a singleton OTU). This leaves the true community $(p_j)$ unchanged but inflates the observed alphabet through errors, as expected in high-throughput amplicon data.

\begin{figure}[t]
    \centering
    \includegraphics[width=\linewidth]{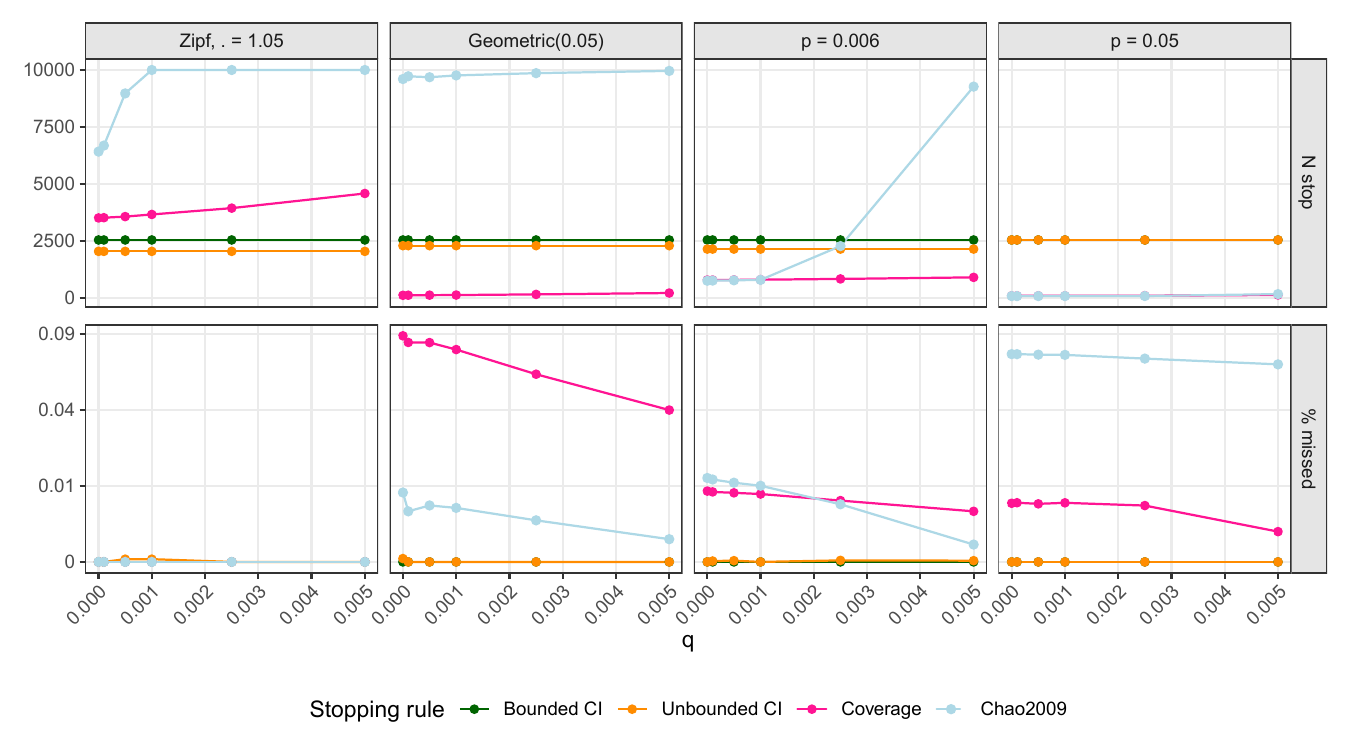}
    \caption{Stopping time $N_{\mathrm{stop}}$ (top row) and propotion of missed relevant species (bottom row) for the simulation in Section \ref{sec:SS4stopping}.}
    \label{fig:stopping_rules}
\end{figure}

We compare three stopping rules. The first is a coverage-based rule, following \citet{Montes2021}, which targets sample coverage (the proportion of individuals belonging to species already observed) using the estimator of \citet{Coverage2014}, the default choice in biodiversity applications \citep{ChaoJost2012,Coverage2014}. At step $n$ we stop when the estimated coverage exceeds $0.99$, as suggested by \citet{Coverage2014}. The other two rules are based on our confidence intervals for $M_{\max}$: one uses the bounded and the other the unbounded-alphabet interval. In both cases we stop as soon as the $(1-\alpha)$ upper endpoint, with $\alpha=0.05$, falls below $\varepsilon$.

For each scenario $(p_j)$, contamination level $q$, and stopping rule, we simulate $n_{\mathrm{reps}}=200$ independent datasets, with a cap of $n_{\max}=10{,}000$ samples. For each, we monitor: ($i$) the stopping sample size $N_{\mathrm{stop}}$, ($ii$) the proportion of missed relevant species (species with $p_j\ge\varepsilon$ that were not observed before stopping, divided by their total number); and ($iii$) the type I error, defined as the probability of stopping while at least one species with $p_j\ge\varepsilon$ is still unobserved.

Figure~\ref{fig:stopping_rules} reports, for each scenario and contamination level $q$, the mean stopping sample size and the mean proportion of missed relevant species; Tables~\ref{tab:type1} in the appendix reports the estimated type I error. By construction, both $M_{\max}$-based rules control the type I error at level $\alpha$, so that the mean proportion of missed species with $p_j\ge\varepsilon$ is essentially zero across all scenarios. The unbounded-alphabet rule is slightly more efficient than the bounded one, especially in heavy-tailed settings, leading to slightly smaller stopping times for comparable error control.

The coverage-based rule behaves quite differently. In light-tailed scenarios it tends to stop very early, yielding type I error close to one in most cases (Table~\ref{tab:type1}) and a non-negligible fraction of missed relevant species (for instance, around $10\%$ in the geometric case), while producing relatively few extra species because sampling rarely extends far into the tail or the error process. In the heavy-tailed Zipf regime the picture reverses: the coverage rule can achieve negligible type I error, but only by sampling far beyond the stopping times of our $M_{\max}$-based rules, which substantially inflates the number of extra species and the overall sampling effort.

Finally, note that the coverage rule cannot be made robust to contamination simply by raising the target coverage. If $C_{\text{target}}\ge 1-q$, where $q$ is the contamination probability, the rule never stops, since error singletons keep contributing new low-abundance species. By contrast, the $M_{\max}$-based rules retain their finite-sample guarantees and, in these simulations, continue to stop at moderate sample sizes while keeping type I error at or below the nominal level.

\subsection{Application to BBS dataset}
\label{sec:BBS_analysis}
The North American Breeding Bird Survey (BBS) is a continental monitoring program that has collected standardized data on avian populations annually since $1966$, and is a primary resource for studying large-scale patterns in bird abundance, distribution, and biodiversity across North America. Surveys are conducted by trained volunteer observers along randomly selected roadside routes, each consisting of $50$ fixed stops spaced approximately $0.8$ km apart. At each stop, observers perform a three-minute point count and record all bird species seen or heard within a radius of $400$ meters. The resulting observations are collected under a highly standardized protocol and naturally yield presence--absence information.

We focus on BBS data collected in $2019$. We aggregate observations at the route level, declaring a species present on a route if it is detected at least once among the $50$ stops. This yields $n=3194$ surveyed routes and $K_n=634$ distinct observed species. The data are publicly available through the USGS ScienceBase repository; the most recent release is documented in \citet{BBSdata} and was accessed using the \texttt{R} package \texttt{bbsAssistant} \citep{Burnett2019}.

\begin{figure}[t]
    \centering
    \includegraphics[width=0.49\linewidth]{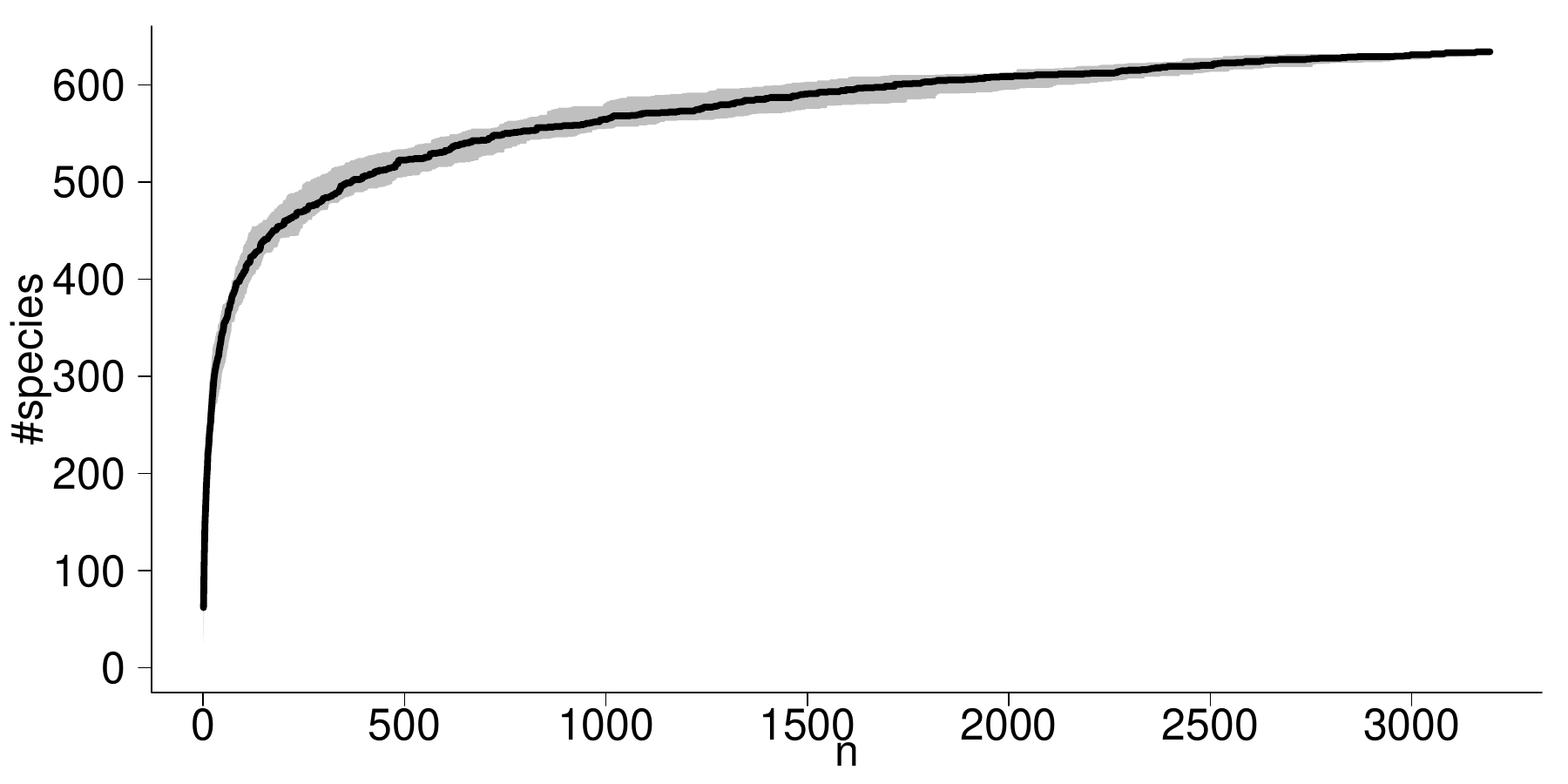}
    \hfill
    \includegraphics[width=0.49\linewidth]{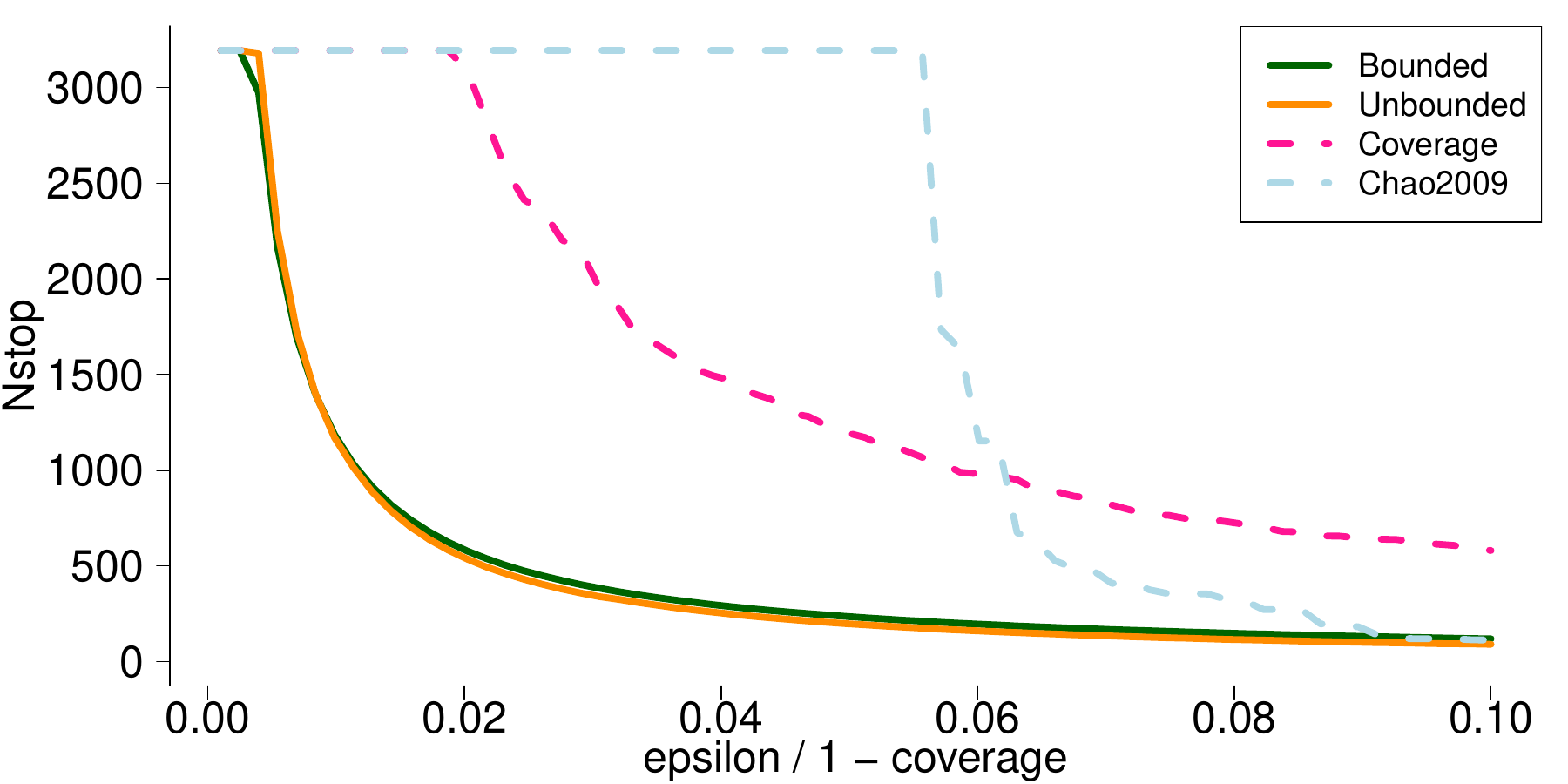}
    \caption{Left panel: accumulation curve of number of discovered birds species in BBS dataset, data refer to $2019$ only.
    Right panel: stopping sample size as a function of the tolerance level $\varepsilon$ (solid lines) or $1-\text{coverage}$ (dashed lines).}
    \label{fig:BBSdata}
\end{figure}

BBS surveys are costly in time and volunteer effort, motivating the question of whether sampling could be stopped earlier without substantial loss of information. For the $2019$ dataset, the accumulation curve (left panel of Figure~\ref{fig:BBSdata}) increases rapidly within the first $1000$ routes and then flattens, with only slightly more than $50$ new species discovered over the remaining $2000$ routes. To quantify this trade-off, we apply the stopping rule of Section~\ref{sec:stopping} and compute the smallest sample size $N_{\mathrm{stop}}<n$ such that the probability of encountering any previously unseen species is guaranteed to be below a tolerance level $\varepsilon$. Since the choice of $\varepsilon$ is application-dependent, we report $N_{\mathrm{stop}}$ over a grid of plausible values (right panel of Figure~\ref{fig:BBSdata}). For comparison, the same figure reports the corresponding stopping sample sizes obtained using the coverage-based criteria of Section~\ref{sec:SS4stopping}, evaluated over a grid of target coverage levels; for ease of comparison, the horizontal axis is reported in terms of $1-\text{coverage}$.

\section{Discussion}

The bounds derived in this paper are optimal in a minimax sense, as they match the lower bounds we exhibit in Theorems \ref{thm:lower_bound_finite} and \ref{thm:lower_bound_infinite}. 
However, such lower bounds are attained by worst-case distributions that erase most structural information.
In many practical situations, the prevalence sequence is far from least-favourable, and one should expect that substantially sharper uncertainty quantification is possible by exploiting more of the observed incidence spectrum. 



A natural direction, which we are currently pursuing, is to move from worst-case analysis to distribution-adaptive inference by introducing weak structural assumptions on the prevalence sequence. 
One pragmatic route is to treat the prevalences as random variables and use a working model to localize inference in a Bayesian setting (either nonparametric or empirical).
Developing such methods, and understanding precisely when and how they improve over minimax procedures without sacrificing robustness, seems especially promising in the unbounded setting, where uniform validity forces conservative rates but typical instances may be far more benign.

\FloatBarrier

\bibliographystyle{chicago}
\bibliography{references}

\clearpage

\newpage
\setcounter{page}{1} 
\setcounter{equation}{0}
\setcounter{section}{0}
\setcounter{table}{0}
\setcounter{figure}{0}
\renewcommand\thesection{A\arabic{section}}
\renewcommand\thetable{A\arabic{table}}
\renewcommand\thefigure{A\arabic{figure}}
\renewcommand\theequation{A\arabic{equation}}

\begin{center}
   \LARGE Appendix for:\\
    "Confidence intervals for maximum unseen probabilities, with application to sequential sampling design"
\end{center}

\section{Proofs for Section \ref{sec:unbounded}}

\begin{proof}[Proof of Proposition \ref{prop:fea_worstcase_fails}]
Fix \(r\ge 1\) and \(n\ge 1\). For any \(p\in\mathcal F\), the counts \(N_j \sim \mathrm{Bin}(n,p_j)\) are independent and
\[
E_{r,n}(p) = \sum_{j\ge 1} p_j^r \mathbb{P}_p(N_j = 0) = \sum_{j\ge 1} p_j^r (1-p_j)^n.
\]
Now consider, for each integer \(m\ge 1\), the sequence \(p^{(m)} = (p^{(m)}_j)_{j\ge 1}\) defined by
\[
p^{(m)}_j =
\begin{cases}
1/2, & j = 1,\ldots,m,\\
0, & j \ge m+1.
\end{cases}
\]
Each \(p^{(m)}\) belongs to \(\mathcal F\), since \(\sum_{j\ge 1} p^{(m)}_j = m/2 < \infty\). For this choice,
\[
E_{r,n}(p^{(m)}) = \sum_{j=1}^m \left(\frac{1}{2}\right)^r \left(1 - \frac{1}{2}\right)^n = m \left(\frac{1}{2}\right)^{r+n}.
\]
Letting \(m\to\infty\) shows that \(\sup_{p\in\mathcal F} E_{r,n}(p) \ge \sup_{m\ge 1} E_{r,n}(p^{(m)}) = \infty\), as claimed.
\end{proof}

\begin{proof}[Proof of Theorem \ref{thm:fea_no_U_n}]
    We assume that the alphabet is unbounded $(M=\infty)$. Then, for any distribution $p\in\mathcal F$, we have 
\begin{equation}
    \label{eqn:freqfeat_claim2_proof1}
    \begin{aligned}
        &\P\left(M_{\max}(N,p) \geq U_n(\alpha)\right)  \\
        &\quad =   
        1-\P\left(
        \max_{j \geq 1}\left\{p_j \indic(N_{j}=0)\right\} < U(\alpha,n)\right) \\
        &\quad =  
        1-\prod_{j \geq 1} \P\left(p_j \indic(N_{j}=0)<U(\alpha,n)\right)  
    \end{aligned}
\end{equation}
In the following, it is useful to note that 
$\P\left(p_j \indic(N_{j}=0)<U(\alpha,n)\right) = 1$ for any $p_j=0$ (this is trivial) as well as for any any $p_j = 1$, since $\indic(N_{j}=0)$ would be equal to zero almost surely.

We now fix some positive integer $m\geq1$ and define $p^\star_m   =   \left(p_{m,j}^\star\right)_{j\geq1} \in\mathcal F$ as
\begin{equation*}
    \label{eqn:freqfeat_claim2_proof2}
    p^\star_{m,j}   =   
    \begin{cases}
          1,   & j=1 \ldots m-1 \\ 
          x,   & j=m \\
          0,   & j \geq m+1 ,
    \end{cases}
\end{equation*}
for some $x\in(0,1)$. 
$p^\star$ belongs to $\mathcal F$ since 
$\sum_{j\geq1} p^\star_{m,j} = x + (m-1)<\infty$. 
Then, plugging $p^\star_m$ into Equation \eqref{eqn:freqfeat_claim2_proof1}, we get 
\begin{equation}
    \label{eqn:freqfeat_claim2_proof3}
    \begin{aligned}
        \P\left(M_{\max}(N,p^\star_m) \geq U(\alpha,n)\right)  
          =   
        1-\P\left(x\indic(N_{j}=0)<U(\alpha,n)\right) 
          =  
        1-\P\left(\indic(N_{j}=0)<\frac{U(\alpha,n)}{x}\right)  ,
    \end{aligned}
\end{equation}
where the final equality holds since $x>0$. 
$U(\alpha,n)$ implies that there are only two possibilities: 
(i) if $U(\alpha,n)/x \leq 1$, then $\P\left(\indic(N_{j}=0)<\frac{U(\alpha,n)}{x}\right) = 
\P\left(\indic(N_{j}=0)=0\right) = (1-x)^n$, and
(ii) if $U(\alpha,n)/x > 1$, then 
$\P\left(\indic(N_{j}=0)<\frac{U(\alpha,n)}{x}\right) = 1$.
We now assume $x\geq T$ so that the probability we are interested in is
\begin{equation}
    \label{eqn:freqfeat_claim2_proof4}
    \begin{aligned}
        \P\left(M_{\text {max}}\left(p^\star\right) \geq T\right)  
          =   
        1 - (1-x)^n   .
    \end{aligned}
\end{equation}
In the following, we assume there exist some constant $c>1$, possibly depending on $U(\alpha,n)$, such that 
$x = 1 - \left(1 - c \alpha\right)^{1/n}$. 
If so, then we have that
\begin{equation*}
    \label{eqn:freqfeat_claim2_proof5}
    \begin{aligned}
        \P\left(M_{\max}(N,p^\star_m) \geq U(\alpha,n)\right)
          =   
        1 - (1-x)^n 
          =   
        1 - (1 - 1 + \left(1 - c \alpha\right)^{1/n})^n 
          =   
        c \alpha > \alpha
          ,
    \end{aligned}
\end{equation*}
which ends the proof, showing that for any upper bound $U(\alpha,n)$ it is possible to define some distribution $p^\star_m$, which depends on the value of $U(\alpha,n)$, such that the required probability exceeds the $\alpha$ confidence level. 
We are only left to prove that the assumptions we made can actually hold simultaneously. Namely, we must verify that it is possible to find some $c>1$ such that, for any $n\geq 1$, $x\in(0,1)$ and $x\geq U(\alpha,n)$.
\begin{equation*}
    \label{eqn:freqfeat_claim2_proof6}
    \begin{aligned}
        & x > 0  \iff  
        1 - \left(1 - c\alpha\right)^{1/n} > 0  \iff  
        c > 0   . \\
        & x > 1 \iff 
        1 - \left(1 - c\alpha\right)^{1/n} > 1  \iff  
        c < \frac{1}{\alpha}   . \\
        & x \geq U(\alpha,n) \iff 
        1 - \left(1 - c\alpha\right)^{1/n} \geq U(\alpha,n)  \iff  
        c \geq \frac{1}{\alpha}\left[1 - (1-U(\alpha,n))^n\right]   . \\
    \end{aligned}
\end{equation*}
Hence, we have that 
$\max\left\{1,\dfrac{1}{\alpha}\left[1 - (1-U(\alpha,n))^n\right]\right\} < c < \dfrac{1}{\alpha}$.
This set is well defined since $\alpha < 1$ implies $\dfrac{1}{\alpha} > 1$ and $U(\alpha,n)\in(0,1)$ implies $\dfrac{1}{\alpha}\left[1 - (1-U(\alpha,n))^n\right] < \dfrac{1}{\alpha}$.
To conclude, note that if $U(\alpha,n)$ is only function of $n$ and $\alpha$, then $c$ also depends only on the sample size and the confidence level.
\end{proof}

\begin{proof}[Proof of Lemma \ref{lem:rnorm-infinite}]
Recall that \(M_r(N) = \sum_{j\ge 1} p_j^r \mathbbm{1}\{N_j = 0\}\) and that \(M_{\max}(N) \le (M_r(N))^{1/r}\) for every \(r\ge 1\). Thus, for any \(t>0\),
\[
\mathbb{P}_p\bigl(M_{\max}(N) > t\bigr) \le \mathbb{P}_p\bigl(M_r(N) > t^r\bigr).
\]
By Markov's inequality,
\[
\mathbb{P}_p\bigl(M_r(N) > t^r\bigr) \le \frac{\mathbb{E}_p[M_r(N)]}{t^r},
\]
so it suffices to control \(\mathbb{E}_p[M_r(N)]\). Since \(N_j \sim \mathrm{Bin}(n,p_j)\),
\[
\mathbb{P}_p(N_j = 0) = (1-p_j)^n,
\]
and therefore
\[
\mathbb{E}_p[M_r(N)] = \sum_{j\ge 1} p_j^r \mathbb{P}_p(N_j = 0) = \sum_{j\ge 1} p_j^r (1-p_j)^n.
\]
We bound this expectation using the total mass \(S\) by writing
\[
p_j^r (1-p_j)^n = p_j \cdot p_j^{r-1} (1-p_j)^n,
\]
so that
\[
\mathbb{E}_p[M_r(N)] \le \left(\sup_{x\in[0,1]} x^{r-1} (1-x)^n\right) \sum_{j\ge 1} p_j = S \sup_{x\in[0,1]} x^{r-1} (1-x)^n.
\]
The function \(x\mapsto x^{r-1}(1-x)^n\) attains its maximum at \(x^\star = (r-1)/(n+r-1)\), where
\[
x^{\star\,r-1} (1-x^\star)^n = \left(\frac{r-1}{n+r-1}\right)^{r-1} \left(\frac{n}{n+r-1}\right)^n.
\]
Thus
\[
\mathbb{E}_p[M_r(N)] \le S \left(\frac{r-1}{n+r-1}\right)^{r-1} \left(\frac{n}{n+r-1}\right)^n.
\]
Let
\[
A_r \coloneqq S \left(\frac{r-1}{n+r-1}\right)^{r-1} \left(\frac{n}{n+r-1}\right)^n.
\]
Then for any \(t>0\),
\[
\mathbb{P}_p\bigl(M_r(N) > t^r\bigr) \le \frac{A_r}{t^r}.
\]
Choosing \(t = U_n(r,S;\alpha) = (A_r/\alpha)^{1/r}\) gives \(\mathbb{P}_p(M_r(N) > U_n(r,S;\alpha)^r) \le \alpha\), and hence \(\mathbb{P}_p(M_{\max}(N) \le U_n(r,S;\alpha)) \ge 1-\alpha\).
\end{proof}

\begin{proof}[Proof of Lemma \ref{lem:Ur-asymptotic}]
For brevity write \(r\) for \(r_n\). From the definition,
\[
\log U_n(r,S;\alpha) = \frac{1}{r} \log\frac{S}{\alpha} + \frac{r-1}{r} \bigl(\log(r-1) - \log(n+r-1)\bigr) + \frac{n}{r} \bigl(\log n - \log(n+r-1)\bigr).
\]
Set \(\delta = (r-1)/n\). Then \(\delta \to 0\) and \(r \sim n\delta\). We have \(n+r-1 = n(1+\delta)\) and \(r-1 = n\delta\), so
\[
\log(r-1) = \log n + \log\delta + o(1),
\qquad
\log(n+r-1) = \log n + \log(1+\delta).
\]
Substituting,
\[
\log U_n(r,S;\alpha) = \frac{1}{r} \log\frac{S}{\alpha} + \frac{r-1}{r} \bigl(\log\delta - \log(1+\delta) + o(1)\bigr) + \frac{n}{r} \bigl(-\log(1+\delta)\bigr).
\]
Since \((r-1)/r \to 1\) and \(n/r \sim 1/\delta\),
\[
\log U_n(r,S;\alpha) = \frac{1}{r} \log\frac{S}{\alpha} + \bigl(\log\delta - \log(1+\delta)\bigr) - \frac{1}{\delta} \log(1+\delta) + o(1).
\]
Using \(\log(1+\delta) = \delta - \delta^2/2 + O(\delta^3)\),
\[
-\frac{1}{\delta} \log(1+\delta) = -1 + \frac{\delta}{2} + O(\delta^2),
\qquad
-\log(1+\delta) = -\delta + \frac{\delta^2}{2} + O(\delta^3).
\]
Hence
\[
\log U_n(r,S;\alpha) = \frac{1}{r} \log\frac{S}{\alpha} + \log\delta - 1 - \frac{\delta}{2} + O(\delta^2) + o(1).
\]
Since \(\delta = (r-1)/n \sim r/n\), we have \(\log\delta = \log r - \log n + o(1)\) and \(\delta \to 0\), so
\[
\log U_n(r,S;\alpha) = \log r - \log n - 1 + \frac{1}{r} \log\frac{S}{\alpha} + o(1).
\]
Exponentiating gives
\[
U_n(r,S;\alpha) = \frac{r}{en} \left(\frac{S}{\alpha}\right)^{1/r} \bigl(1 + o(1)\bigr),
\]
which is as in Equation \eqref{eq:asymp_U}.
\end{proof}

\begin{proof}[Proof of Theorem \ref{thm:rnorm_unbounded}]
\emph{Step 1: an upper confidence bound for \(S\).} The estimator \(\hat S = n^{-1} \sum_{j\ge 1} N_j\) is unbiased for \(S\). By \citet[Equation (2.7)]{Ayed19}, for every \(\beta\in(0,1)\),
\[
\mathbb{P}_p\left( S > \left( \sqrt{\frac{\log(1/\beta)}{2n}} + \sqrt{\frac{\log(1/\beta)}{2n} + \hat S} \right)^2 \right) \le \beta.
\]
By definition of \(S^\star\), this implies \(\mathbb{P}_p(S > S^\star) \le \beta\) and \(\mathbb{P}_p(S \le S^\star) \ge 1-\beta\). Let \(A = \{S \le S^\star\}\).

\emph{Step 2: oracle bound under known \(S\).} For any fixed \(S\) and any \(r\ge 1\), Lemma~\ref{lem:rnorm-infinite} with confidence level \(\alpha - \beta\) yields
\[
\mathbb{P}_p\bigl(M_{\max}(N) \le U_r(n,S;\alpha - \beta)\bigr) \ge 1 - (\alpha - \beta),
\]
so that
\[
\mathbb{P}_p\bigl(M_{\max}(N) > U_r(n,S;\alpha - \beta)\bigr) \le \alpha - \beta.
\]
In particular, this holds for the oracle choice \(r = r^\star\) in Equation \eqref{eq:r_star}, with \(S\) the true total mass.

\emph{Step 3: replacing \(S\) and \(r^\star\) by \(S^\star\) and \(R^\star\).} The map \(S \mapsto U_r(n,S;\gamma)\) is nondecreasing for fixed \(r\) and \(\gamma\), so on \(A\),
\[
U_{r^\star}(n,S;\alpha - \beta) \le U_{r^\star}(n,S^\star;\alpha - \beta).
\]
We now compare \(r^\star\) and \(R^\star\). Writing \(c = \log n - \log\log n\) and
\[
F(r) = \exp\left(-\frac{c}{r}\right) \left(\frac{r-1}{n+r-1}\right)^{(r-1)/r} \left(\frac{n}{n+r-1}\right)^{n/r},
\]
we may rewrite
\[
U_r(n,S;\gamma) = \left(\frac{S}{\gamma}\right)^{1/r} e F(r).
\]
For fixed \(n\), a direct computation shows that \(\frac{\mathrm{d}}{\mathrm{d}r} \log F(r) \ge 0\) whenever
\[
\log(r-1) + (r-1) \ge \log\log n + 1,
\]
that is, whenever condition \eqref{eq:cond_rstar} holds. In particular, for all \(r \ge r^\star\), \(F(r) \ge F(r^\star)\).

On the event \(A\) we have \(S^\star \ge S\), and by definition of \(R^\star\),
\[
R^\star = \log\left(\frac{S^\star}{\alpha - \beta}\right) + \log n - \log\log n \ge \log\left(\frac{S}{\alpha - \beta}\right) + \log n - \log\log n = r^\star.
\]
As both \((S/\gamma)^{1/r}\) and \(F(r)\) are nondecreasing as we move from \((S,r^\star)\) to \((S^\star,R^\star)\), it follows that on \(A\),
\[
U_{r^\star}(n,S;\alpha - \beta) \le U_{R^\star}(n,S^\star;\alpha - \beta).
\]
Hence
\[
\{M_{\max}(N) > U_{R^\star}(n,S^\star;\alpha - \beta)\} \cap A \subseteq \{M_{\max}(N) > U_{r^\star}(n,S;\alpha - \beta)\}.
\]
Taking probabilities,
\[
\mathbb{P}_p\bigl(M_{\max}(N) > U_{R^\star}(n,S^\star;\alpha - \beta), A\bigr) \le \mathbb{P}_p\bigl(M_{\max}(N) > U_{r^\star}(n,S;\alpha - \beta)\bigr) \le \alpha - \beta.
\]

Finally, on the complement \(A^c\) we have \(\mathbb{P}_p(A^c) \le \beta\). By the law of total probability,
\begin{align*}
\mathbb{P}_p\bigl(&M_{\max}(N) > U_{R^\star}(n,S^\star;\alpha - \beta)\bigr)
\\
&=\mathbb{P}_p\bigl(M_{\max}(N) > U_{R^\star}(n,S^\star;\alpha - \beta), A\bigr) + \mathbb{P}_p\bigl(M_{\max}(N) > U_{R^\star}(n,S^\star;\alpha - \beta), A^c\bigr)\\
&\le (\alpha - \beta) + \beta = \alpha.
\end{align*}
Thus \(\mathbb{P}_p(M_{\max}(N) \le U_{R^\star}(n,S^\star;\alpha - \beta)) \ge 1-\alpha\), as claimed.
\end{proof}

\begin{proof}[Proof of Theorem \ref{thm:lower_bound_infinite}]
Fix \(S>0\) and \(\alpha\in(0,1)\). For \(\varepsilon>0\), define
\[
K_\varepsilon = \left\lfloor \frac{S}{\varepsilon} \right\rfloor
\]
and consider the prevalence vector \(p^{(\varepsilon)} = (p^{(\varepsilon)}_j)_{j\ge 1}\) with
\[
p^{(\varepsilon)}_j =
\begin{cases}
\varepsilon, & j = 1,\ldots,K_\varepsilon,\\
0, & j \ge K_\varepsilon + 1,
\end{cases}
\]
together with a possible adjustment on the last coordinate to ensure \(\sum_j p^{(\varepsilon)}_j = S\). Each such \(p^{(\varepsilon)}\) belongs to \(\mathcal F\). Let \(\mathbb{P}^{(\varepsilon)}\) denote probabilities under \(p^{(\varepsilon)}\).

Under \(p^{(\varepsilon)}\), the random variables \(N_j \sim \mathrm{Bin}(n,\varepsilon)\) are i.i.d. for \(j \le K_\varepsilon\), and all remaining coordinates have probability at most \(\varepsilon\). For any \(\varepsilon>0\),
\[
\mathbb{P}^{(\varepsilon)}(M_{\max}(N) \le \varepsilon) = \mathbb{P}^{(\varepsilon)}\bigl(N_j > 0 \text{ for all } j=1,\ldots,K_\varepsilon\bigr) = \bigl(1 - (1-\varepsilon)^n\bigr)^{K_\varepsilon}.
\]
Define
\[
\Phi(\varepsilon) = \bigl(1 - (1-\varepsilon)^n\bigr)^{K_\varepsilon}, \qquad K_\varepsilon = \left\lfloor \frac{S}{\varepsilon} \right\rfloor.
\]
If \(T(N)\) is a valid upper \((1-\alpha)\)-level confidence bound for \(M_{\max}(N)\) for all \(p\in\mathcal F\), then, under \(p^{(\varepsilon)}\),
\[
\mathbb{P}^{(\varepsilon)}\bigl(T(N) \ge \varepsilon\bigr) \ge \mathbb{P}^{(\varepsilon)}\bigl(M_{\max}(N) \le \varepsilon\bigr) = \Phi(\varepsilon) \quad \text{for all } \varepsilon>0.
\]
In particular, define the critical value
\[
\varepsilon^\star = \inf\left\{\varepsilon>0 \colon \Phi(\varepsilon) \ge 1-\alpha\right\}.
\]
Then any uniformly valid procedure must satisfy \(\mathbb{P}^{(\varepsilon^\star)}(T(N) \ge \varepsilon^\star) \ge 1-\alpha\).

We derive the asymptotic behaviour of \(\varepsilon^\star\) as \(n\to\infty\). Using \((1-y)^m \le \exp(-my)\) for \(y\in[0,1]\),
\[
\Phi(\varepsilon) = \bigl(1 - (1-\varepsilon)^n\bigr)^{K_\varepsilon} \le \exp\bigl(-K_\varepsilon (1-\varepsilon)^n\bigr).
\]
If \(\Phi(\varepsilon) \ge 1-\alpha\), this implies
\[
K_\varepsilon (1-\varepsilon)^n \le -\log(1-\alpha).
\]
For \(\varepsilon\) tending to zero with \(n\), \((1-\varepsilon)^n = e^{-n\varepsilon} (1+o(1))\), so
\[
K_\varepsilon e^{-n\varepsilon} \le -\log(1-\alpha) (1+o(1)).
\]
Since \(K_\varepsilon \ge S/\varepsilon - 1\) and \(K_\varepsilon \to \infty\) in the relevant regime, we may ignore the \(-1\) and obtain
\[
\frac{S}{\varepsilon} e^{-n\varepsilon} \le -\log(1-\alpha) (1+o(1)).
\]
Set \(\gamma = -\log(1-\alpha) \in (0,\infty)\) and \(x_n = Sn/\gamma\). The inequality above can be rewritten as
\[
x_n \le n\varepsilon e^{n\varepsilon} (1+o(1)).
\]
As \(x_n \to \infty\), this is asymptotically inverted by the Lambert \(W\) function, defined by \(W(x)e^{W(x)} = x\). Neglecting the \(1+o(1)\) factor for a lower bound, any \(\varepsilon\) with \(\Phi(\varepsilon) \ge 1-\alpha\) must satisfy
\[
n\varepsilon \ge W\bigl(x_n (1+o(1))\bigr) = W(x_n) + o(1).
\]
For large \(x\),
\[
W(x) = \log x - \log\log x + o(1),
\]
so that
\[
n\varepsilon \ge \log x_n - \log\log x_n + o(1).
\]
Recalling that \(x_n = Sn/\gamma\),
\[
\varepsilon \ge \frac{1}{n} \left( \log\frac{Sn}{\gamma} - \log\log\frac{Sn}{\gamma} \right) + o\left(\frac{1}{n}\right).
\]
Expanding the logarithms,
\[
\log\frac{Sn}{\gamma} = \log\frac{S}{\gamma} + \log n,
\qquad
\log\log\frac{Sn}{\gamma} = \log\left( \log n + \log\frac{S}{\gamma} \right) = \log\log n + o(1),
\]
so that
\[
\varepsilon \ge \frac{1}{n} \left( \log\frac{S}{\gamma} + \log n - \log\log n \right) + o\left(\frac{1}{n}\right).
\]
Since \(\gamma = -\log(1-\alpha)\), this yields
\[
\varepsilon^\star \ge \frac{1}{n} \left( \log\frac{S}{-\log(1-\alpha)} + \log n - \log\log n \right) + o\left(\frac{1}{n}\right).
\]
Therefore any uniformly valid confidence procedure must satisfy
\[
T(N) \ge \varepsilon^\star \ge \frac{1}{n} \left( \log\frac{S}{-\log(1-\alpha)} + \log n - \log\log n \right) + o\left(\frac{1}{n}\right)
\]
on a set of probability at least \(1-\alpha\) under \(\mathbb{P}^{(\varepsilon^\star)}\), which is exactly \eqref{eq:lower_unbounded_rate}.
\end{proof}

\section{Bounded vs Unbounded intervals when upper-bouding $M$}\label{app:m_upperbound}

We now assess the sensitivity of the proposed procedures to misspecification of the alphabet size. The \textit{Bounded} upper bound is derived under the assumption that the alphabet size is finite and known. In practice, while it is often reasonable to assume that the alphabet is bounded, the true value of $M$ is frequently unknown. In such situations, it is natural to rely on a conservative upper bound $M_{\operatorname{guess}}$ in place of the true alphabet size.

\begin{figure}
    \centering
    \includegraphics[width=0.325\linewidth]{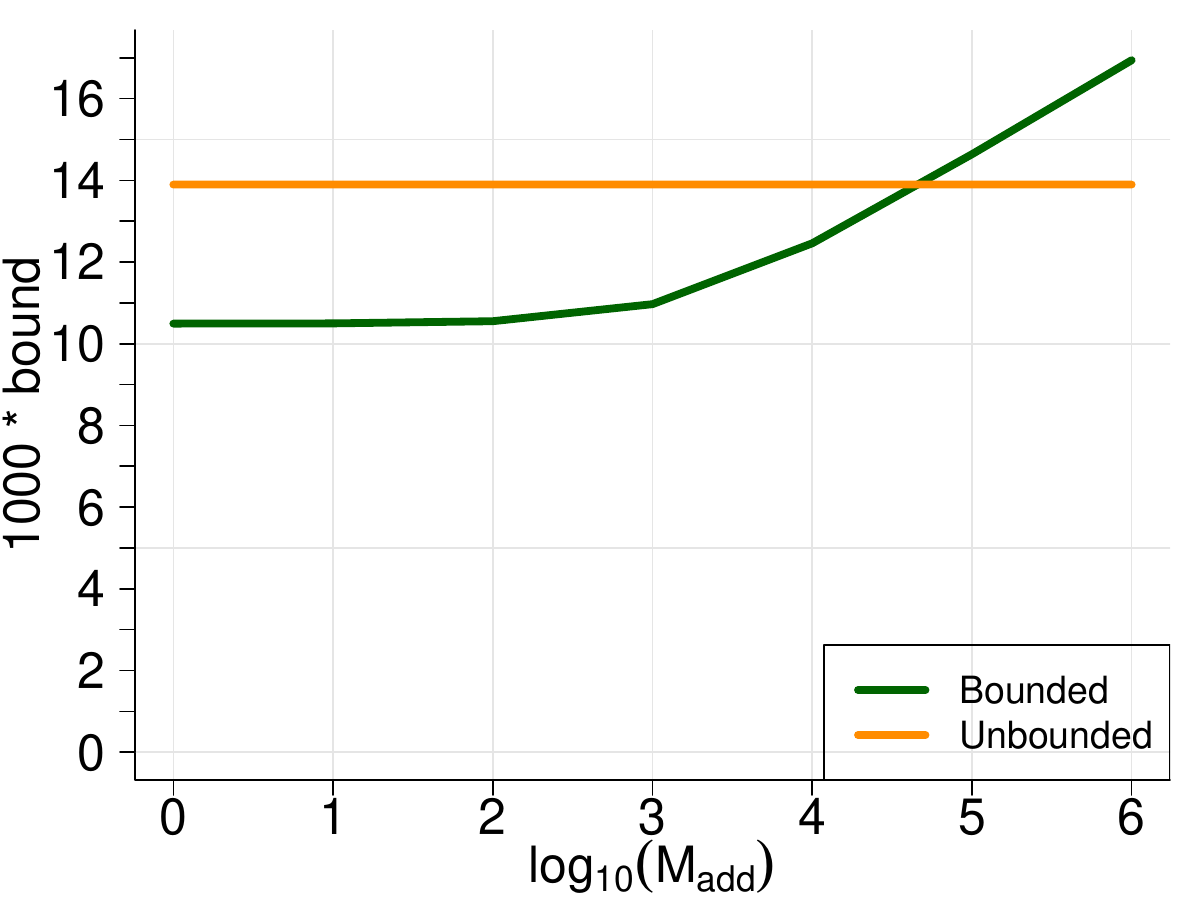}
    \hfill
    \includegraphics[width=0.325\linewidth]{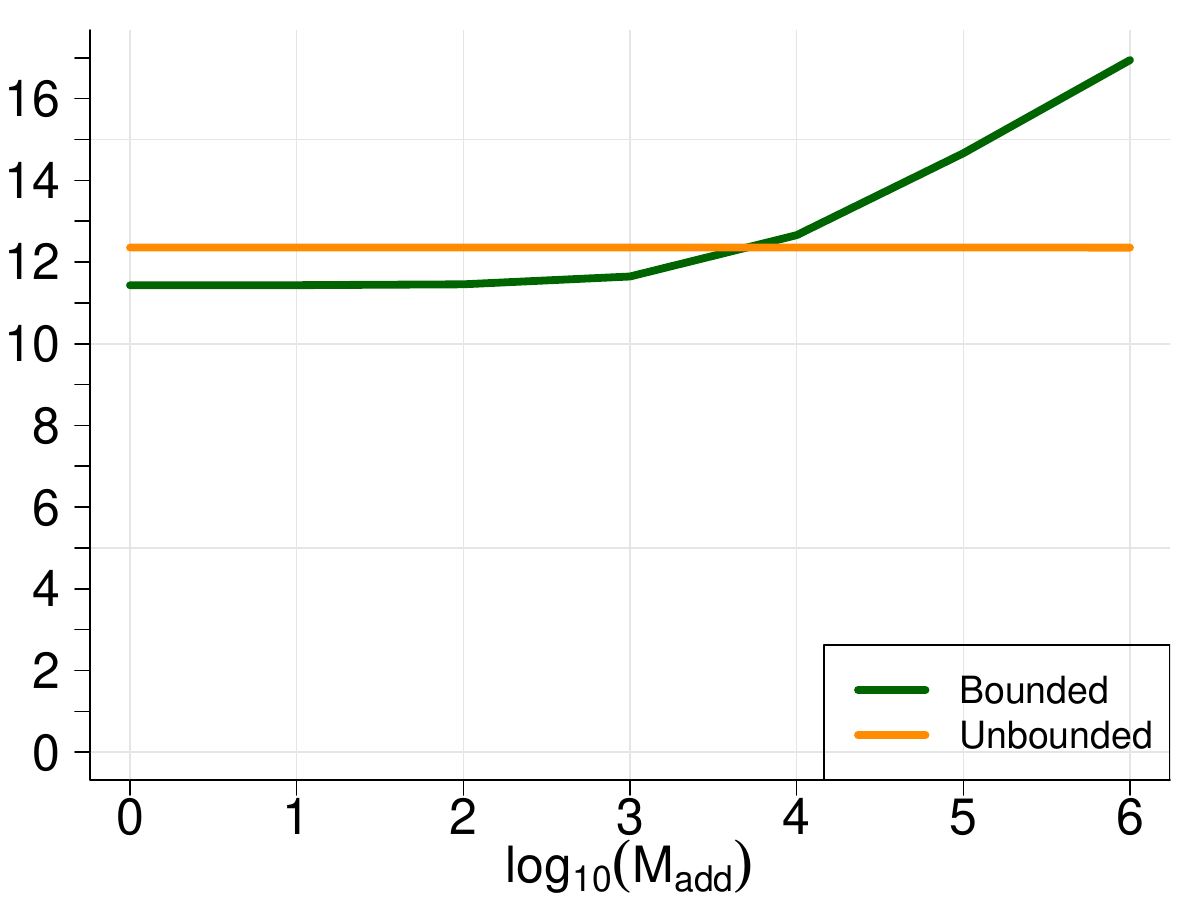}
    \hfill
    \includegraphics[width=0.325\linewidth]{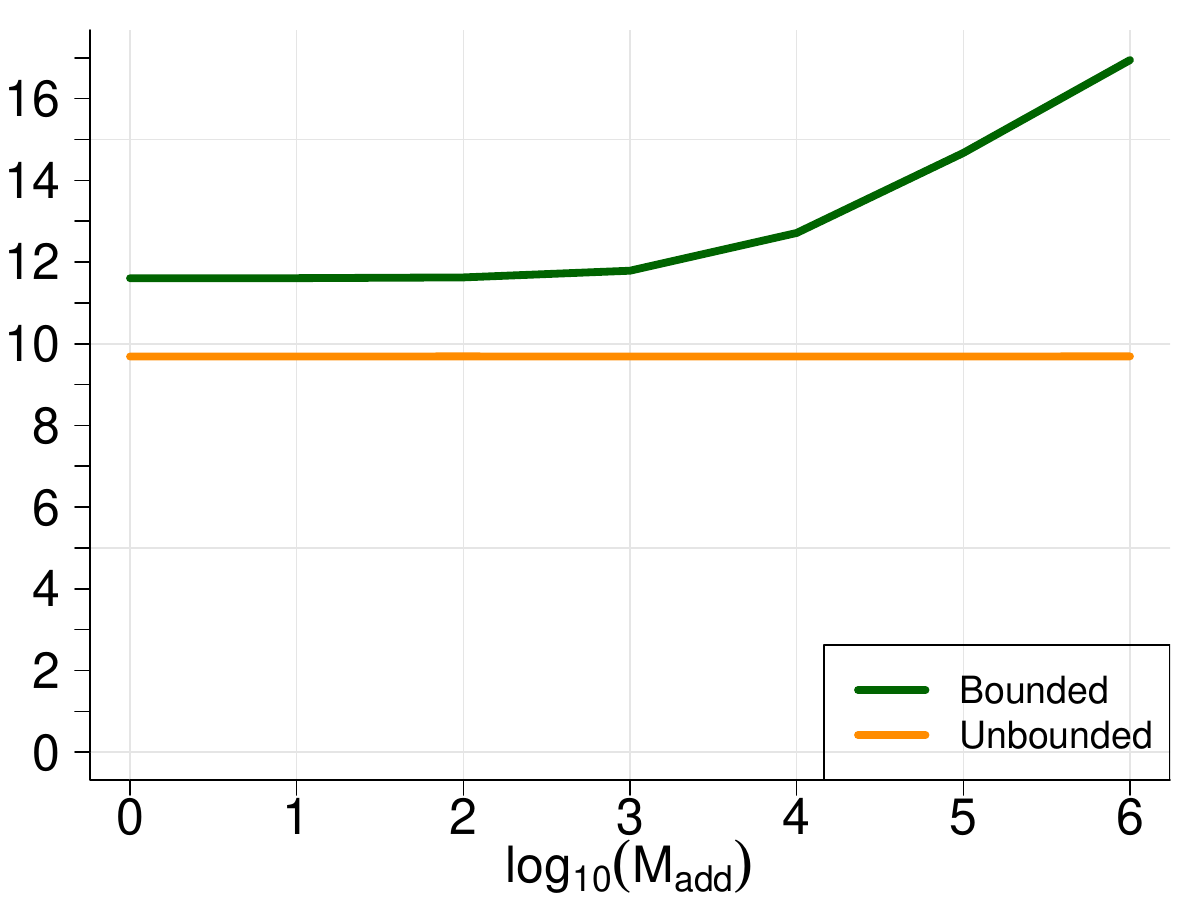}
    \caption{Length of the confidence interval as the alphabet size is overestimated by $M_{\operatorname{add}}$ extra symbols. $x$-axis is in $\log_{10}$ scale.}
    \label{fig:SSovershooting_M}
\end{figure}

In the following experiment, we investigate the robustness of the \textit{Bounded} estimator with respect to this choice, and compare it with the \textit{Unbounded} estimator, which does not depend on any specification of $M$. We consider the same data-generating mechanism as in the previous experiment, focusing on a Zipf-like distribution.
Similar conclusions hold for other benchmark distributions and are therefore omitted. 
We fix $n=1000$ and set the true (but unknown) alphabet size to $M=5000$. We then compute the \textit{Bounded} upper bound assuming an alphabet size $M_{\operatorname{guess}} = M + M_{\operatorname{add}}$, where $M_{\operatorname{add}} = 10^k$ for $k=1,\ldots,6$. For comparison, we also compute the \textit{Unbounded} upper bound. The results are reported in Figure~\ref{fig:SSovershooting_M}.

As expected, the \textit{Unbounded} upper bound is constant in $M_{\operatorname{add}}$, since it does not depend on the alphabet size and is therefore robust to misspecification of $M$. In contrast, the \textit{Bounded} upper bound remains nearly unchanged up to $M_{\operatorname{add}}=1000$, and increases noticeably only for $M_{\operatorname{add}}>10000$, i.e., when the assumed alphabet size exceeds the true one by more than a factor of two. This indicates substantial robustness to moderate overspecification of $M$. This behavior is explained by the analytical form of the \textit{Bounded} upper bound, where the dependence on $M$ enters only through a $\log(\sqrt{M})$ term, which grows very slowly with $M$.

\clearpage

\section{Additional figures and tables}

\begin{figure}[ht!]
    \centering
    \includegraphics[width=0.325\linewidth]{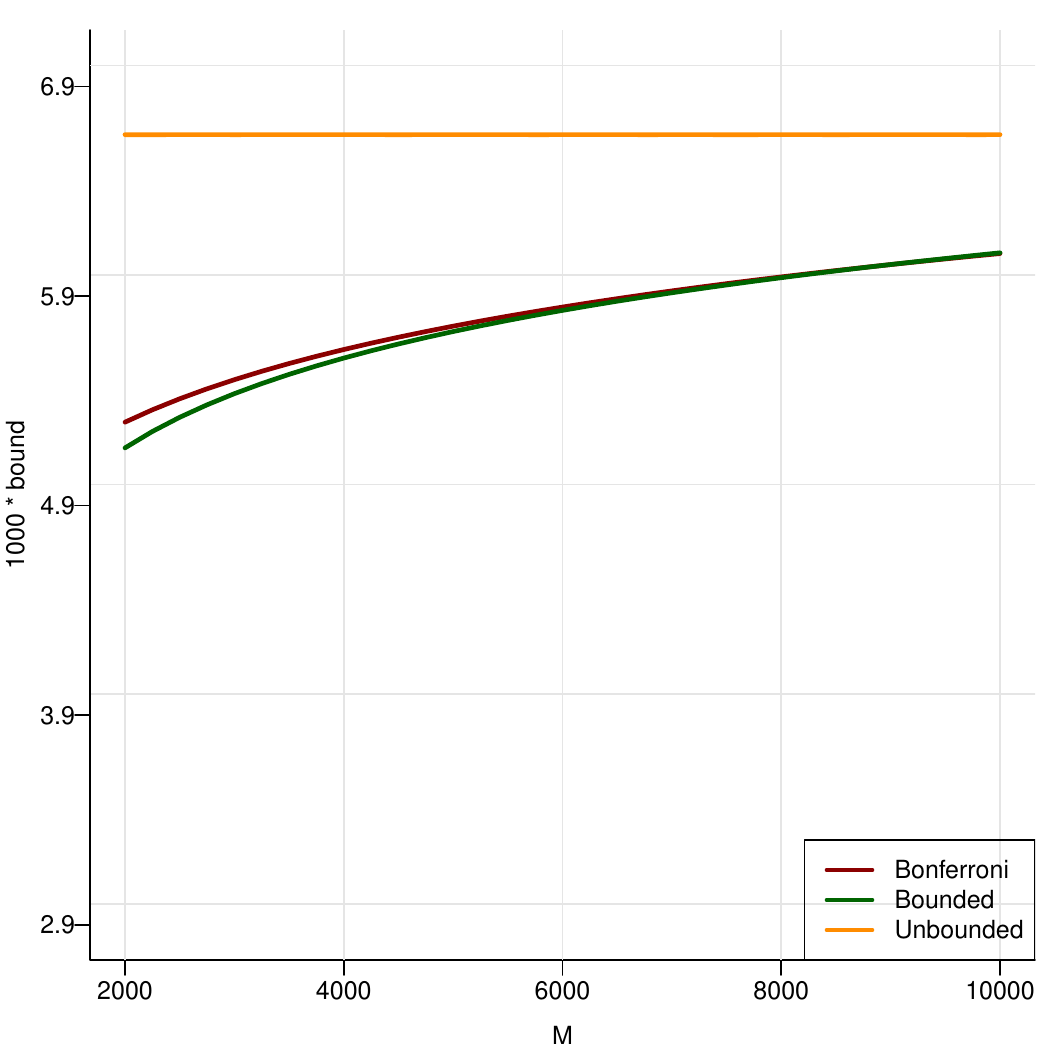}
    \hfill
    \includegraphics[width=0.325\linewidth]{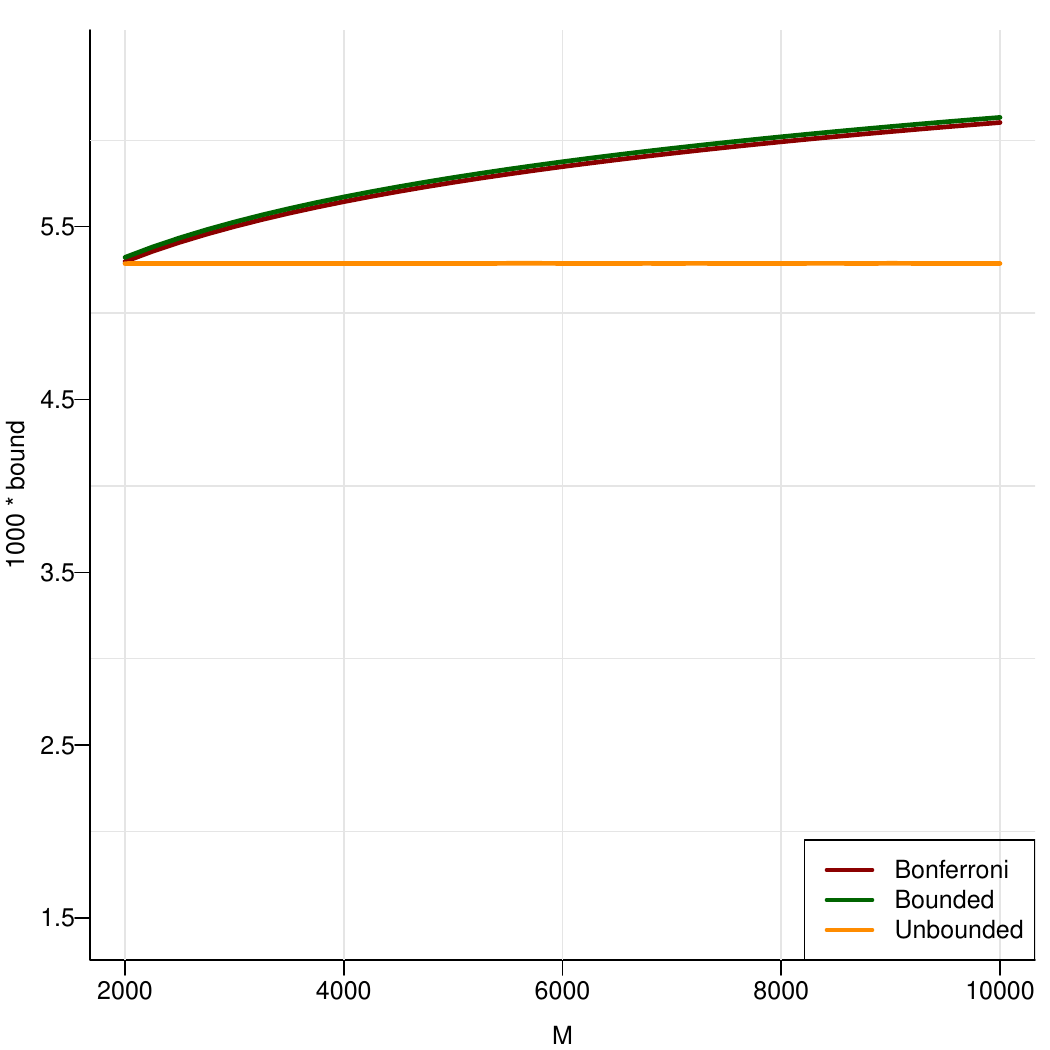}
    \hfill
    \includegraphics[width=0.325\linewidth]{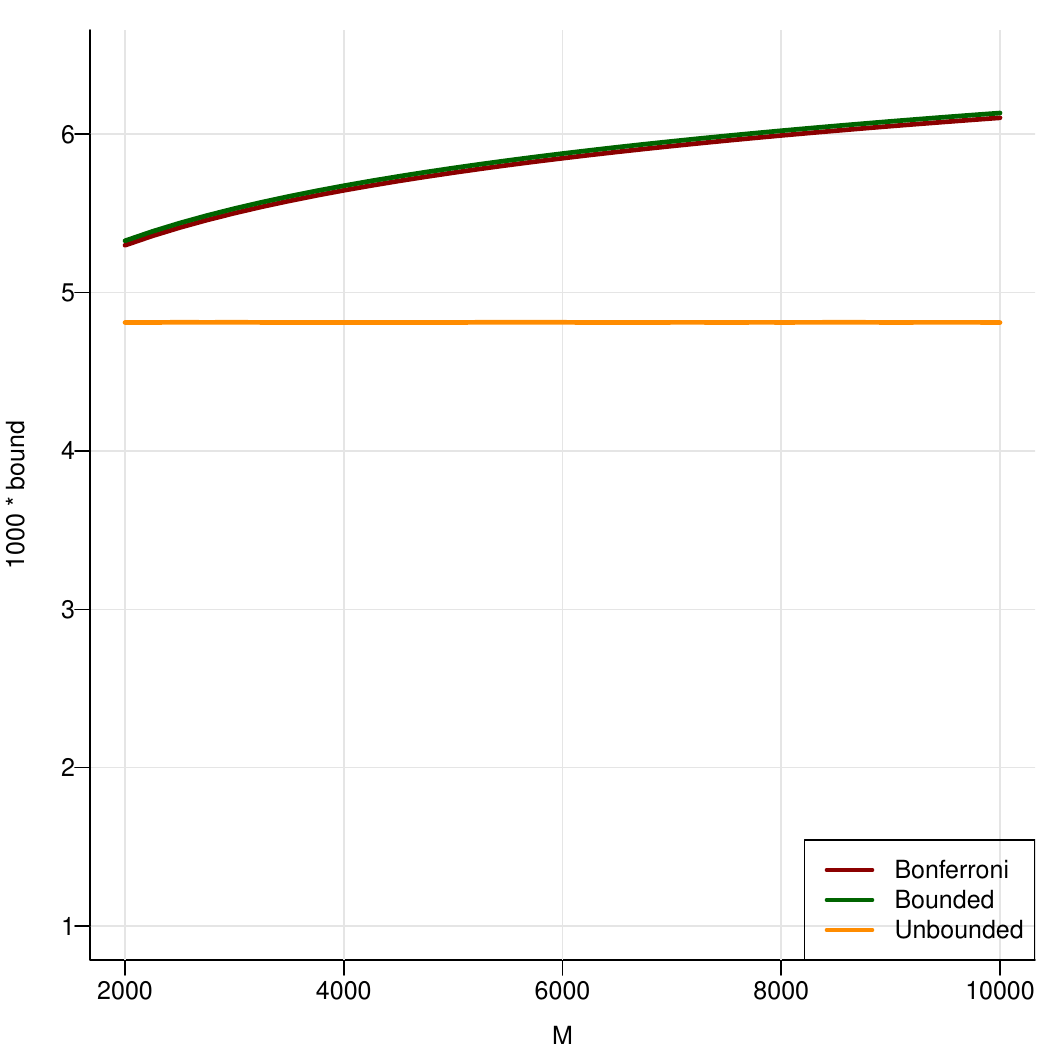}\\
    \includegraphics[width=0.325\linewidth]{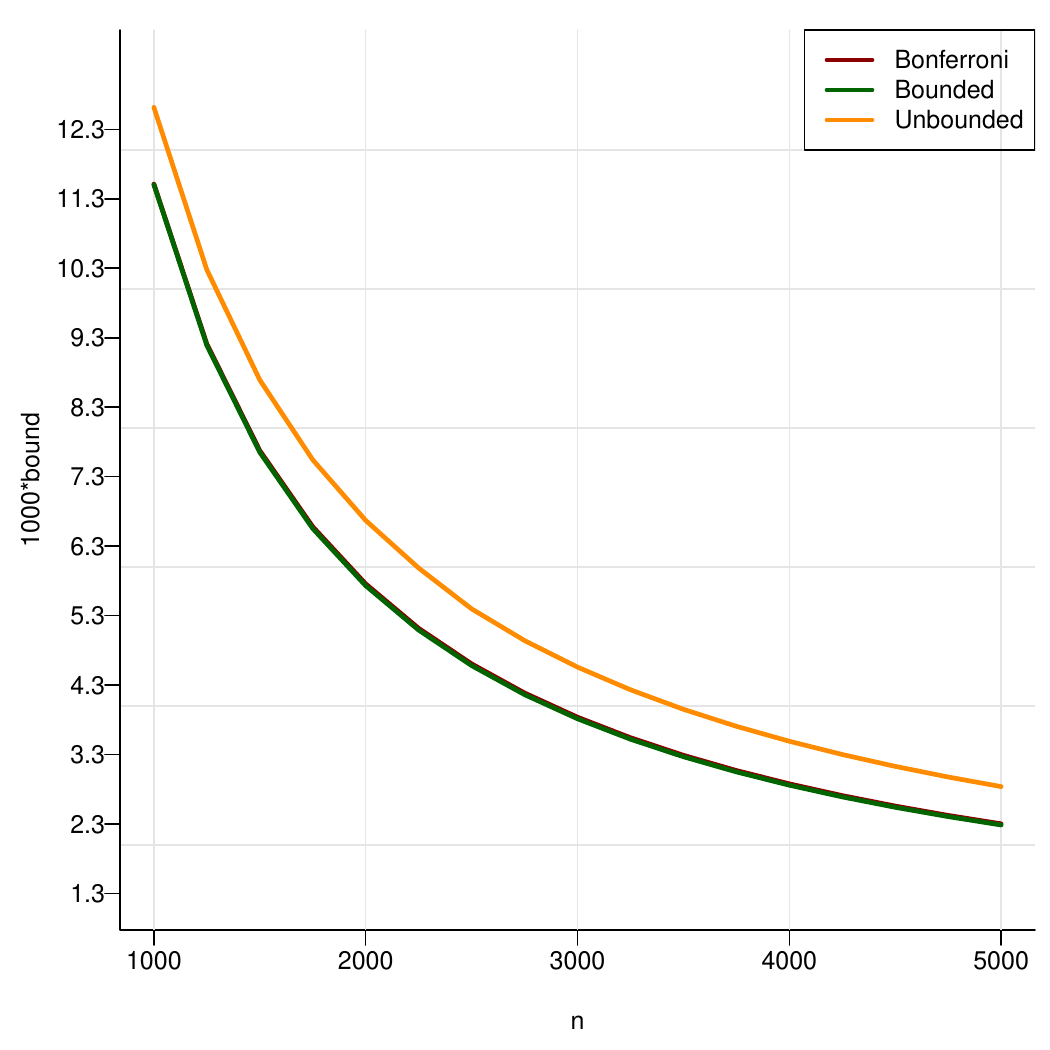}
    \hfill
    \includegraphics[width=0.325\linewidth]{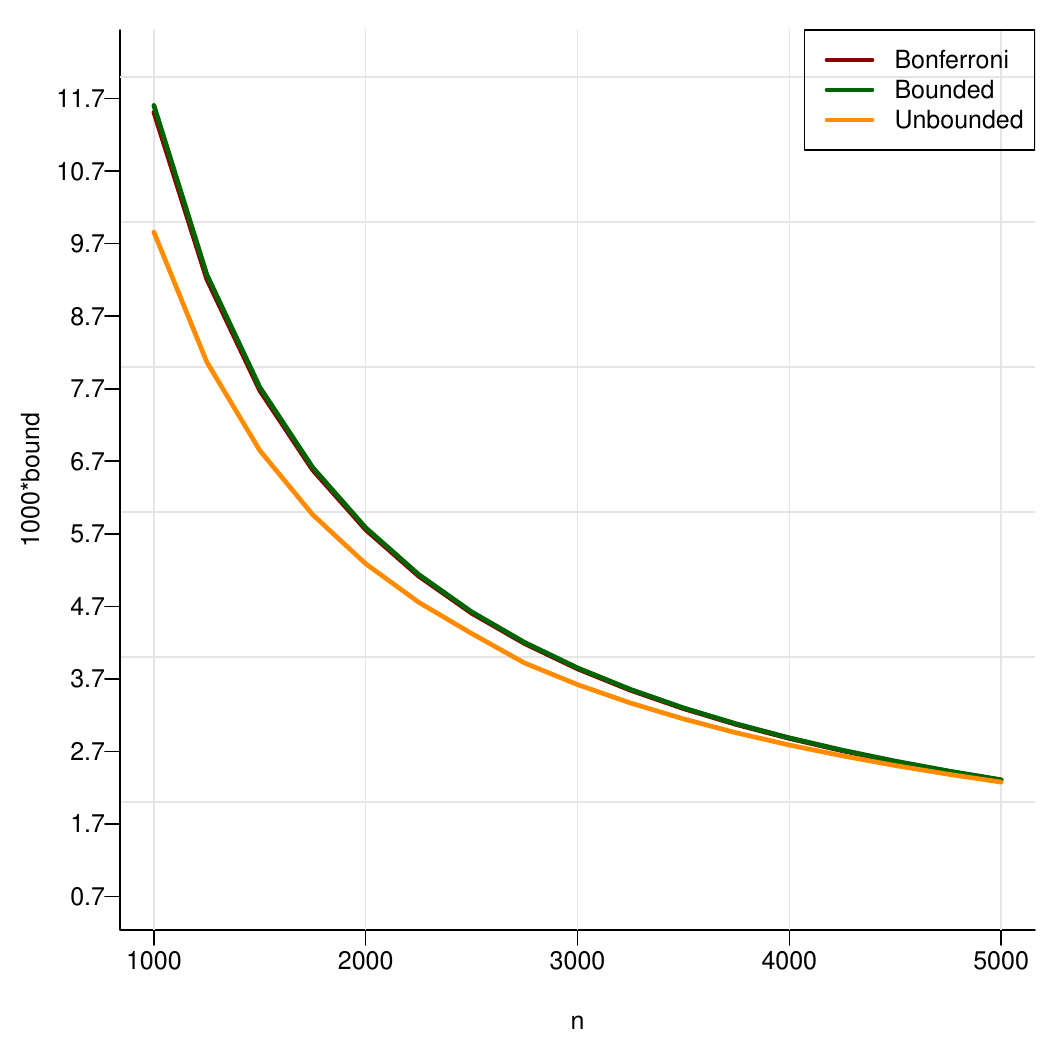}
    \hfill
    \includegraphics[width=0.325\linewidth]{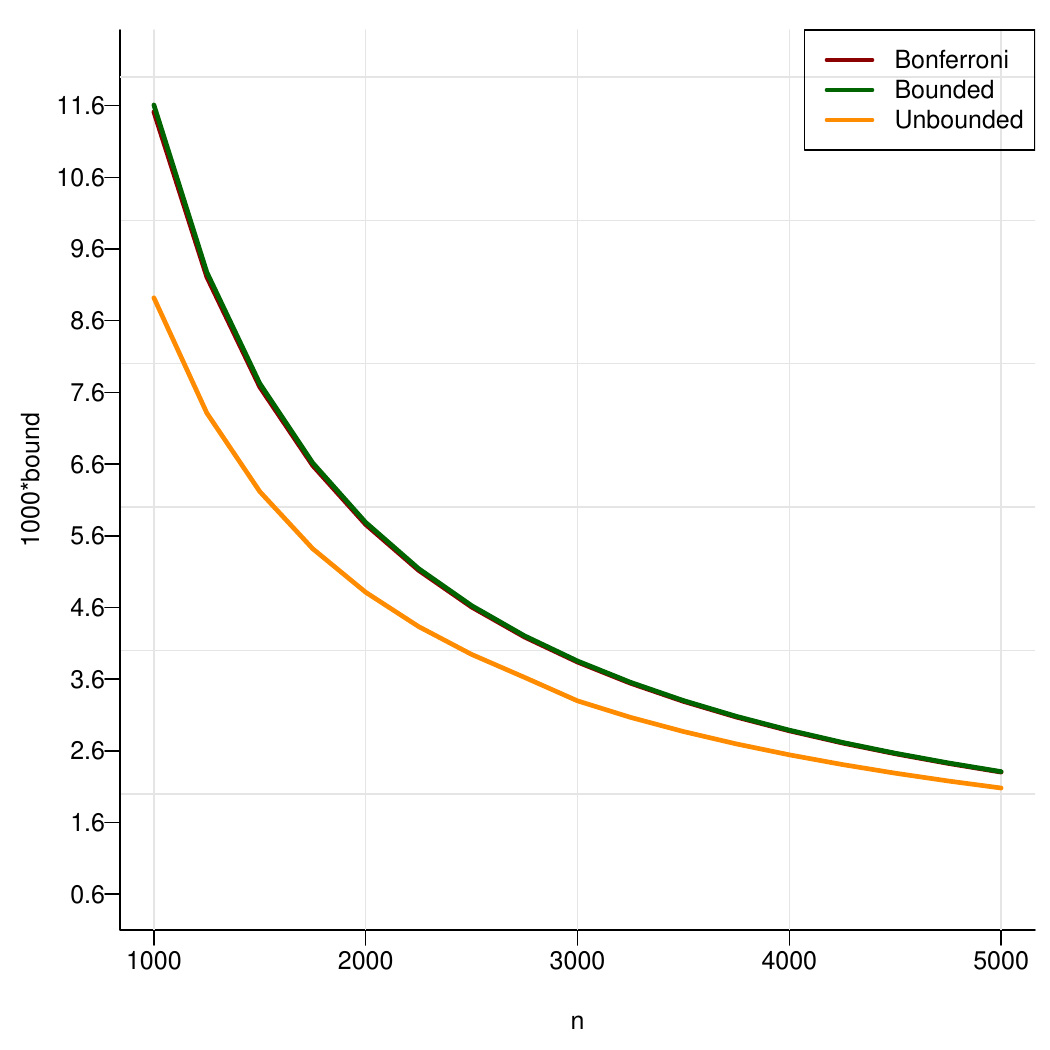}\\
    \caption{Confidence interval length under Geometric Bernoulli probabilities.
    First row: the sample size is fixed while the alphabet size $M$ varies.
    Second row: the alphabet size is fixed while the sample size $n$ increases.
    Each column corresponds to a different value of $a$: $a = 0.005$ (left), $a = 0.1$ (middle), and $a = 0.25$ (right).}
    \label{fig:Bdd_Geom}
\end{figure}

\begin{figure}[ht!]
    \centering
    \includegraphics[width=0.325\linewidth]{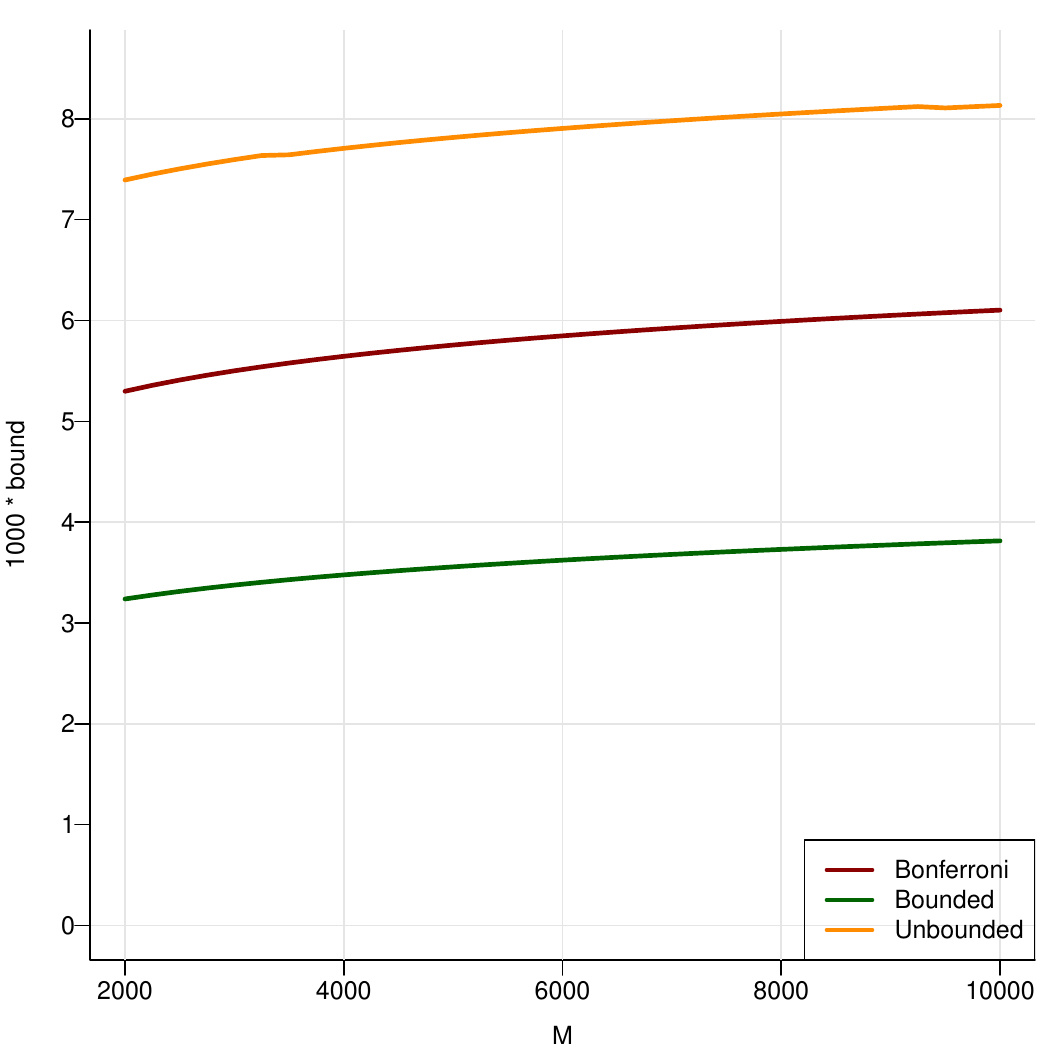}
    \hfill
    \includegraphics[width=0.325\linewidth]{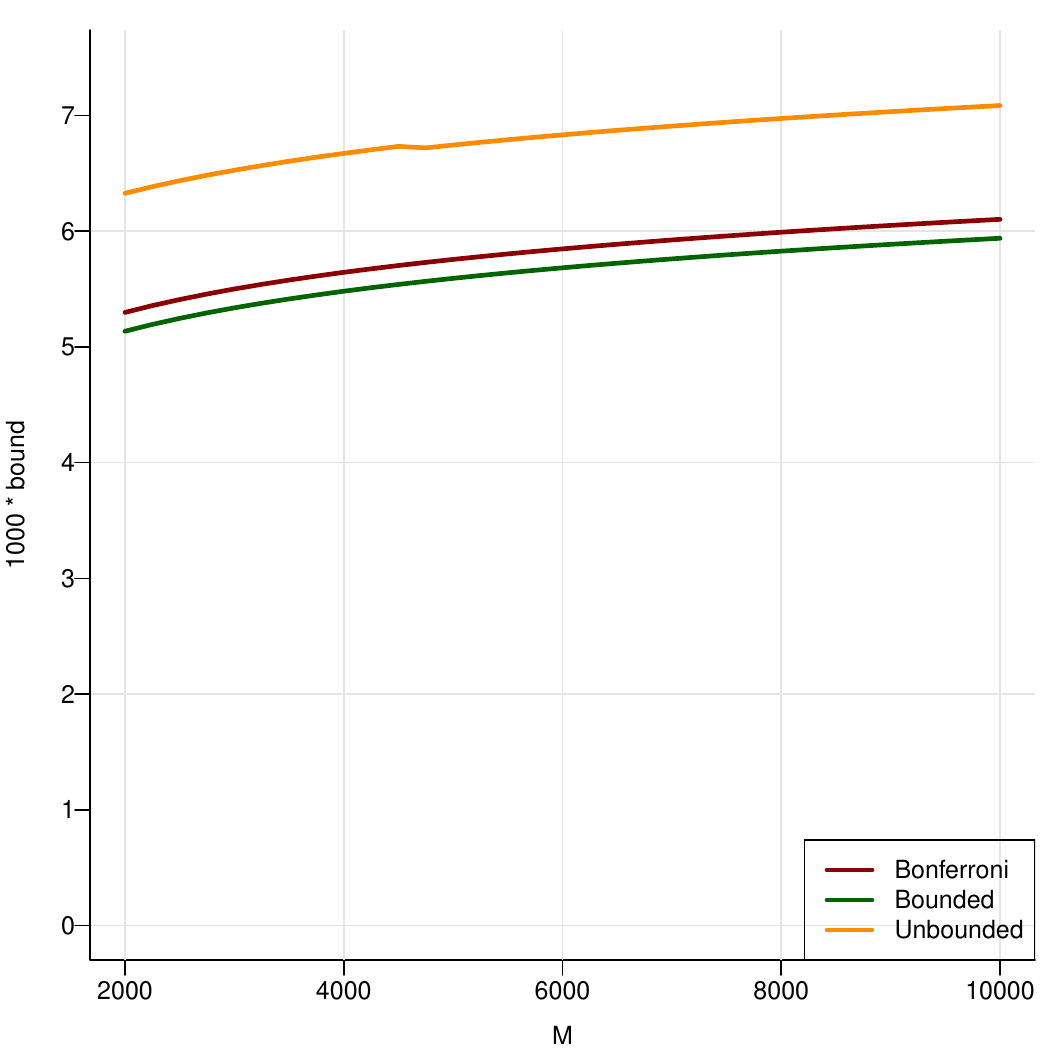}
    \hfill
    \includegraphics[width=0.325\linewidth]{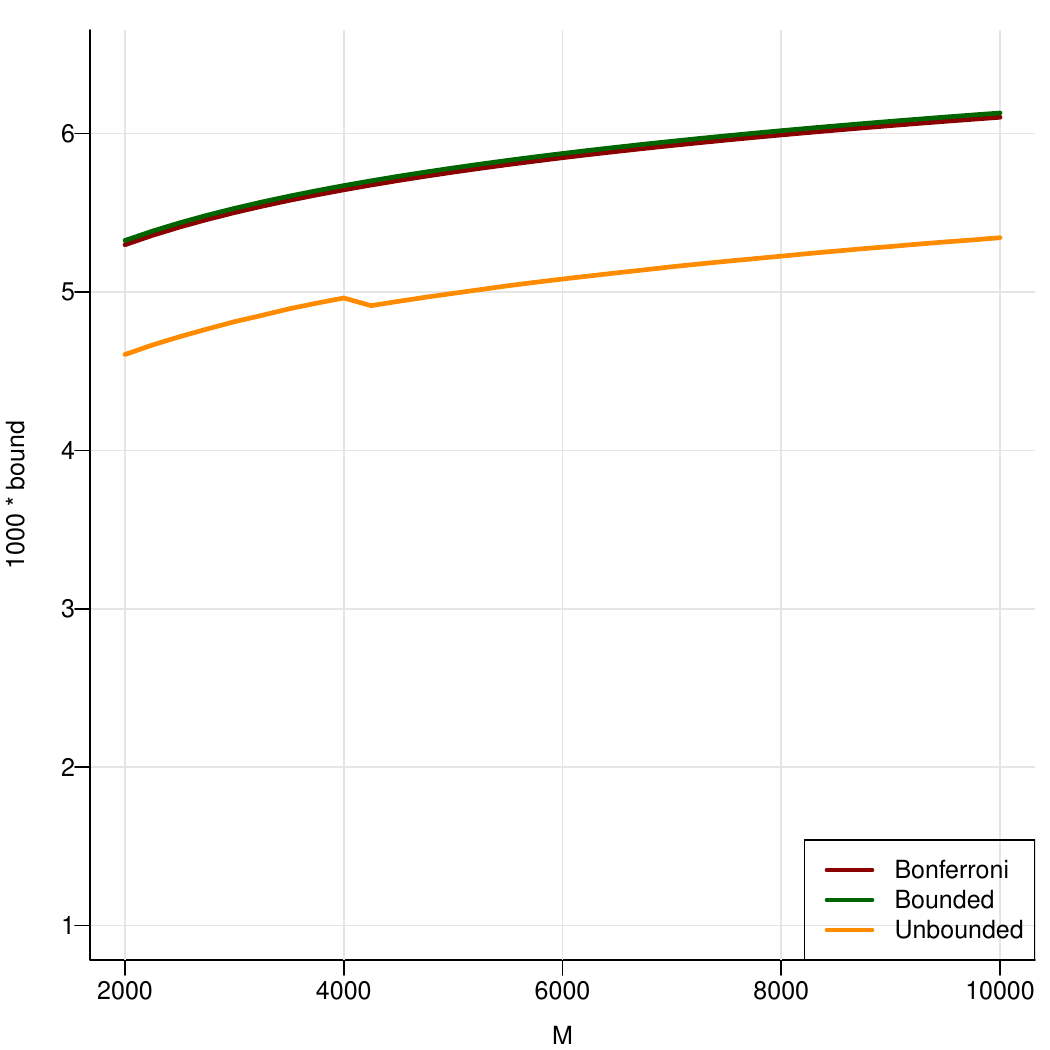}\\
    \includegraphics[width=0.325\linewidth]{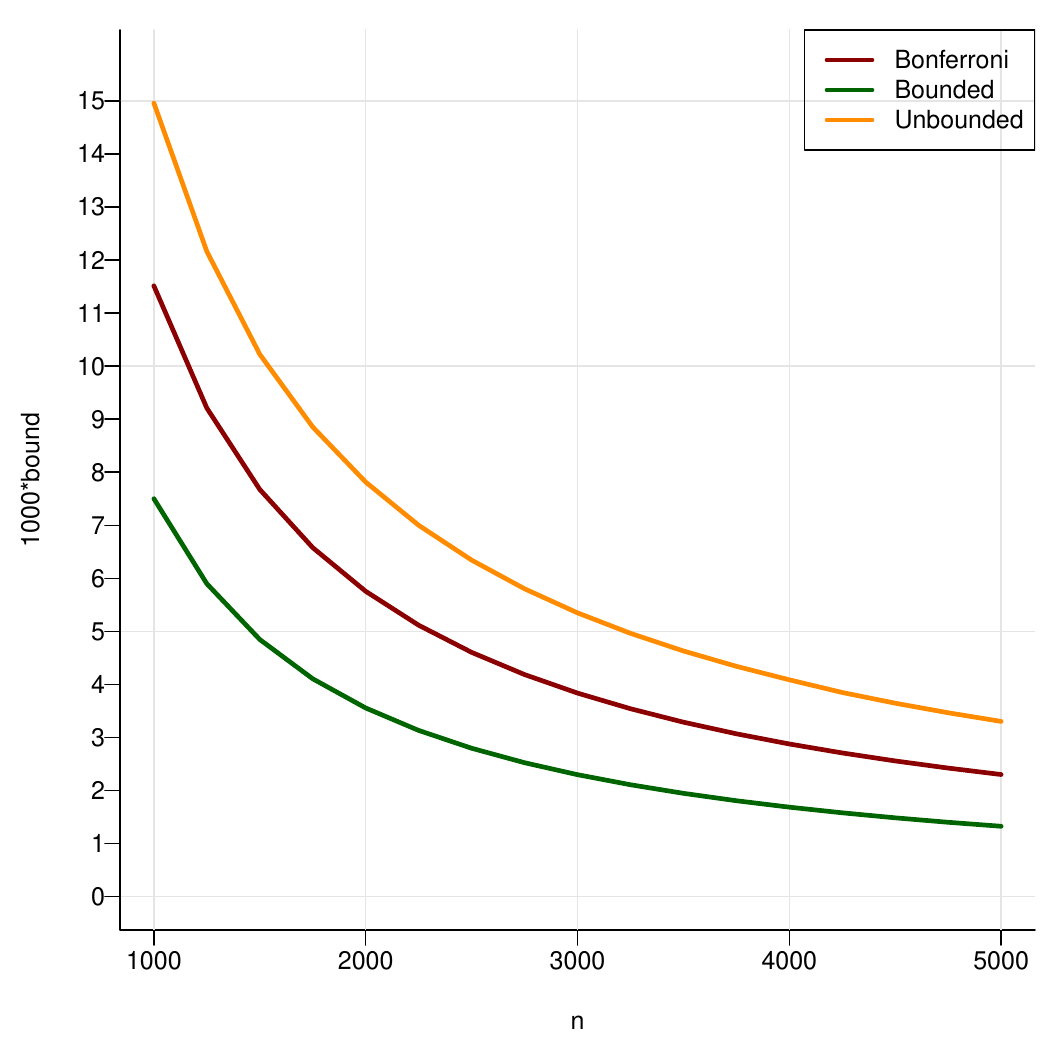}
    \hfill
    \includegraphics[width=0.325\linewidth]{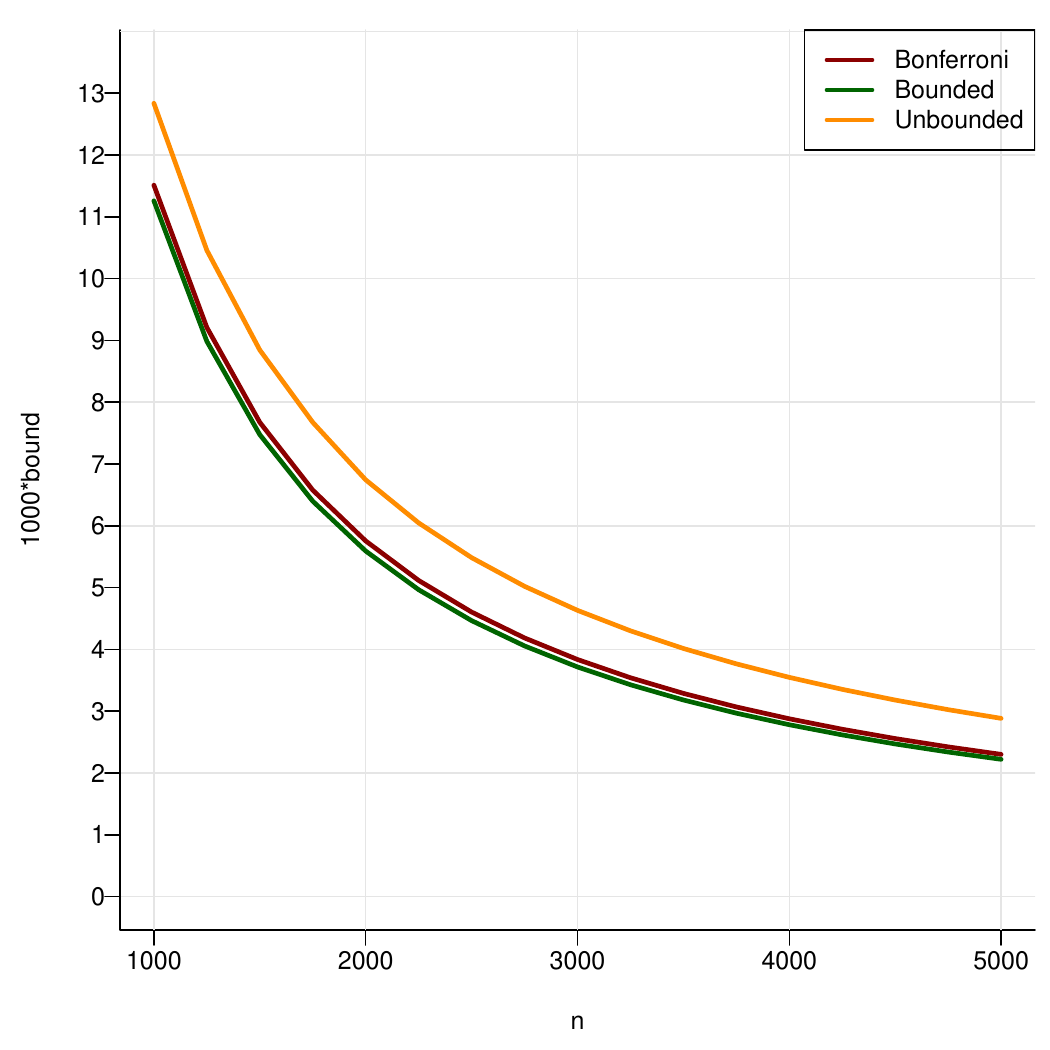}
    \hfill
    \includegraphics[width=0.325\linewidth]{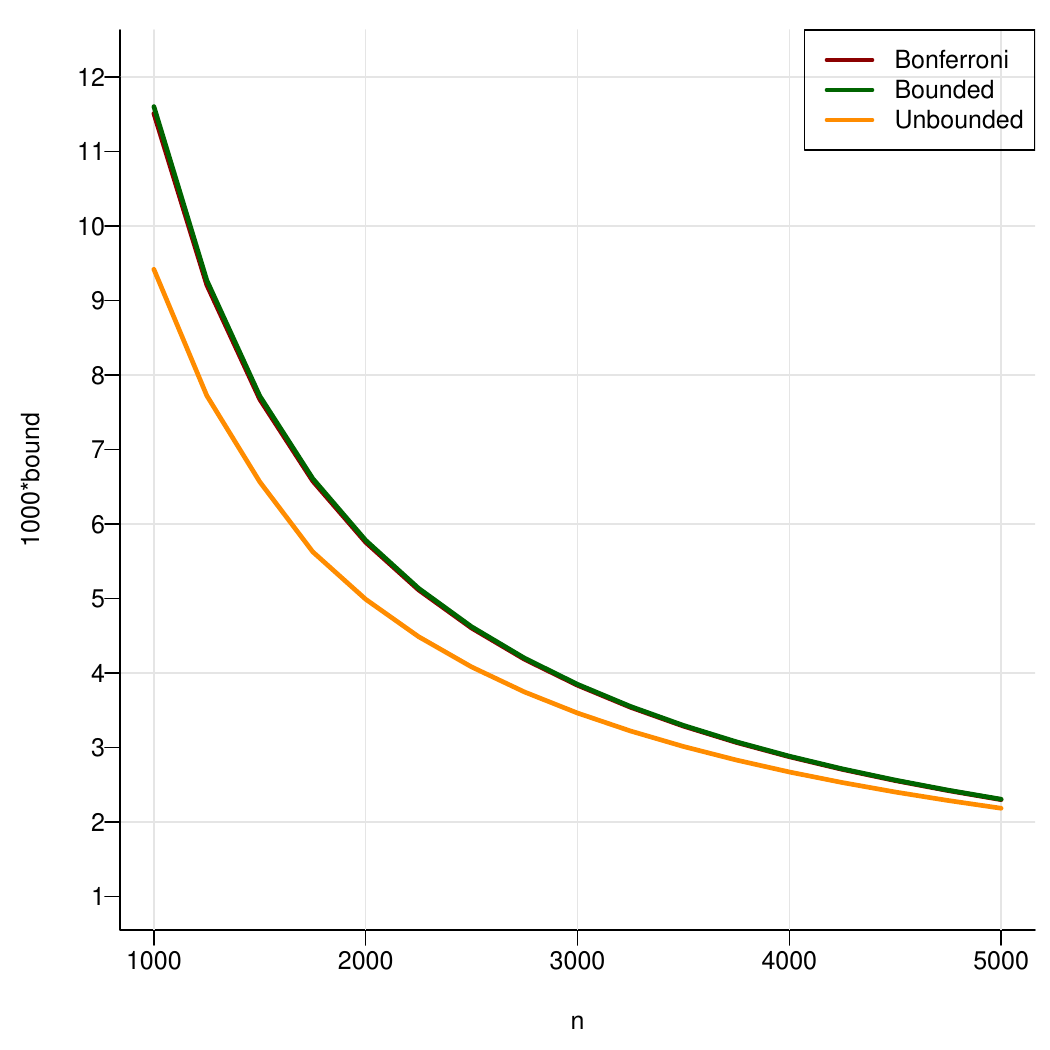}\\
    \caption{Confidence interval length under homogeneous Bernoulli probabilities.
    First row: the sample size is fixed while the alphabet size $M$ varies.
    Second row: the alphabet size is fixed while the sample size $n$ increases.
    Each column corresponds to a different value of $c$: $c = 2$ (left), $c = 20$ (middle), and $c = 1000$ (right).}
    \label{fig:Bdd_Const}
\end{figure}

\begin{figure}[ht!]
    \centering
    \includegraphics[width=0.24\linewidth]{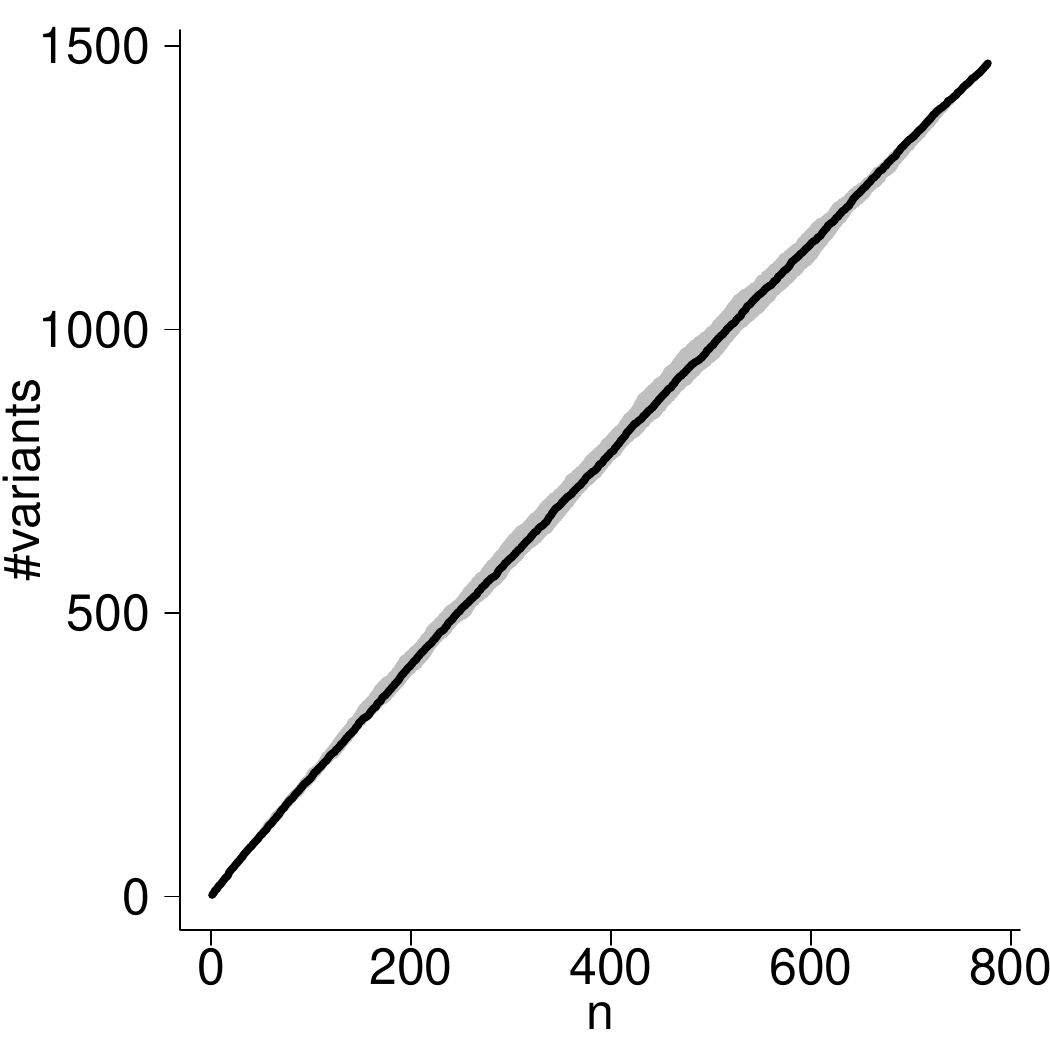}
    \hfill
    \includegraphics[width=0.24\linewidth]{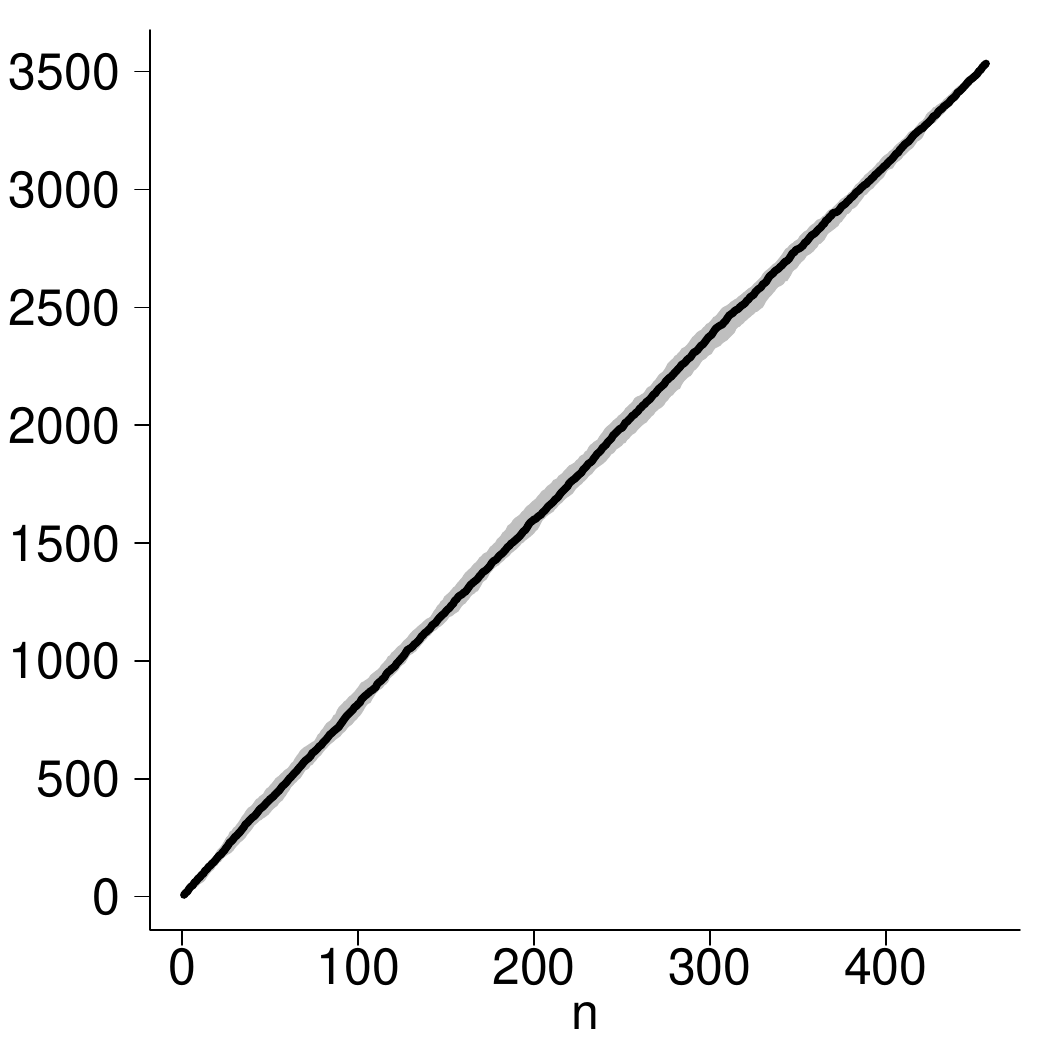}
    \hfill
    \includegraphics[width=0.24\linewidth]{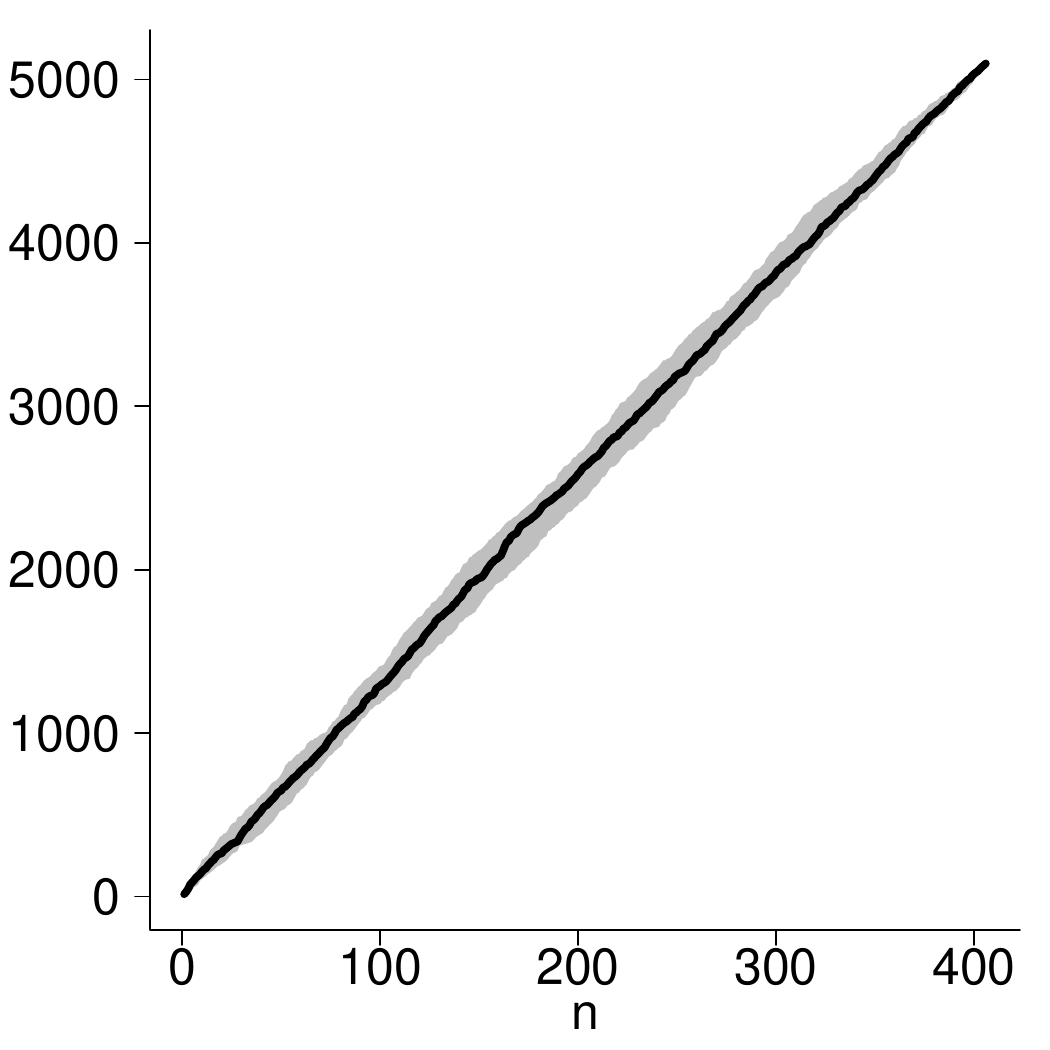}
    \hfill
    \includegraphics[width=0.24\linewidth]{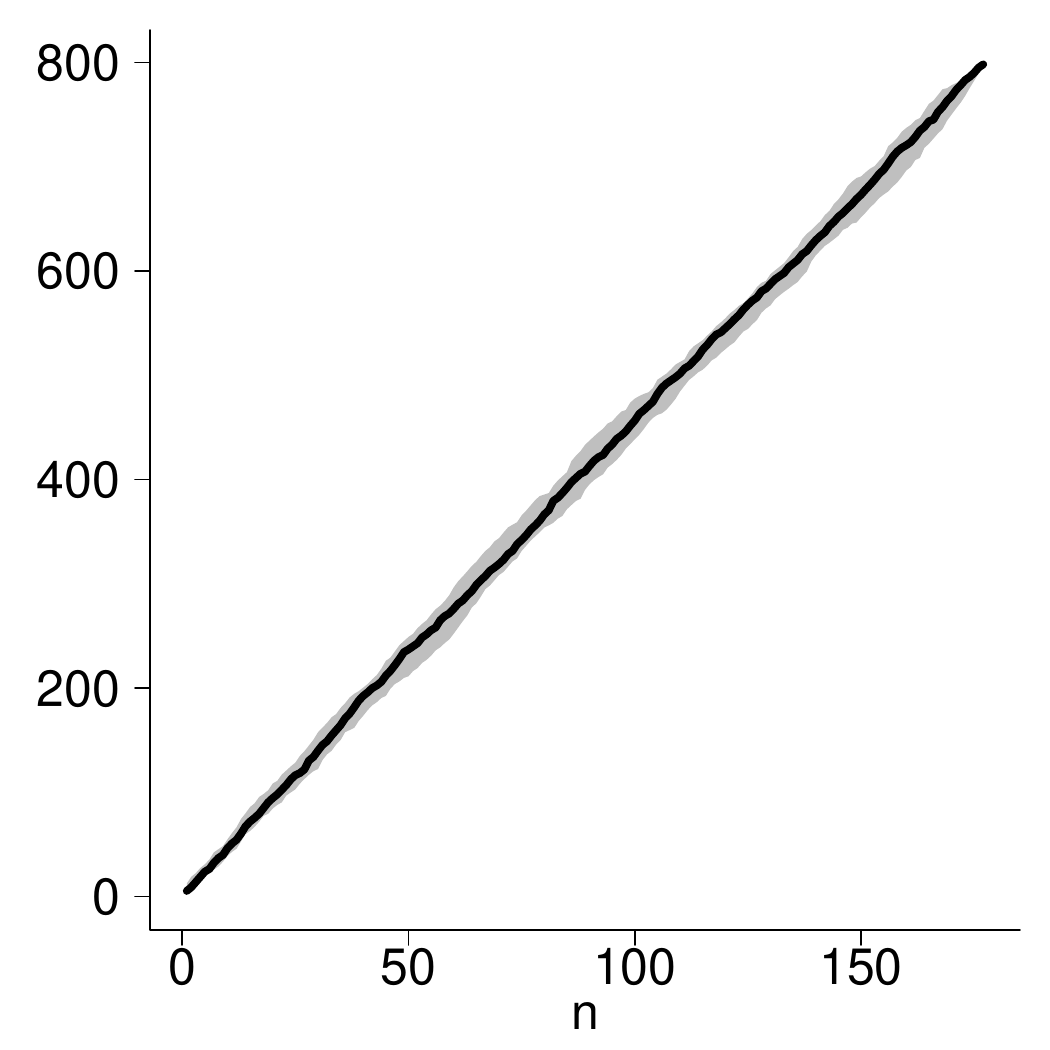}
    \caption{Accumulation curves for four selected TCGA cancer types. Panels correspond to BRCA, LUSC, SKCM, and ESCA, from left to right.}
    \label{fig:TCGA_ExtCurves}
\end{figure}

\begin{figure}[ht!]
    \centering
    \includegraphics[width=\linewidth]{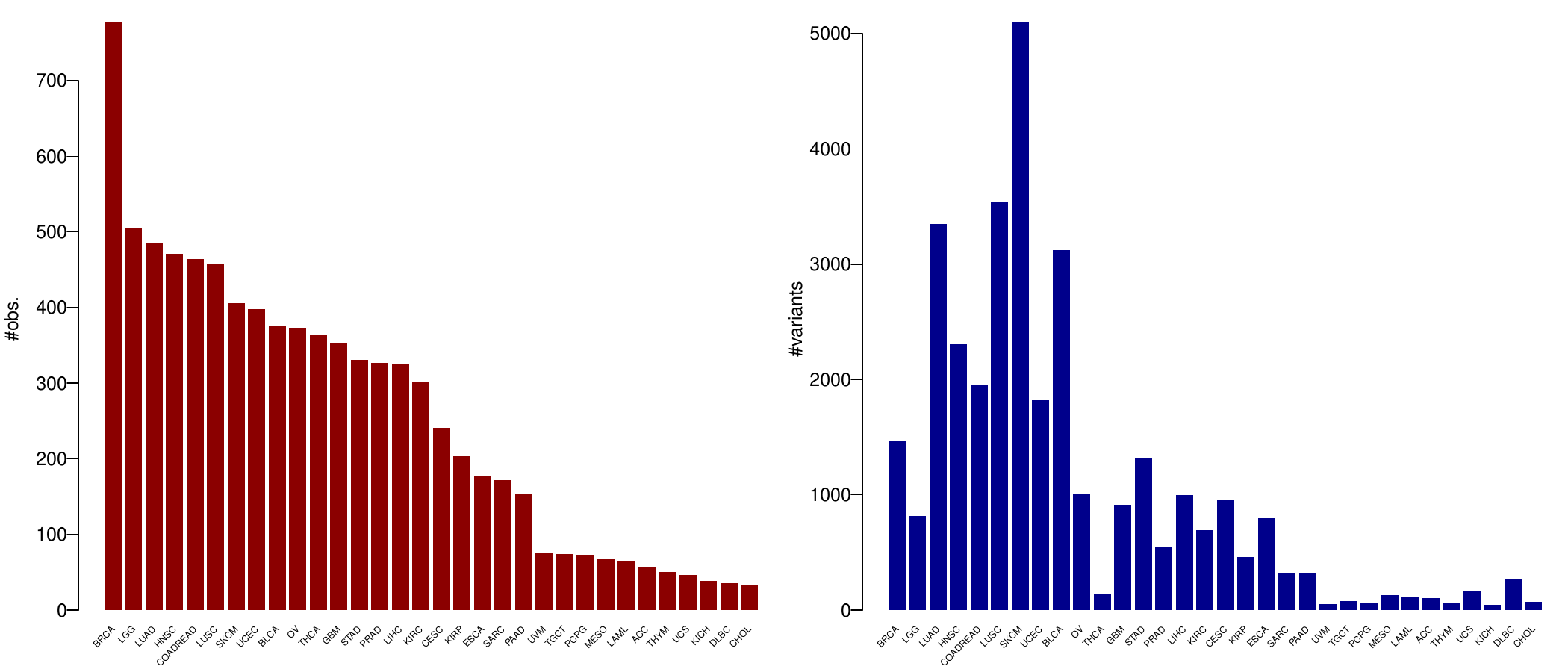}
    \caption{Sample sizes (left panel) and number of discovered variants (right panel) for each type of cancer.
    }
    \label{fig:TCGA_sizes}
\end{figure}

\begin{figure}[ht!]
    \centering
    \includegraphics[width=0.24\linewidth]{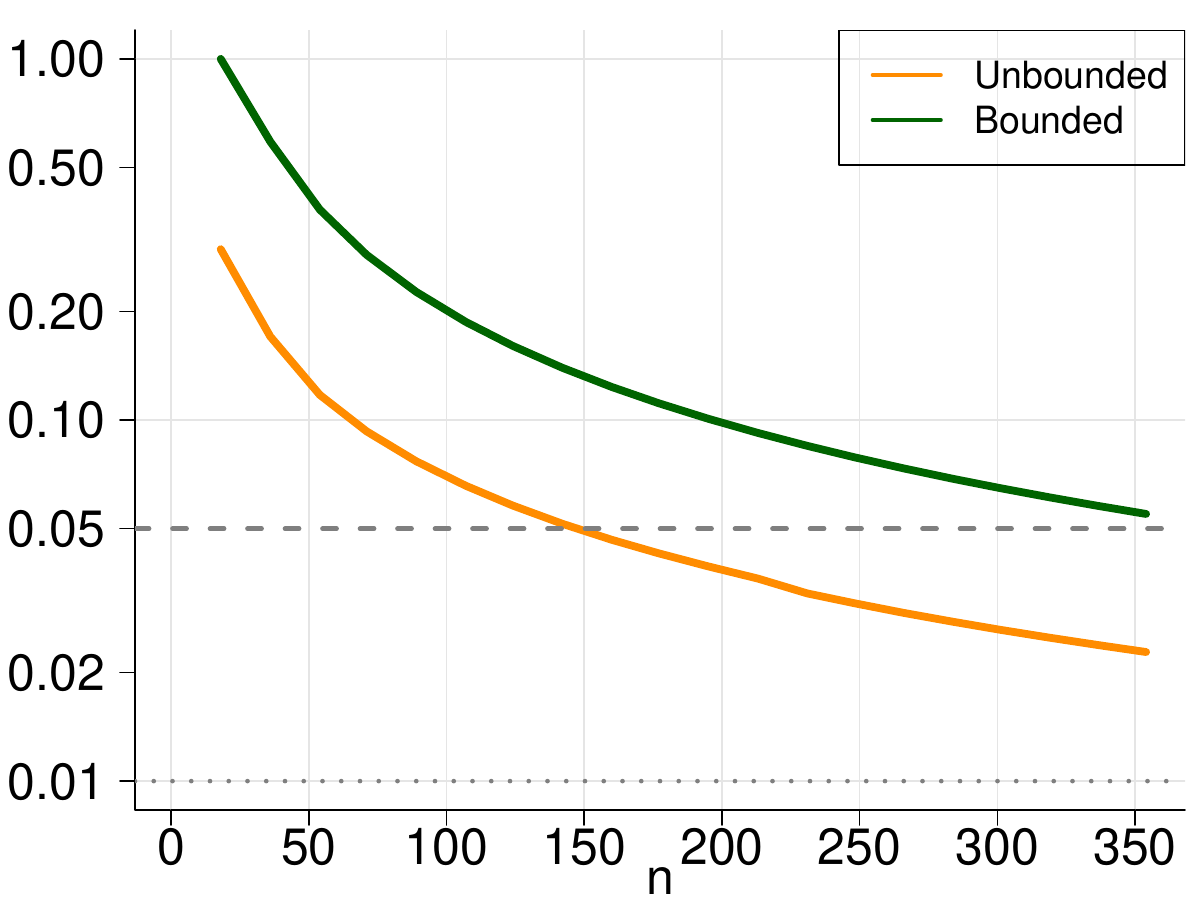}
    \hfill
    \includegraphics[width=0.24\linewidth]{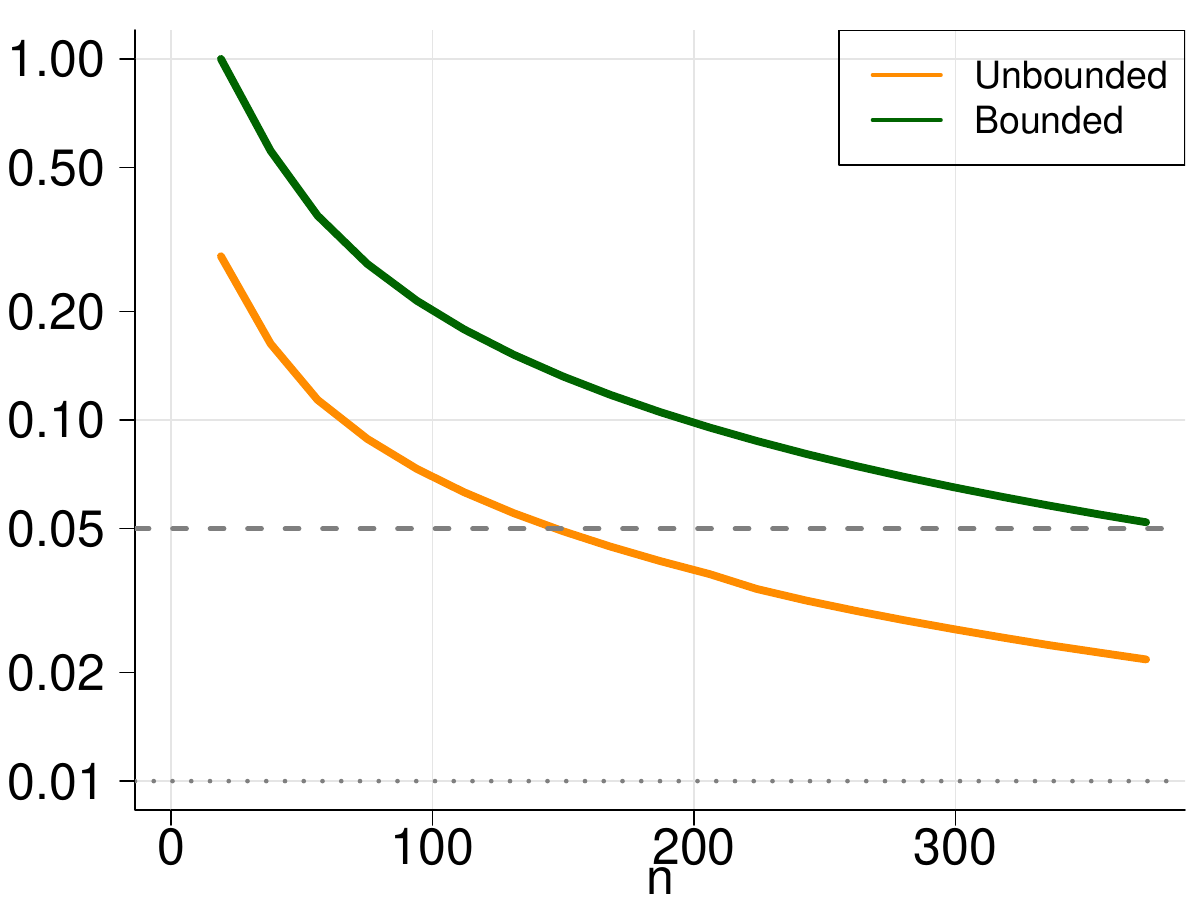}
    \hfill
    \includegraphics[width=0.24\linewidth]{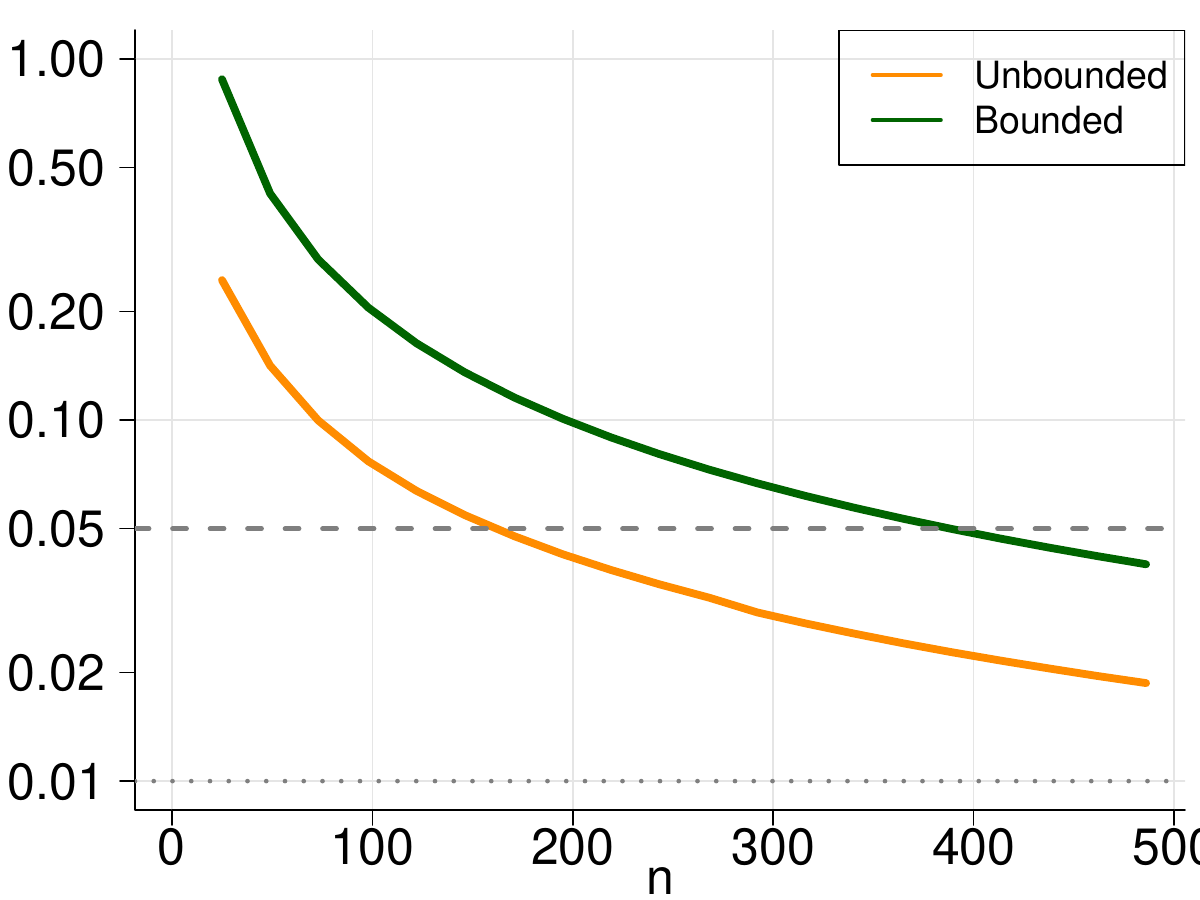}
    \hfill
    \includegraphics[width=0.24\linewidth]{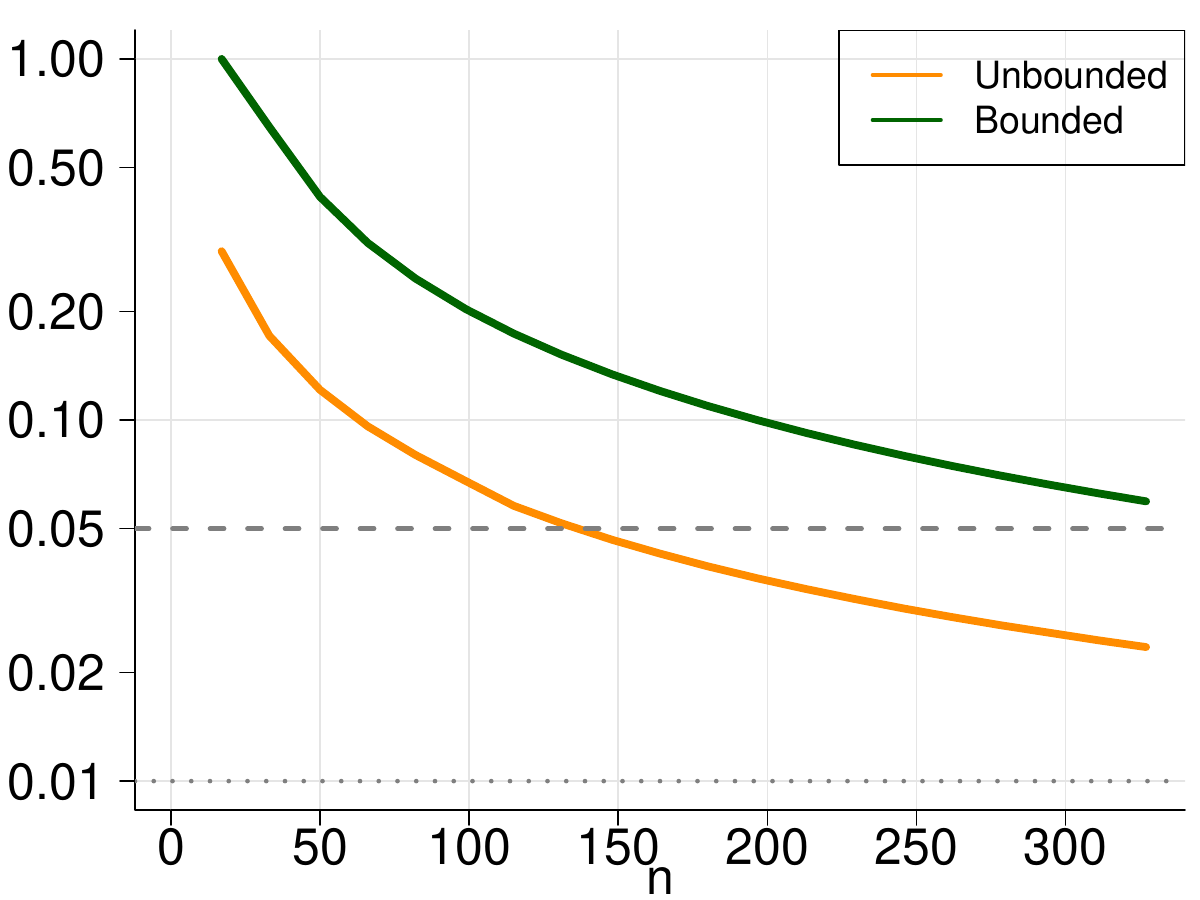} \\
    \includegraphics[width=0.24\linewidth]{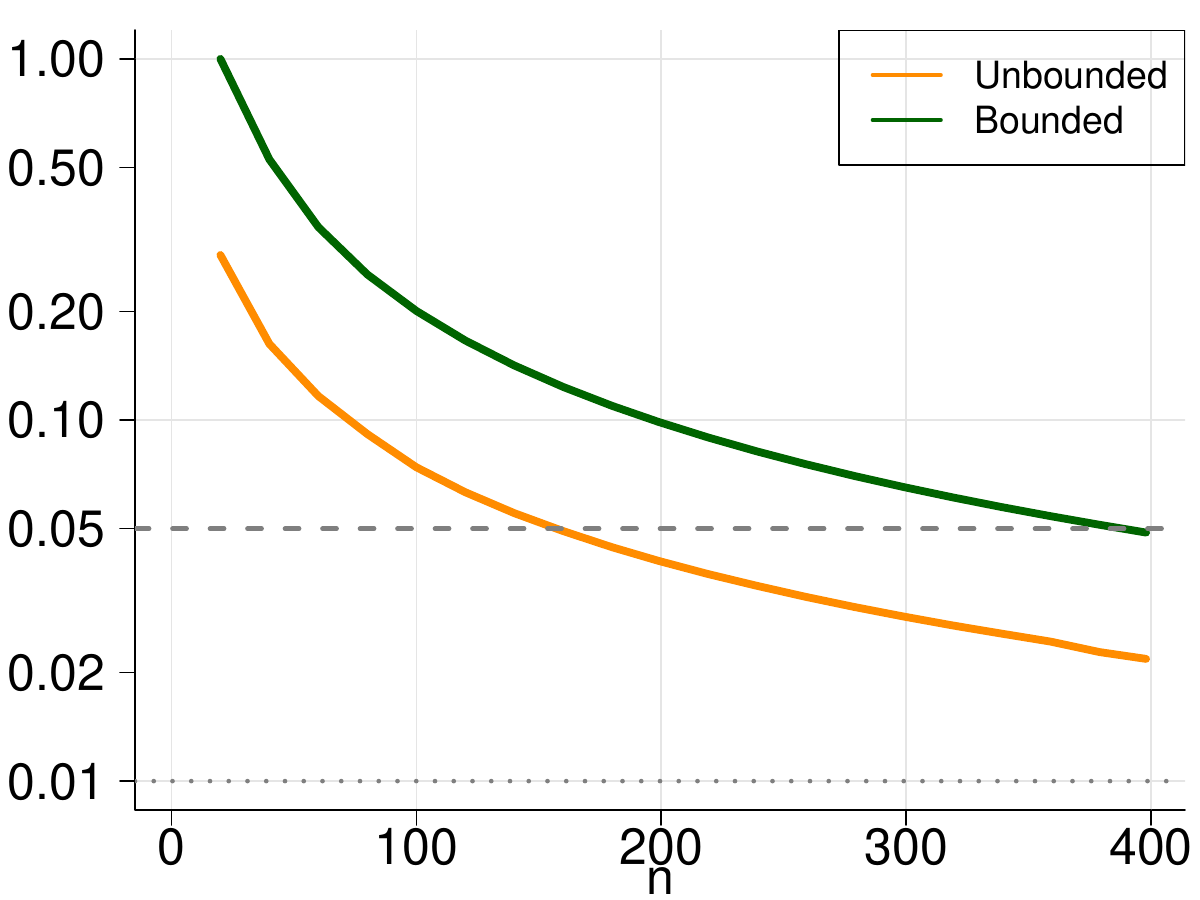}
    \hfill
    \includegraphics[width=0.24\linewidth]{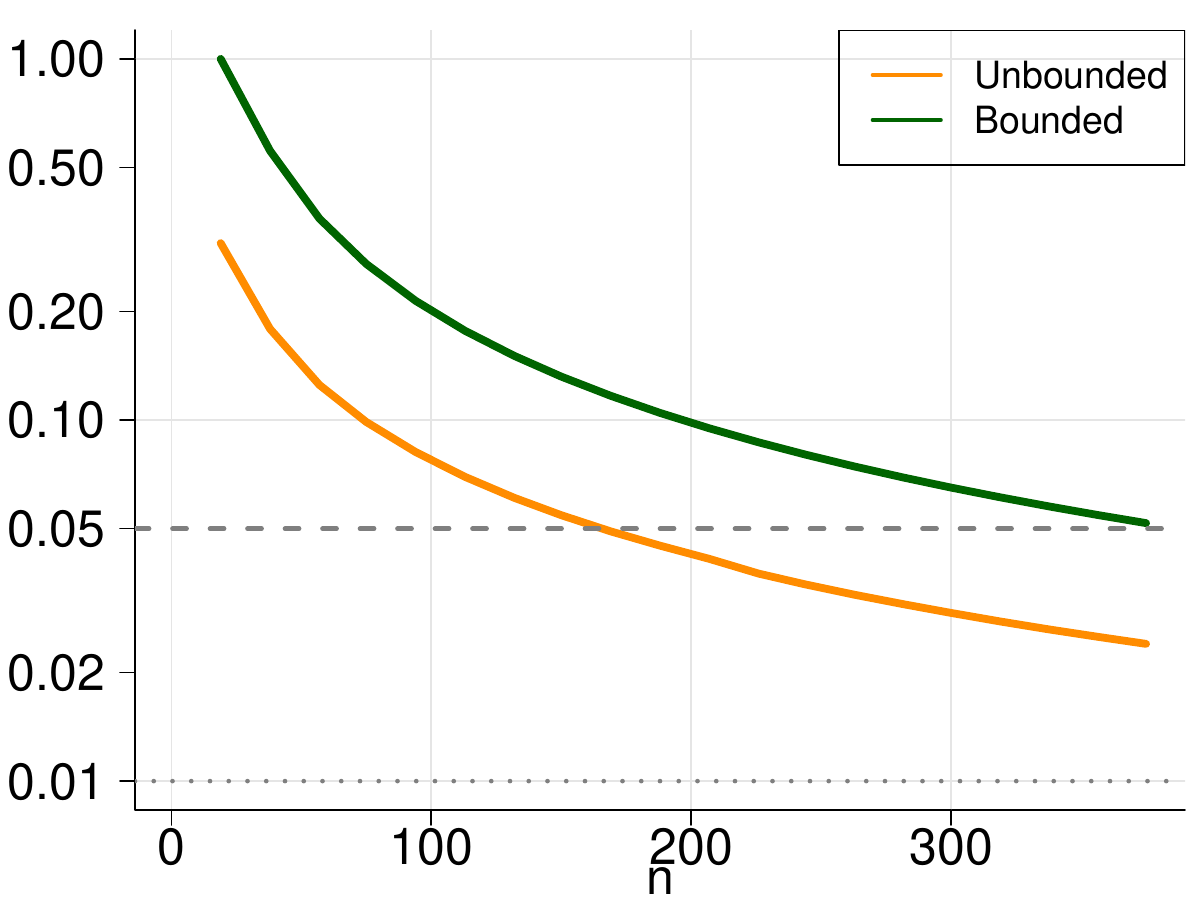}
    \hfill
    \includegraphics[width=0.24\linewidth]{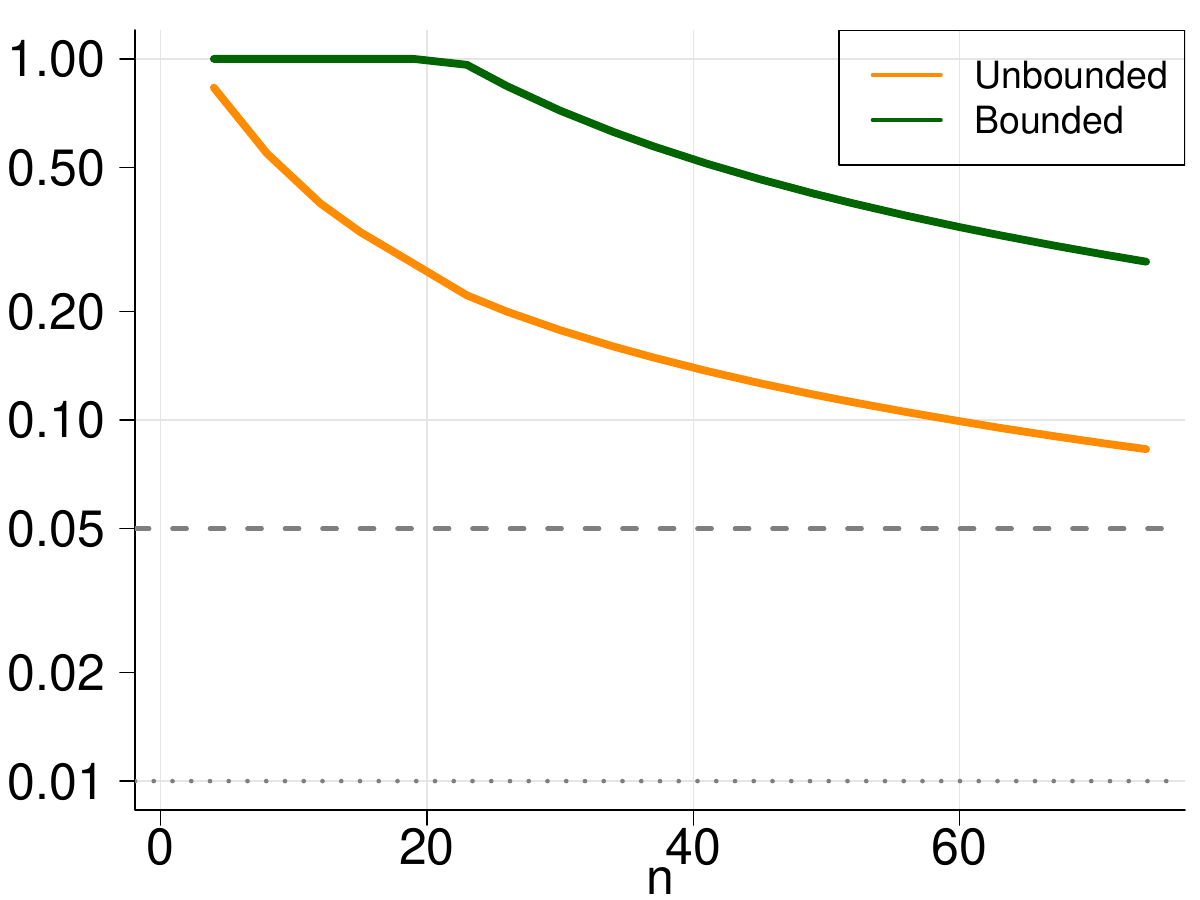}
    \hfill
    \includegraphics[width=0.24\linewidth]{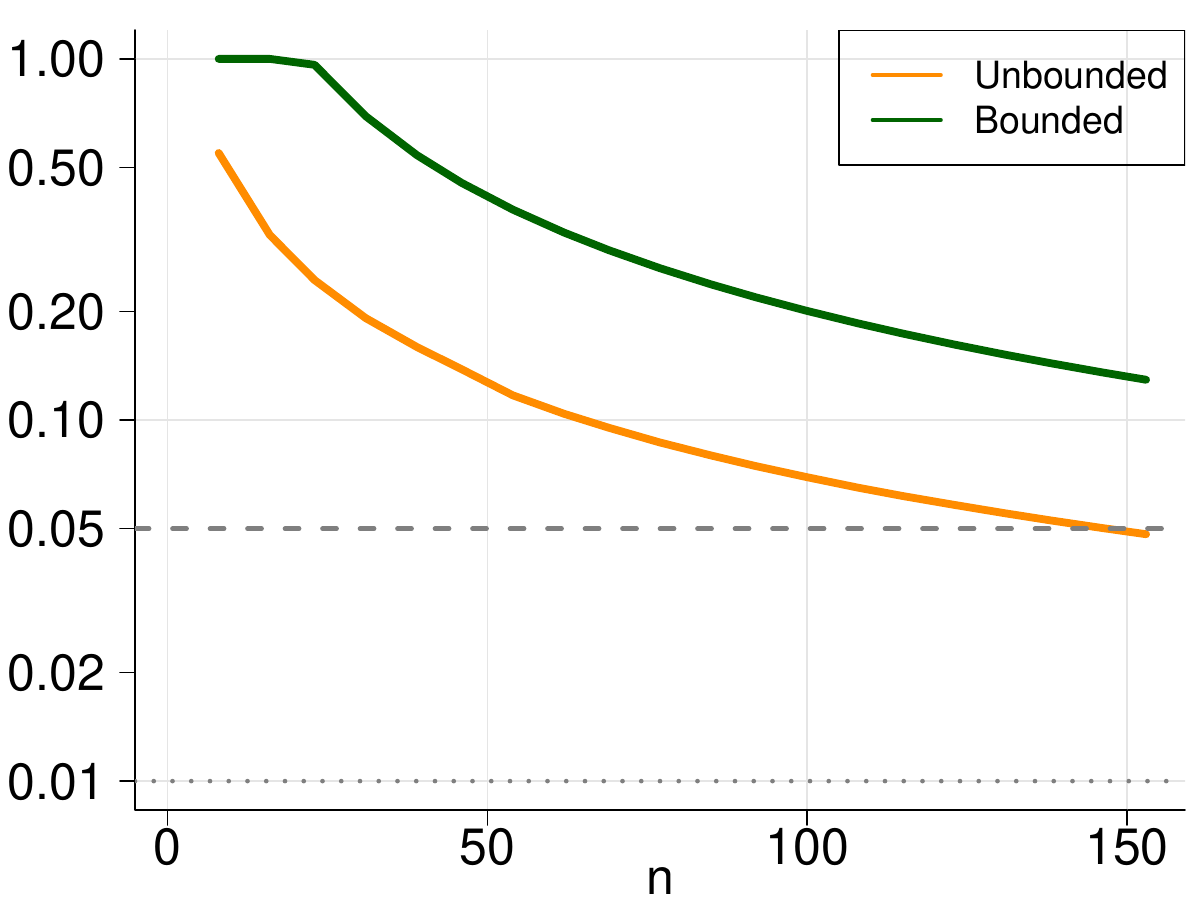} \\
    \includegraphics[width=0.24\linewidth]{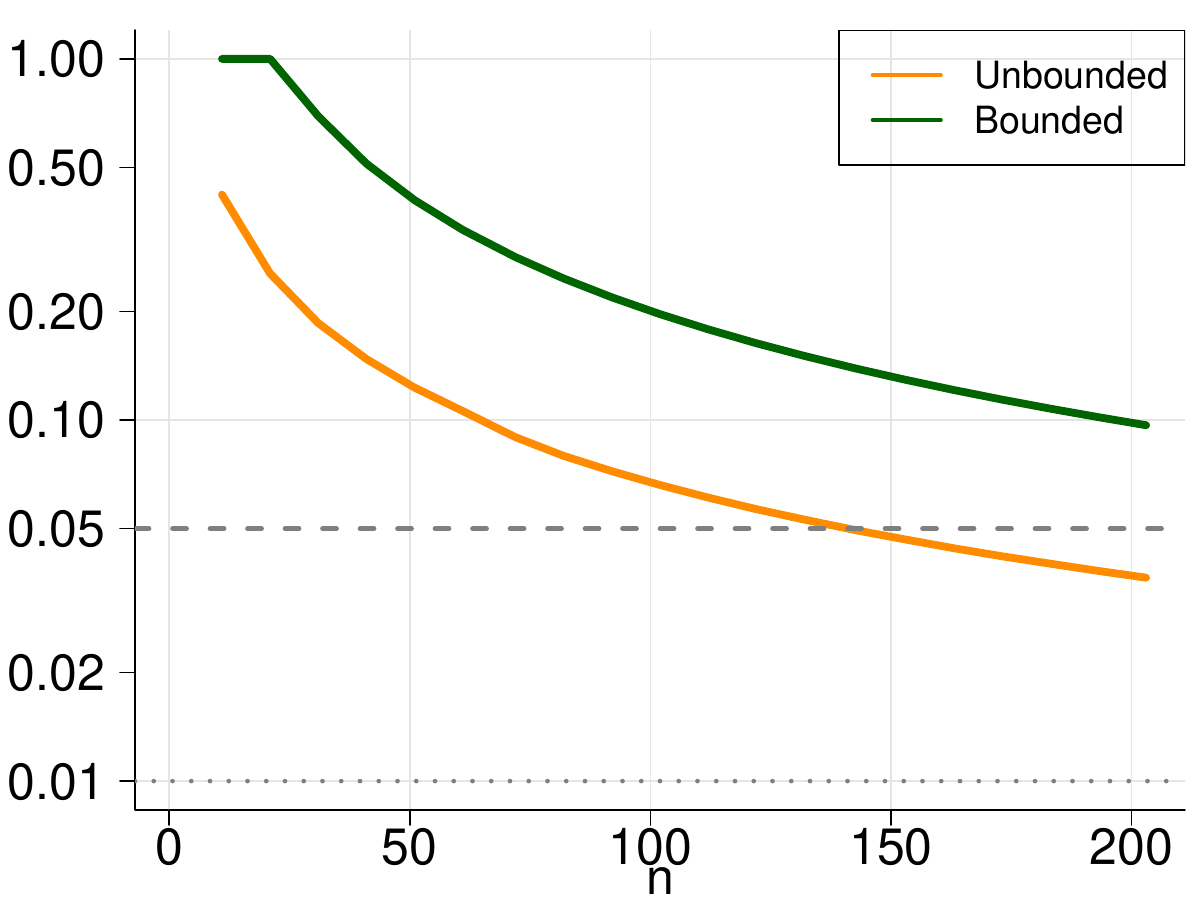}
    \hfill
    \includegraphics[width=0.24\linewidth]{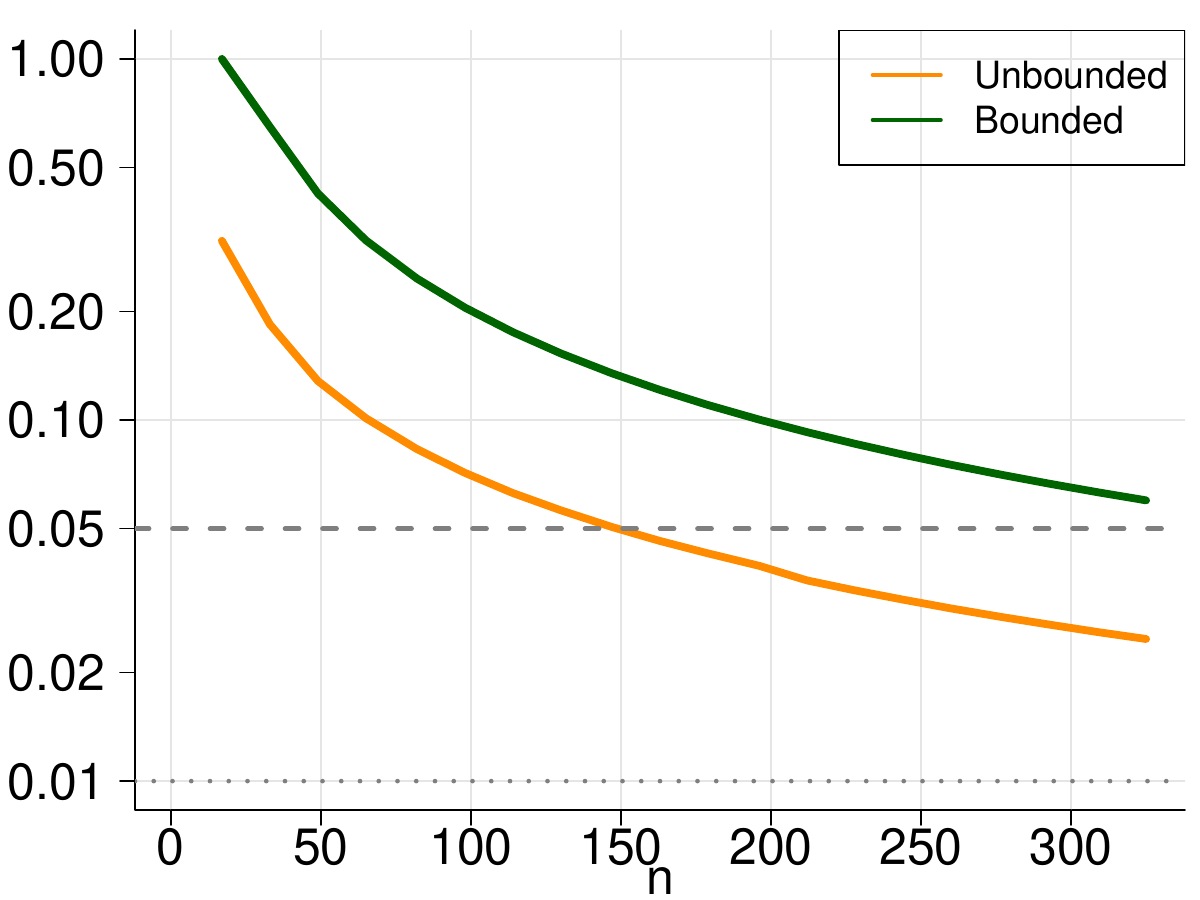}
    \hfill
    \includegraphics[width=0.24\linewidth]{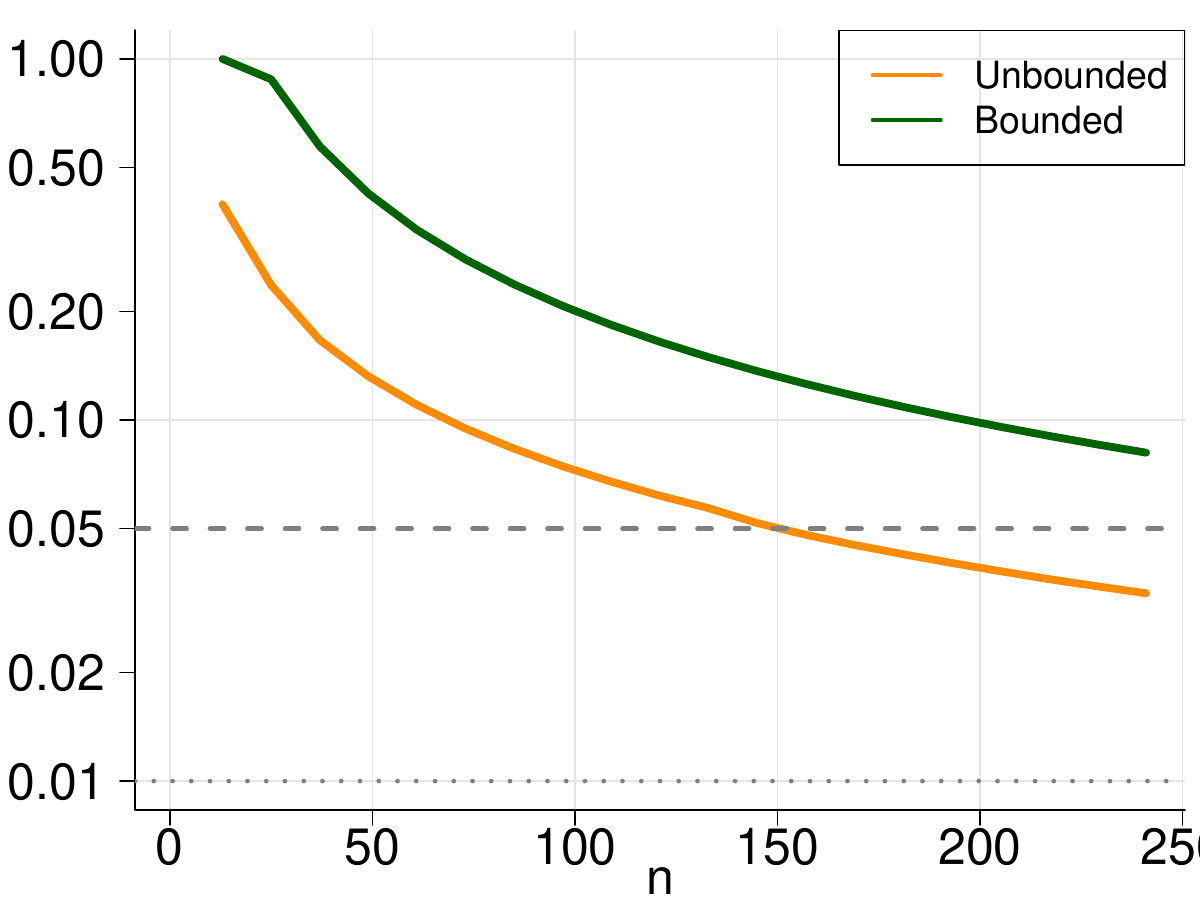}
    \hfill
    \includegraphics[width=0.24\linewidth]{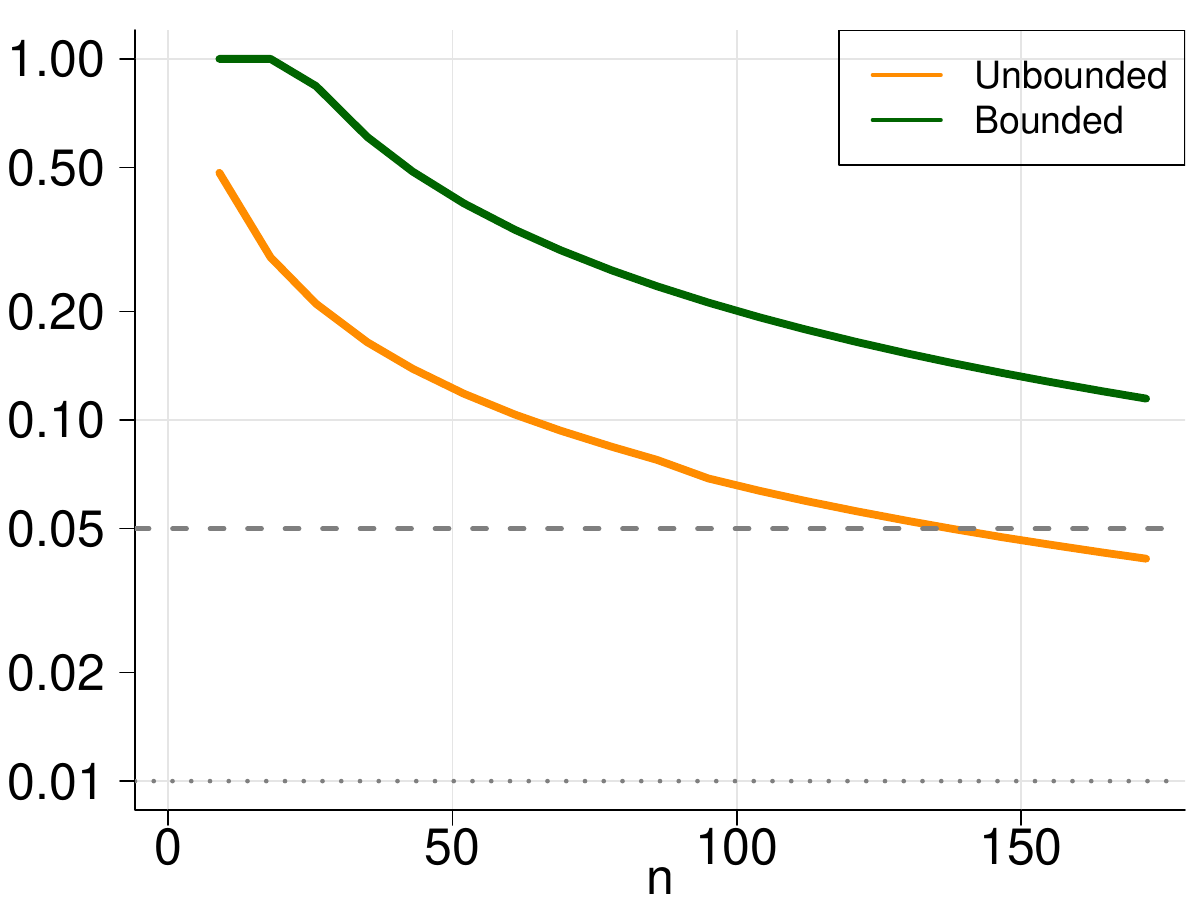} 
    \caption{Length of confidence intervals for  TCGA cancer types as a function of the sample size $n$.
    First row: Glioblastoma Multiforme (GBM), Ovarian Serous Cystadenocarcinoma (OV), Lung Adenocarcinoma (LUAD), Prostate Adenocarcinoma (PRAD), from left to right.
    Second row: Uterine Corpus Endometrial Carcinoma (UCEC), Bladder Urothelial Carcinoma (BLCA), Testicular Germ Cell Tumors (TGCT), Pancreatic Adenocarcinoma (PAAD), from left to right.
    Third row: Kidney Renal Papillary Cell Carcinoma (KIRP), Liver Hepatocellular Carcinoma (LIHC), Cervical Squamous Cell Carcinoma and Endocervical Adenocarcinoma (CESC), Sarcoma (SARC), from left to right.}
    \label{fig:TCGA_fit_others1}
\end{figure}

\begin{figure}[ht!]
    \centering
    \includegraphics[width=0.24\linewidth]{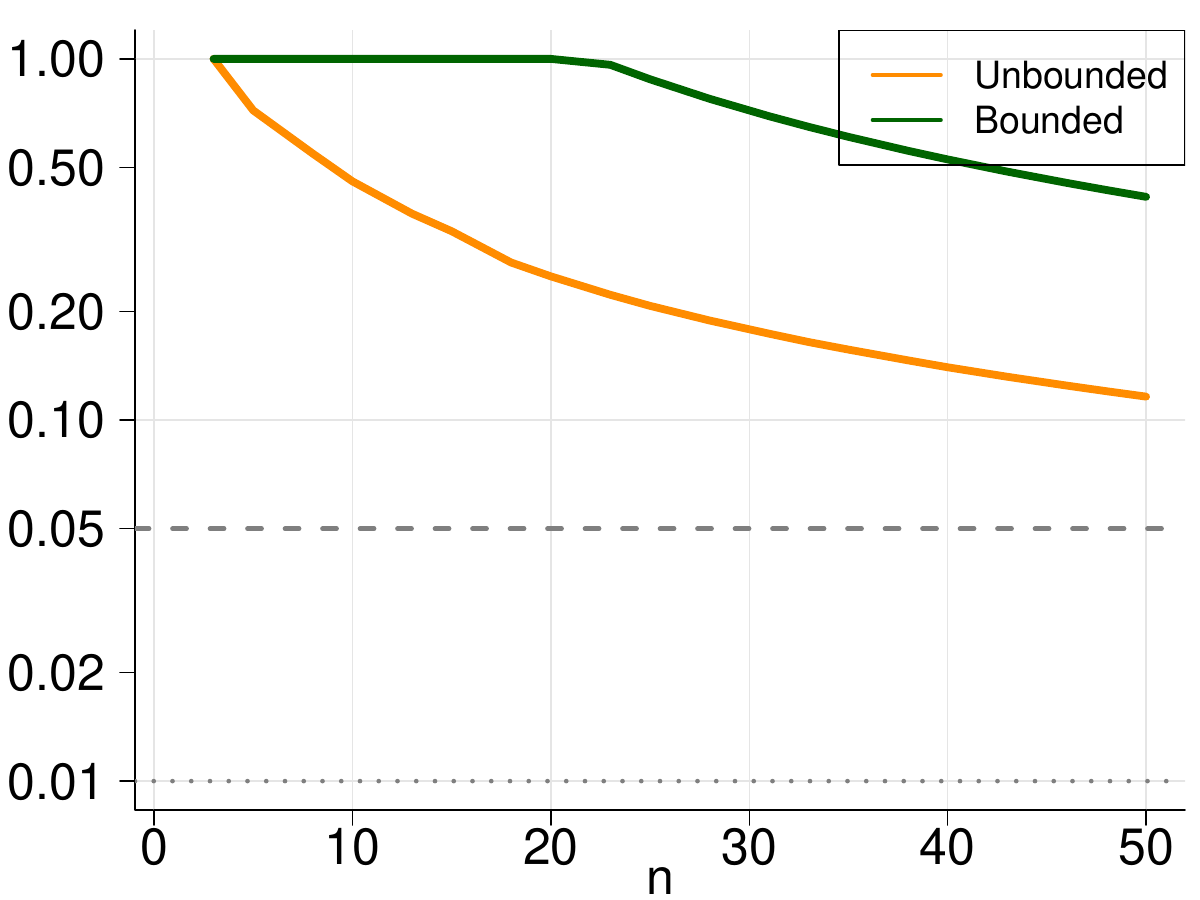}
    \hfill
    \includegraphics[width=0.24\linewidth]{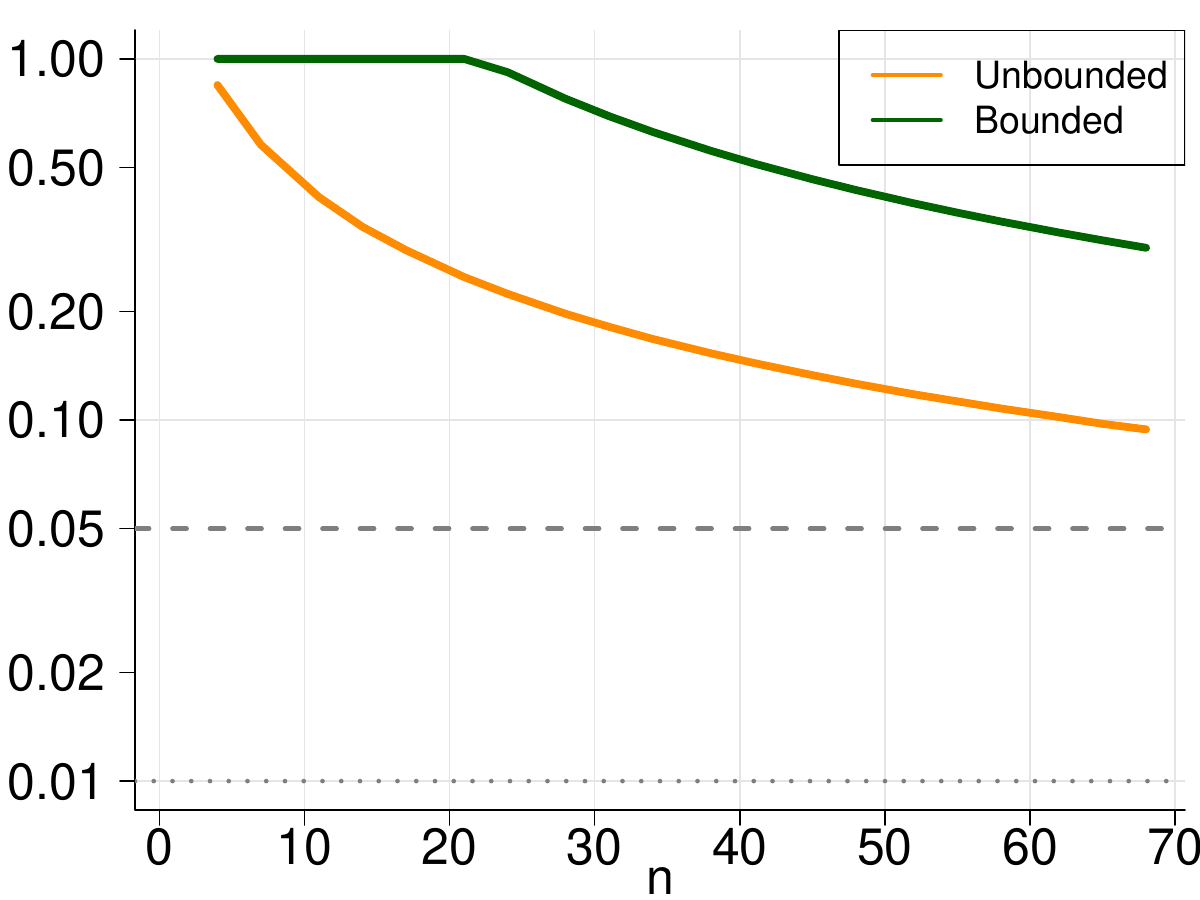}
    \hfill
    \includegraphics[width=0.24\linewidth]{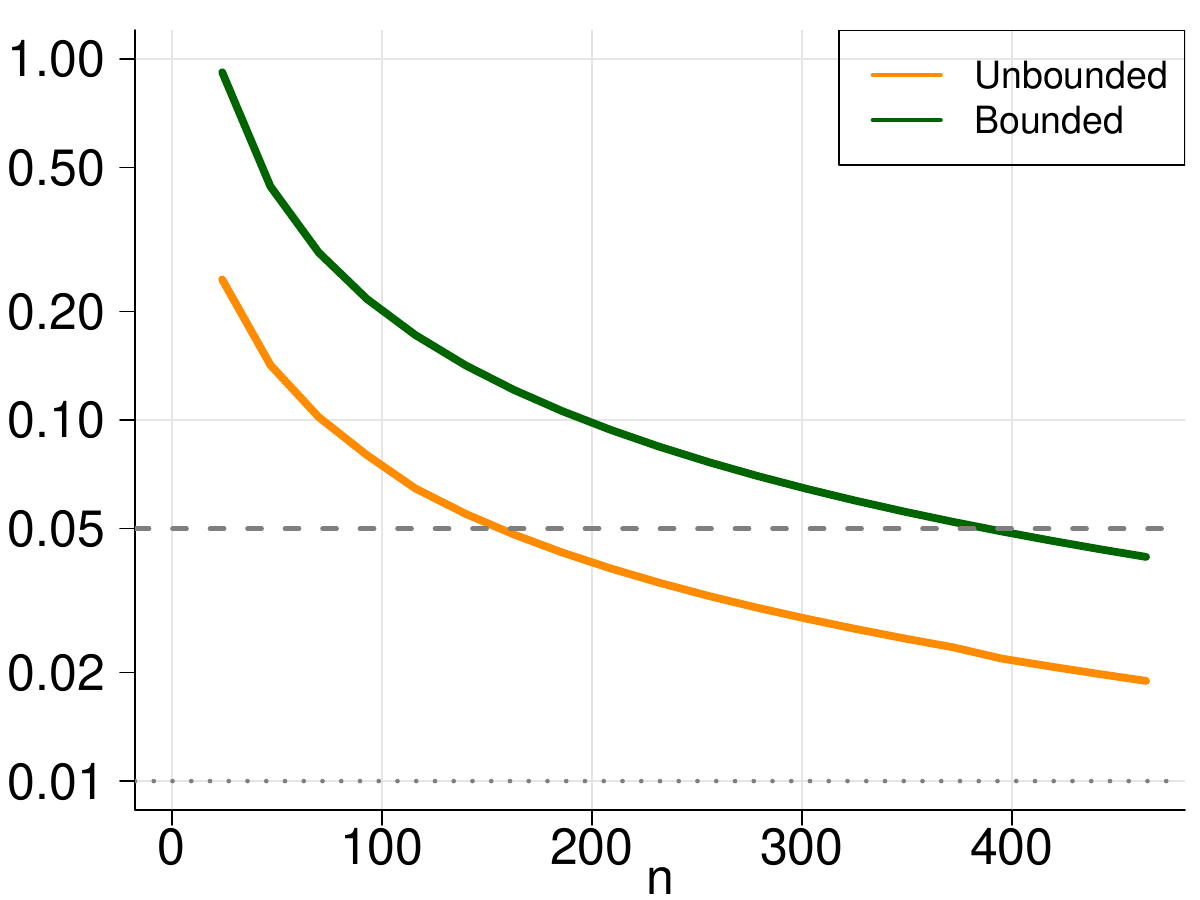}
    \hfill
    \includegraphics[width=0.24\linewidth]{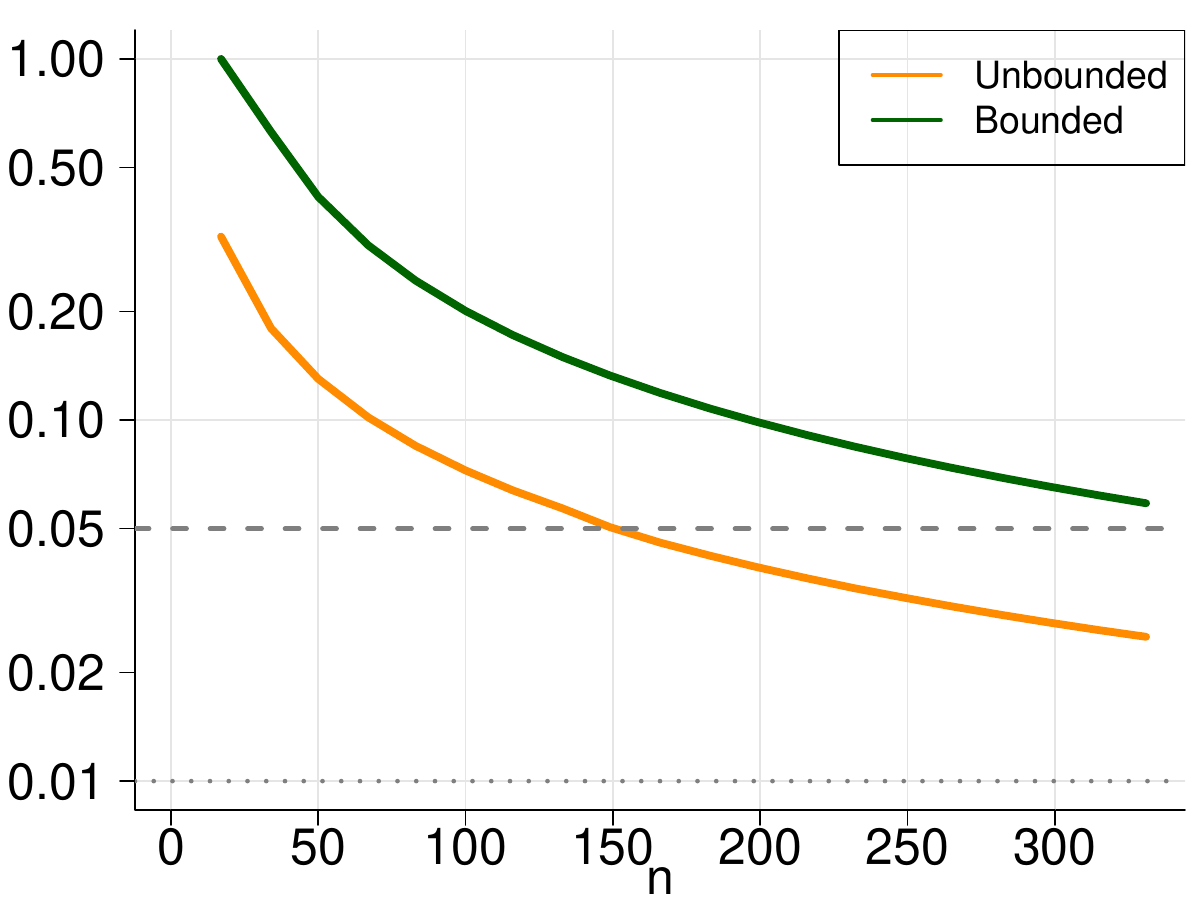} \\
    \includegraphics[width=0.24\linewidth]{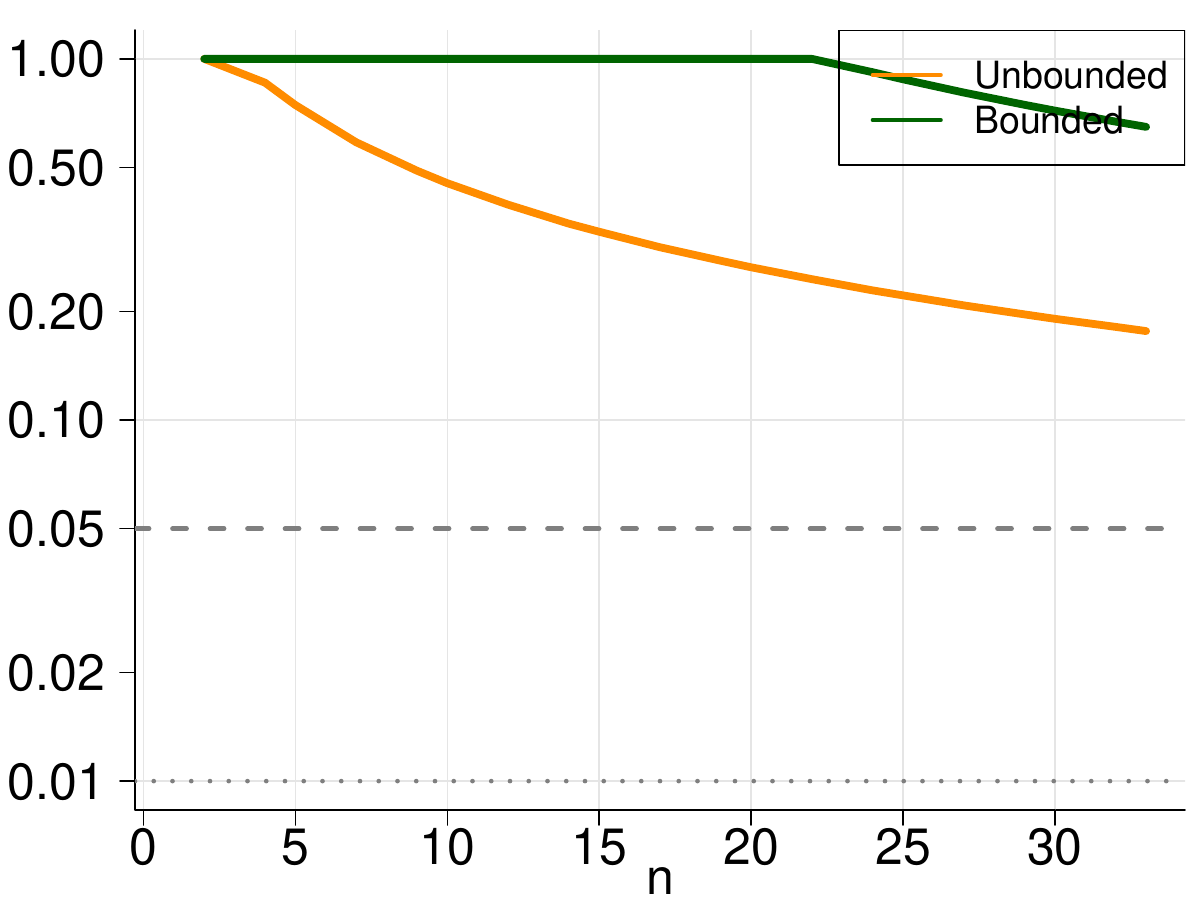}
    \hfill
    \includegraphics[width=0.24\linewidth]{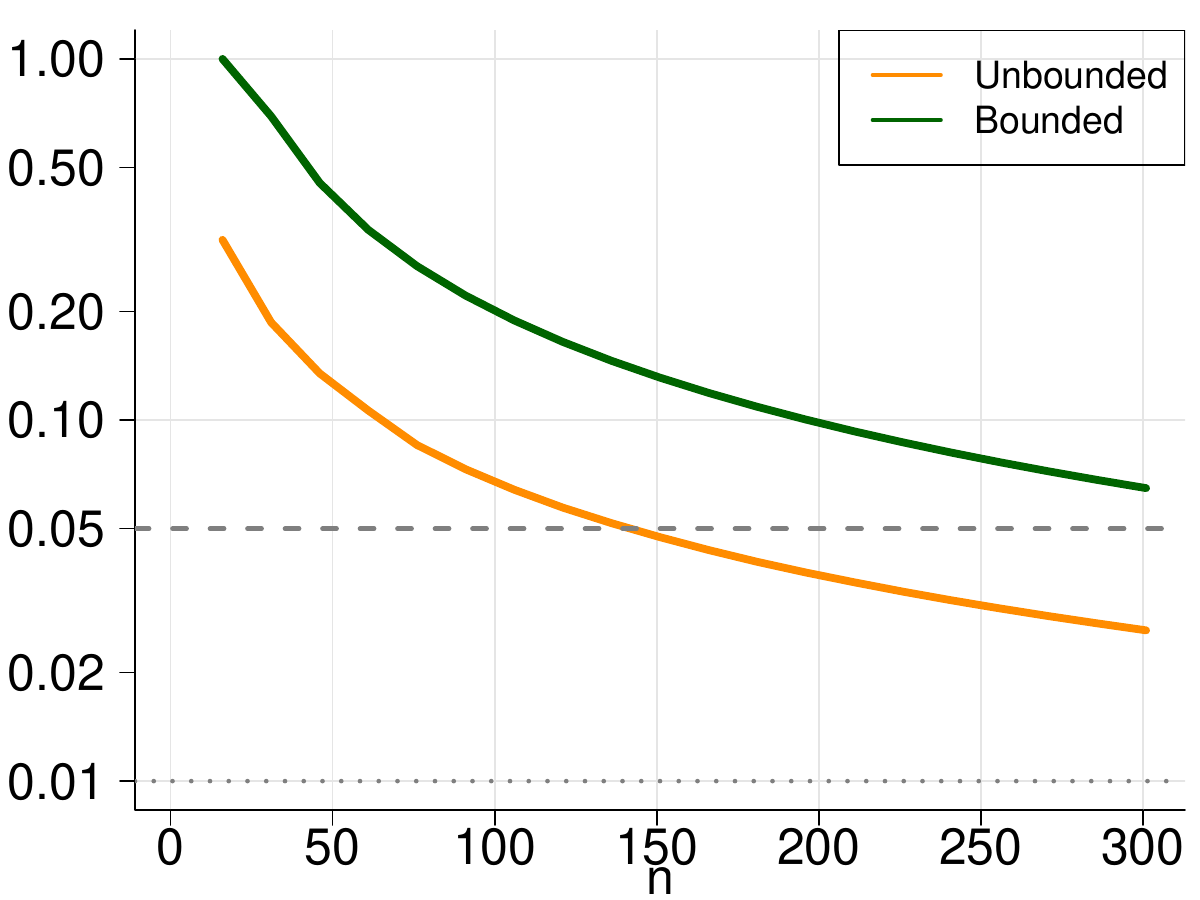}
    \hfill
    \includegraphics[width=0.24\linewidth]{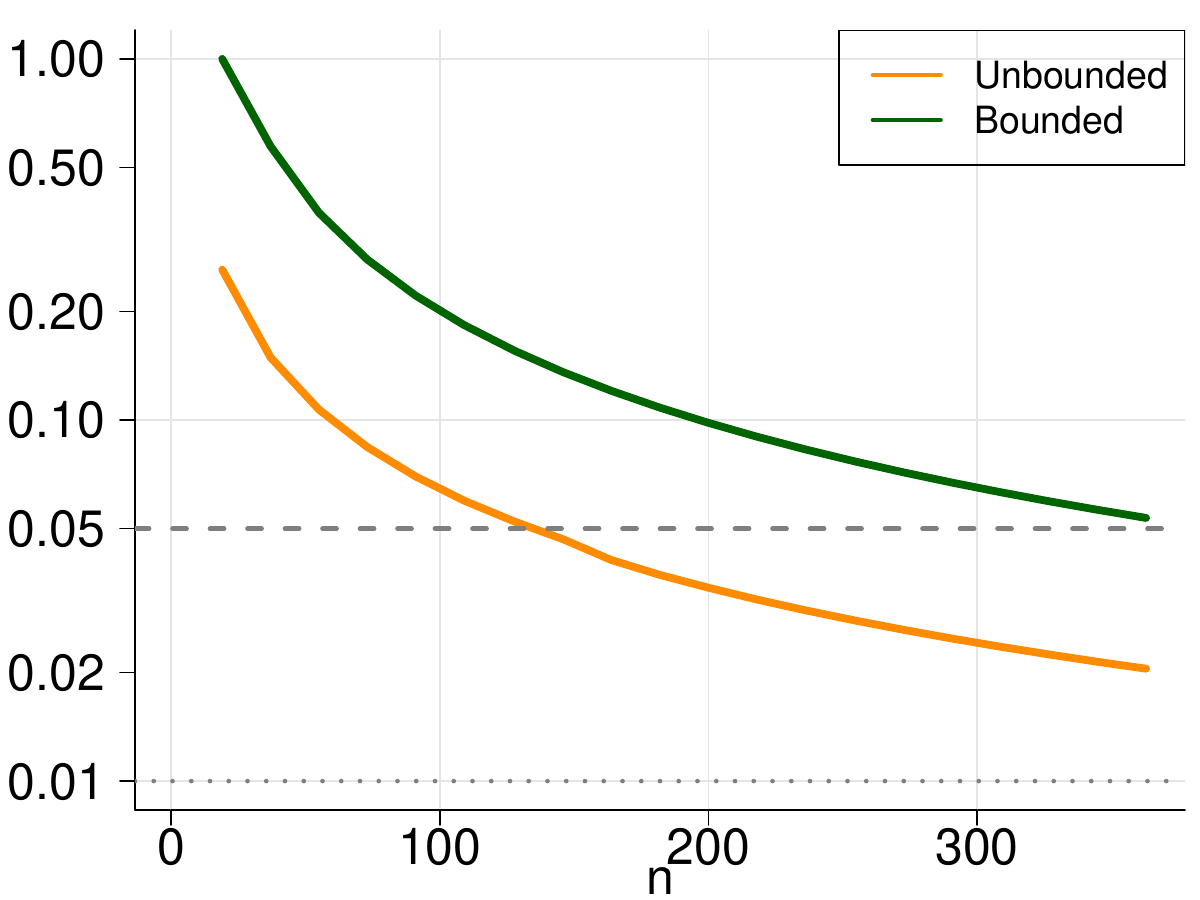}
    \hfill
    \includegraphics[width=0.24\linewidth]{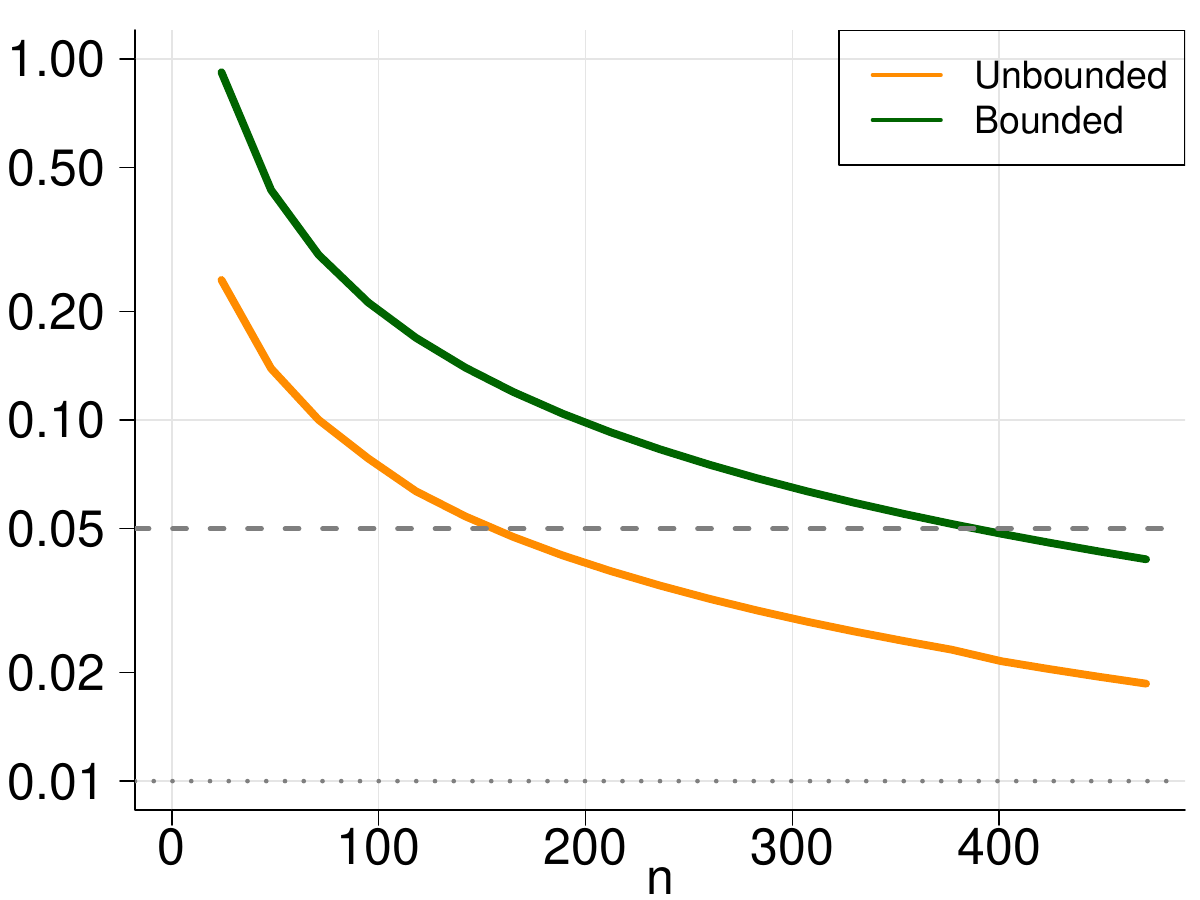} \\
    \includegraphics[width=0.24\linewidth]{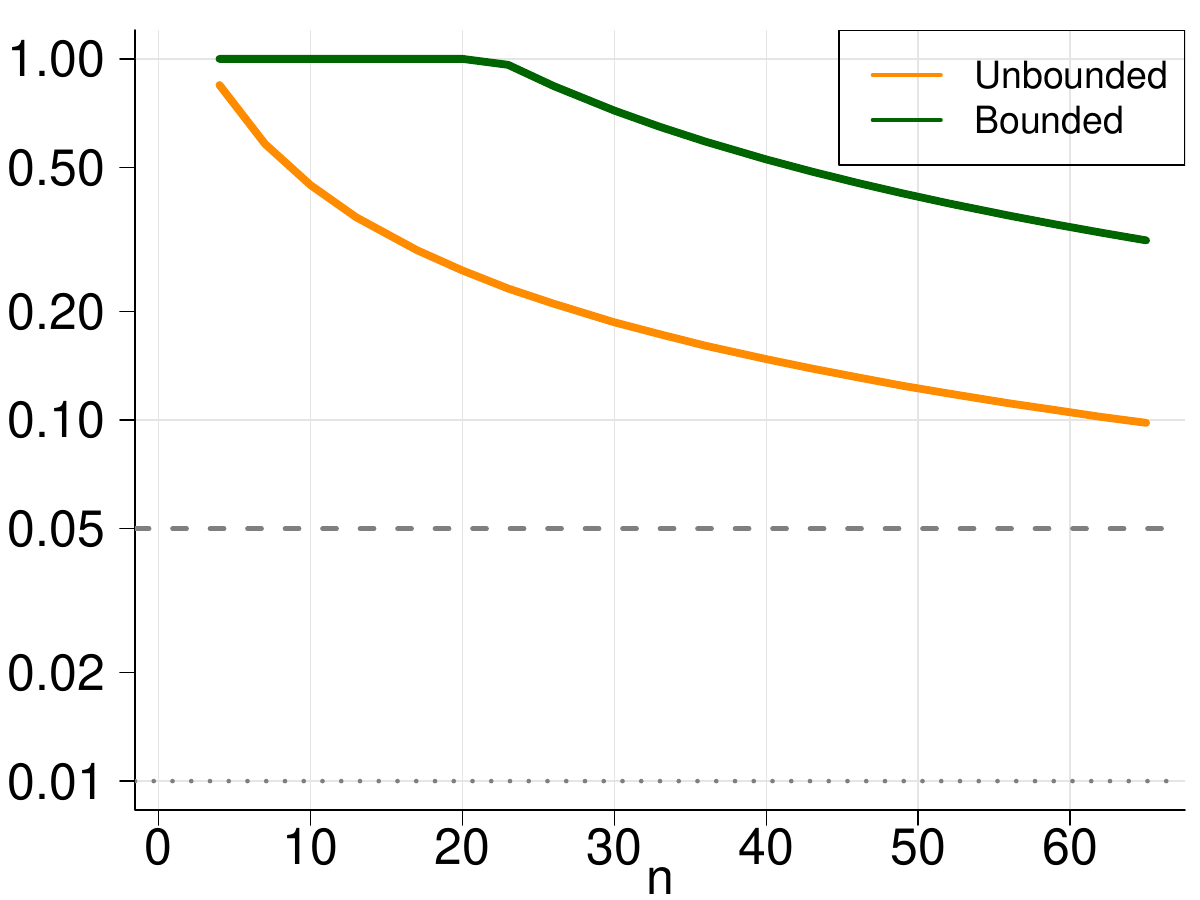}
    \hfill
    \includegraphics[width=0.24\linewidth]{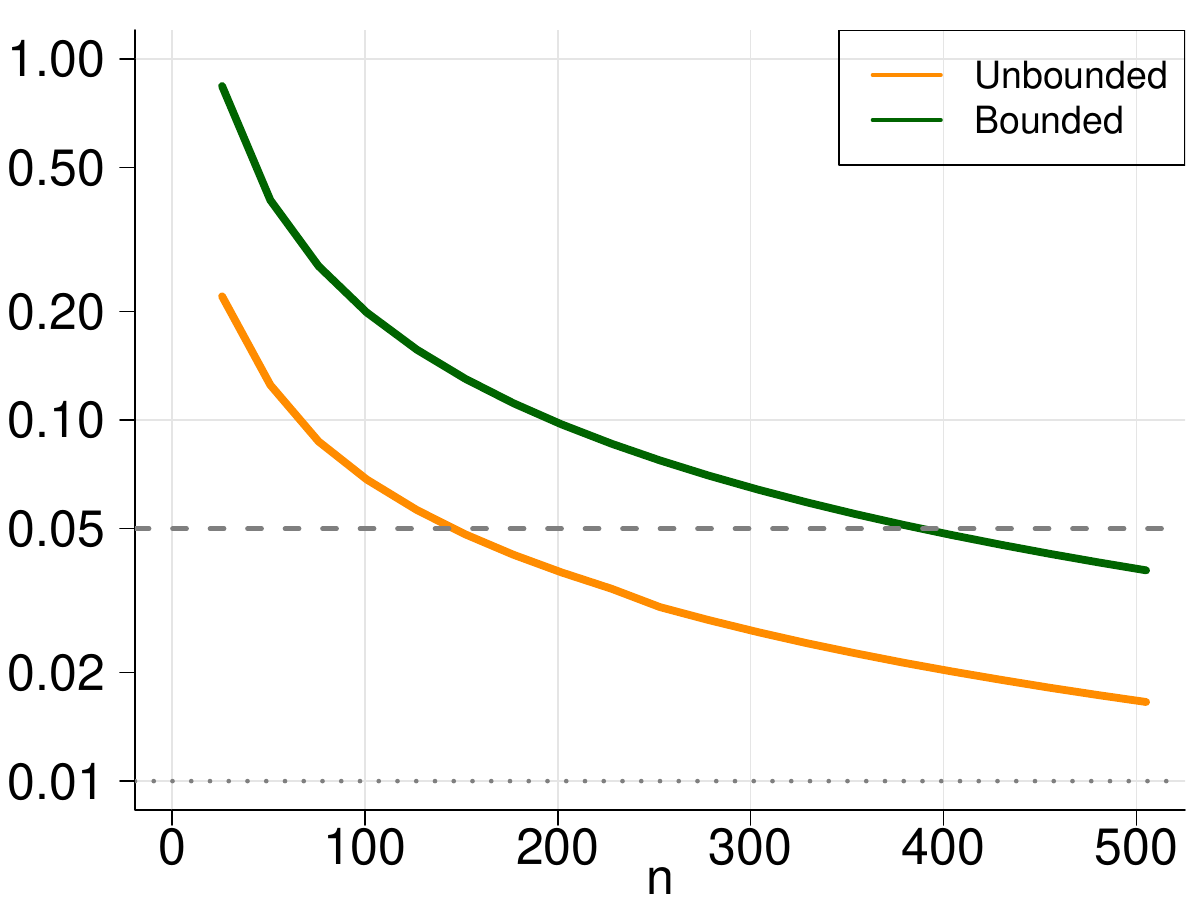}
    \hfill
    \includegraphics[width=0.24\linewidth]{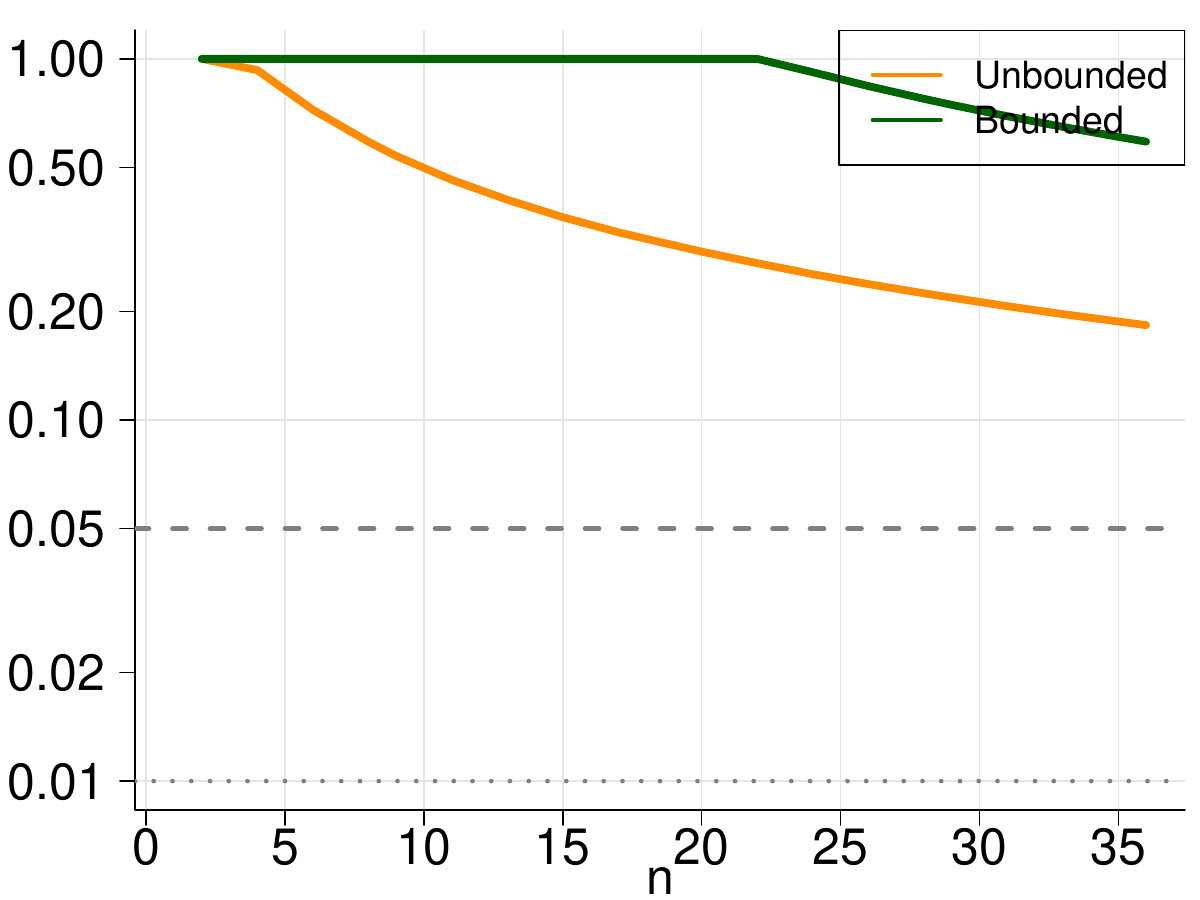}
    \hfill
    \includegraphics[width=0.24\linewidth]{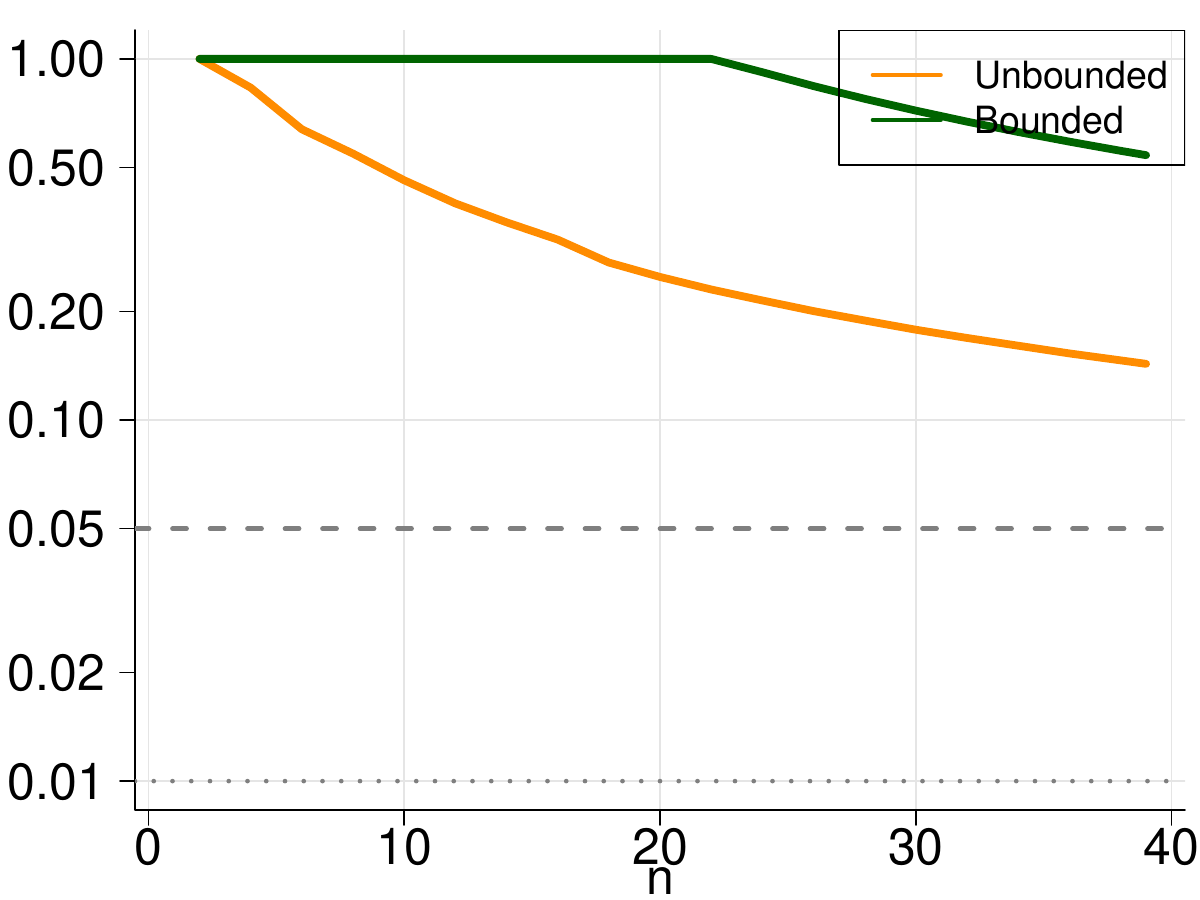} \\
    \includegraphics[width=0.24\linewidth]{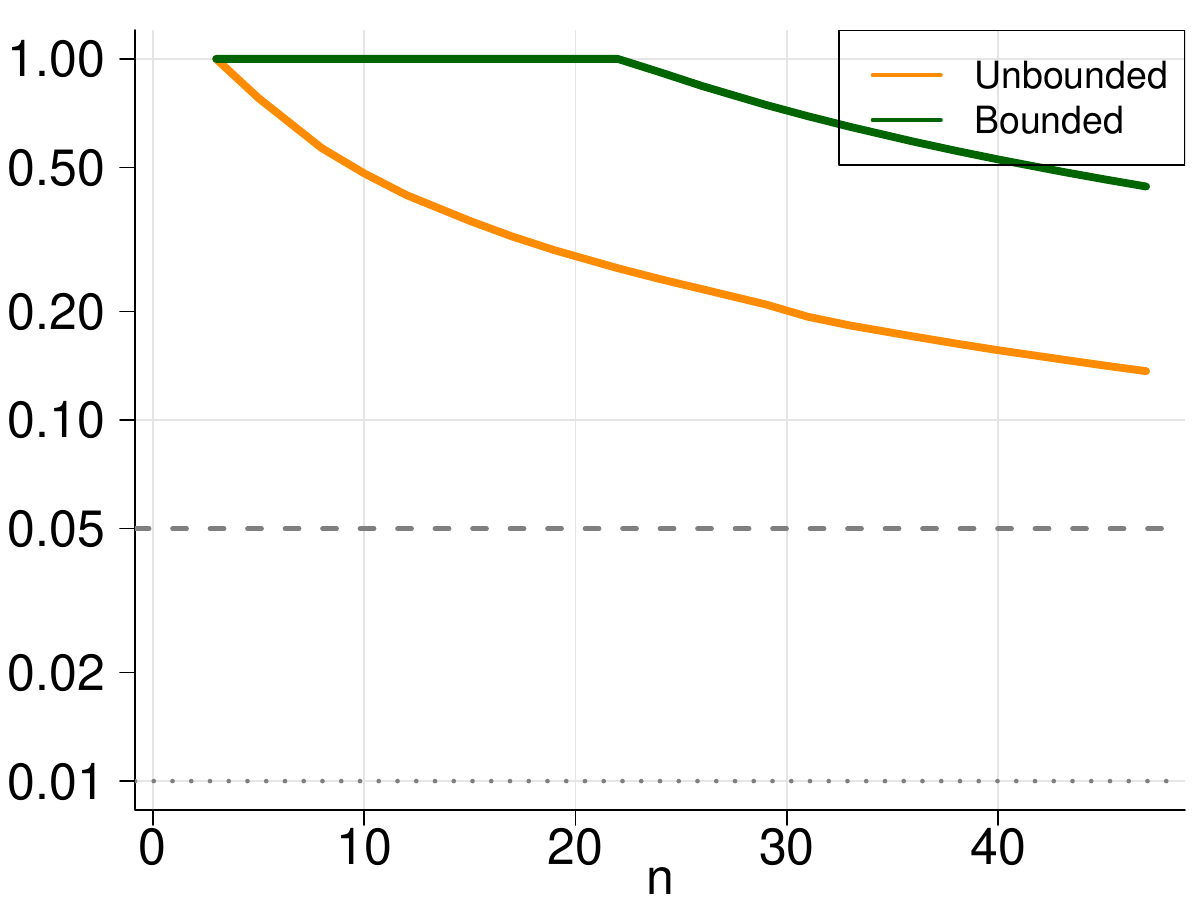}
    \hfill
    \includegraphics[width=0.24\linewidth]{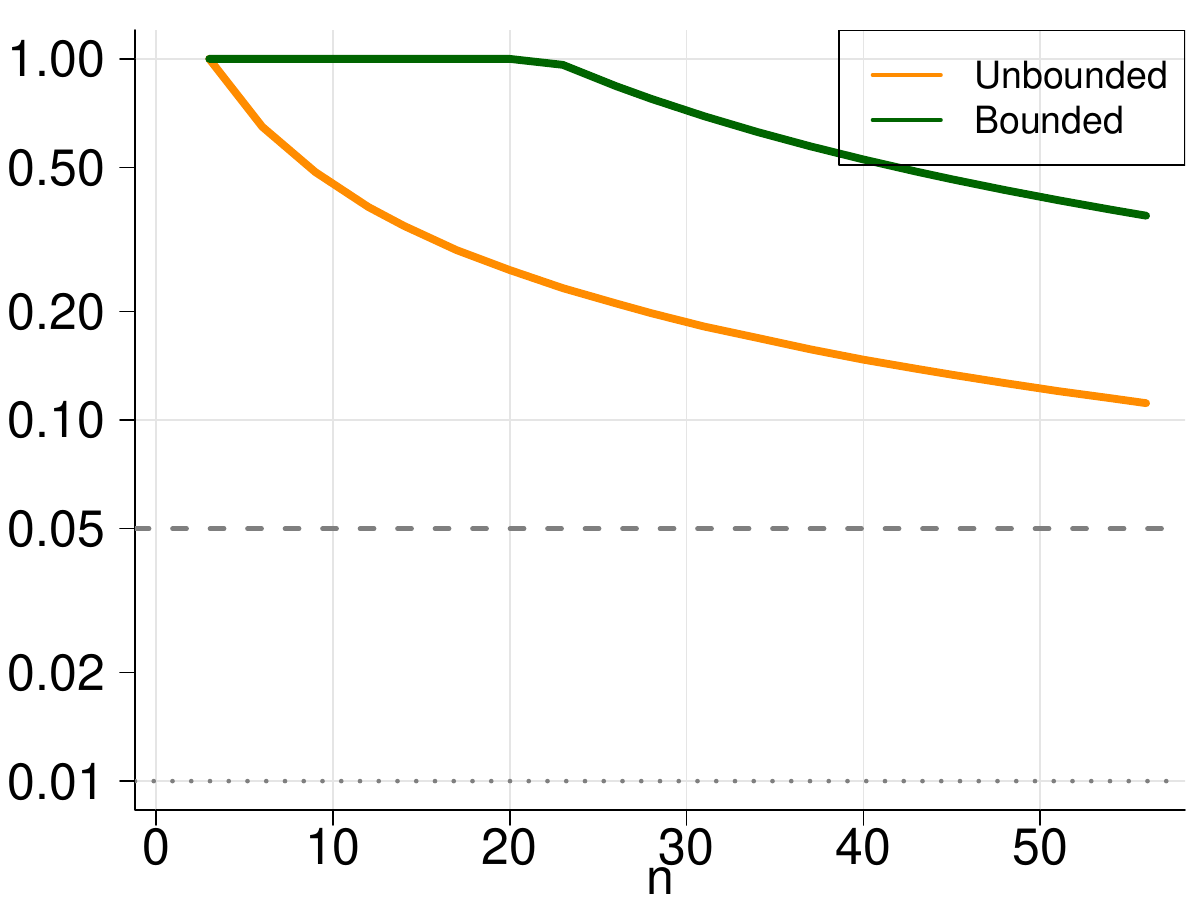}
    \hfill
    \includegraphics[width=0.24\linewidth]{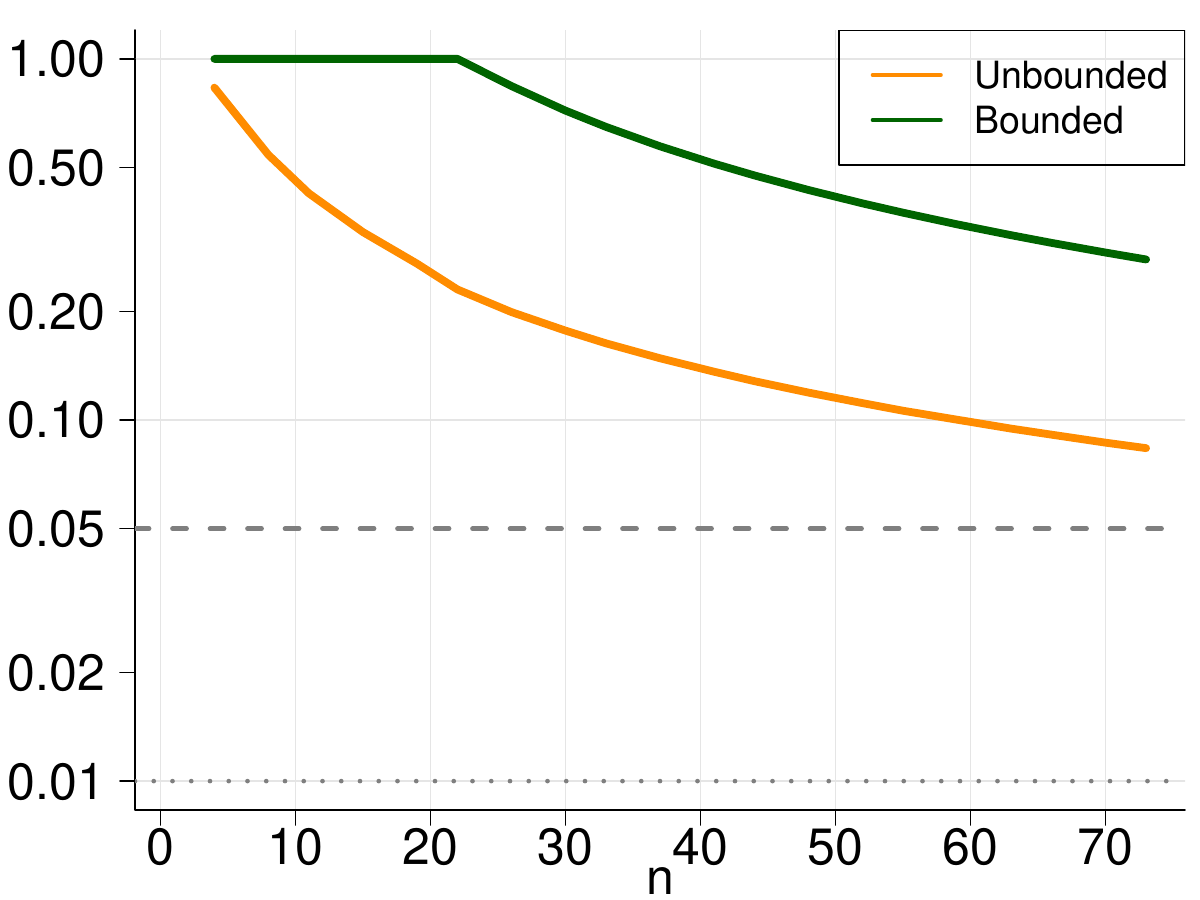}
    \hfill
    \includegraphics[width=0.24\linewidth]{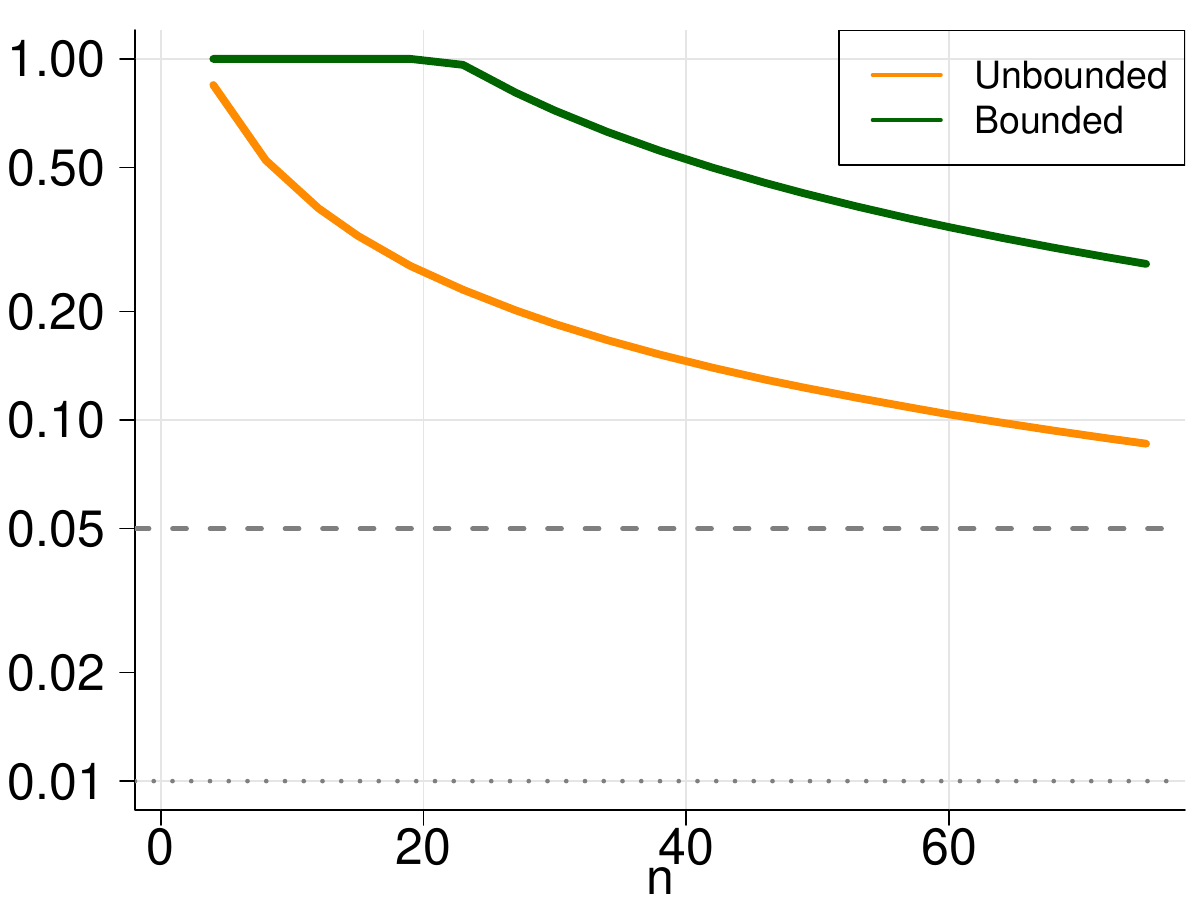} 
    \caption{Length of confidence intervals for TCGA cancer types as a function of the sample size $n$.
    First row: Thymoma (THYM), Mesothelioma (MESO), Colon Adenocarcinoma and Rectum Adenocarcinoma (COADREAD), Stomach Adenocarcinoma (STAD), from left to right.
    Second row: Cholangiocarcinoma (CHOL), Kidney Renal Clear Cell Carcinoma (KIRC), Thyroid Carcinoma (THCA), Head and Neck Squamous Cell Carcinoma (HNSC), from left to right.
    Third row: Acute Myeloid Leukemia (LAML), Brain Lower Grade Glioma (LGG), Diffuse Large B-cell Lymphoma (DLBC), Kidney Chromophobe (KICH), from left to right.
    Fourth row: Uterine Carcinosarcoma (UCS), Adrenocortical Carcinoma (ACC), Pheochromocytoma and Paraganglioma (PCPG), Uveal Melanoma (UVM), from left to right.}
    \label{fig:TCGA_fit_others2}
\end{figure}

\begin{table}[!h]
\centering
\caption{\label{tab:type1}Estimated type I error by scenario, method, and contamination level $q$}
\centering
\begin{tabular}[t]{llrrrrrr}
\toprule
scenario & method & $q=0$ & $q=10^{-4}$ & $q=0.005$ & $q =0.001$ & $q=0.0025$ & $q0.005$ \\
\midrule
$\mathrm{Geom}(0.05)$ & bounded & 0.000 & 0.000 & 0.000 & 0 & 0.000 & 0.000\\
$\mathrm{Geom}(0.05)$ & unbounded & 0.000 & 0.000 & 0.000 & 0 & 0.000 & 0.000\\
$\mathrm{Geom}(0.05)$ & coverage & 1.000 & 1.000 & 1.000 & 1 & 0.995 & 0.970\\
$p_j=0.05$ & bounded & 0.000 & 0.000 & 0.000 & 0 & 0.000 & 0.000\\
$p_j=0.05$ & unbounded & 0.000 & 0.000 & 0.000 & 0 & 0.000 & 0.000\\

$p_j=0.05$ & coverage & 1.000 & 1.000 & 1.000 & 1 & 1.000 & 1.000\\
$p_j=0.006$ & bounded & 0.000 & 0.005 & 0.000 & 0 & 0.000 & 0.000\\
$p_j=0.006$ & unbounded & 0.005 & 0.005 & 0.005 & 0 & 0.005 & 0.005\\
$p_j=0.006$ & coverage & 1.000 & 1.000 & 1.000 & 1 & 1.000 & 1.000\\
Zipf & bounded & 0.000 & 0.000 & 0.000 & 0 & 0.000 & 0.000\\

Zipf & unbounded & 0.000 & 0.000 & 0.005 & 0 & 0.000 & 0.000\\
Zipf & coverage & 0.000 & 0.000 & 0.000 & 0 & 0.000 & 0.000\\
\bottomrule
\end{tabular}
\end{table}

\FloatBarrier


\end{document}